\documentclass[12pt]{article}
\usepackage{graphicx,anysize,amsfonts,amsmath,amsthm,amssymb,enumerate,hyperref,color,setspace,booktabs,enumitem, subcaption}
\usepackage[font = normalsize]{caption}
\usepackage[mathscr]{euscript}
\usepackage[utf8]{inputenc}
\usepackage[english]{babel}
\usepackage{natbib}
\usepackage{moreverb, ragged2e}
\usepackage[font = normalsize]{subcaption}
\newcommand\BibTeX{{\rmfamily B\kern-.05em \textsc{i\kern-.025em b}\kern-.08em
T\kern-.1667em\lower.7ex\hbox{E}\kern-.125emX}}
\newcommand\numberthis{\addtocounter{equation}{1}\tag{\theequation}}
\allowdisplaybreaks
\marginsize{1in}{1in}{1in}{1in}

\usepackage{sectsty}
\sectionfont{\fontsize{12}{15}\selectfont}
\subsectionfont{\fontsize{12}{15}\selectfont}

\usepackage{array}
\newcolumntype{L}[1]{>{\raggedright\let\newline\\\arraybackslash\hspace{0pt}}m{#1}}
\newcolumntype{C}[1]{>{\centering\let\newline\\\arraybackslash\hspace{0pt}}m{#1}}
\newcolumntype{R}[1]{>{\raggedleft\let\newline\\\arraybackslash\hspace{0pt}}m{#1}}
\providecommand{\keywords}[1]
{
  \small	
  \textbf{\textit{Keywords:}} #1
}

\title{\Large \textbf{Rank Intraclass Correlation for Clustered Data}}
\author{\vspace{-1em}\small Shengxin Tu$^{1,*}$, 
Chun Li$^{2}$, 
Donglin Zeng$^{3}$, and
Bryan E. Shepherd$^{1}$\\\vspace{-1em}
\small $^{1}$Department of Biostatistics, Vanderbilt University, Nashville, Tennessee, USA\\\vspace{-1em}
\small $^{2}$Department of Population and Public Health Sciences, University of Southern \small California,\\\vspace{-1em} \small Los Angeles, California, USA\\\vspace{-1em}
\small $^{3}$Department of Biostatistics, University of North Carolina at Chapel Hill, \\ \vspace{-1em}\small Chapel Hill, North Carolina, USA\\ \small $^{*}$shengxin.tu@vanderbilt.edu.}
\date{\vspace{-2ex}}

\begin{document}
\maketitle
\begin{abstract}
Clustered data are common in biomedical research. Observations in the same cluster are often more similar to each other than to observations from other clusters. The intraclass correlation coefficient (ICC), first introduced by R. A. Fisher, is frequently used to measure this degree of similarity. However, the ICC is sensitive to extreme values and skewed distributions, and depends on the scale of the data. It is also not applicable to ordered categorical data. We define the rank ICC as a natural extension of Fisher's ICC to the rank scale, and describe its corresponding population parameter. The rank ICC is simply interpreted as the rank correlation between a random pair of observations from the same cluster. We also extend the definition when the underlying distribution has more than two hierarchies. We describe estimation and inference procedures, show the asymptotic properties of our estimator, conduct simulations to evaluate its performance, and illustrate our method in three real data examples with skewed data, count data, and three-level ordered categorical data.
\end{abstract}
\clearpage 

\pagenumbering{arabic} 

\section{Introduction}
\label{s:intro}

With clustered data, observations in the same cluster are often more similar to each other than to those from other clusters. The degree of similarity is frequently measured by the intraclass correlation coefficient (ICC). R. A. Fisher first introduced the ICC to assess familial resemblance of a trait between siblings \citep{fisher1925}. The ICC has since been used in various disciplines including epidemiology, genetics, and psychology. It also has been employed in clinical trial design \citep{murray2004,hedges2007}. Fisher's ICC measures the correlation between a random pair of observations from a random cluster. When the cluster size is infinite, Fisher's ICC is equal to the variance of cluster means divided by the total variance \citep{Harris1913}. Because of this, the ICC has also been estimated with random effects models, in which it is estimated as the proportion of total variance attributable to the clusters \citep{Shrout1979, Donner1986}.

The ICC is fundamental to the analysis of clustered data. However, similar to Pearson's correlation, it is sensitive to extreme values and skewed distributions, and it depends on the scale of the data. When a variable is transformed to a different scale, the ICC may change. For some non-Gaussian distributions, the ICC might be estimated using generalized linear random effects models. In this case, the ICC is defined on the link function transformed scale and it may be sensitive to the non-normality of random effects or the method used to derive the within-cluster variance \citep{nakagawa2017}. The ICC is also not applicable to ordered categorical data. For ordered categorical data, ordinal regression models with random effects may be used to estimate variance components, but the total variance is undefined unless numbers are assigned to levels of the ordinal response \citep{hallgren2012, denham2016}. %Hence, it is not clear how to derive the ICC from ordinal random effects models.

Several studies have proposed nonparametric measures to evaluate intraclass similarity based on the notion of concordance. One measure is the probability that a random observation from a cluster does not fall between a random pair of observations from a different cluster \citep{Rothery1979}. Another measure is the probability that a random pair of observations from a cluster does not fall between two random observations each from a different cluster \citep{Shirahata1981}. Shirahata (1982) performed comparisons between the two measures and a modification of Kendall's measure of dependence \citep{Shirahata1982}. All three measures are rank-based; however, they are probabilities of concordance and do not share the same spirit as Fisher’s ICC, which is a correlation measure. Methods to estimate the ICC for categorical data have been developed \citep{rishi2021}, but they ignore the order information when applied to ordered categorical data.

In this paper, we define the rank ICC as a natural extension of Fisher's ICC to the rank scale. We provide its population parameter and extend it when the underlying distribution has more than two hierarchies. Our estimator of the rank ICC is insensitive to extreme values and skewed distributions, and does not depend on the scale of the data. It can be used for ordered categorical variables. We also show that our estimator is consistent and asymptotically normal.

This paper is organized as follows. Section 2 introduces population parameters for the rank ICC. Section 3 contains estimation and inference. Section 4 presents simulations evaluating the performance of our estimator. Section 5 illustrates our method in three applications. Section 6 provides a discussion. Proofs of consistency and asymptotic normality are in the Supporting Information. We have developed an R package, \texttt{rankICC}, available on CRAN, which implements our new method. The R script for the three application examples and simulations is on our Github page, https://github.com/shengxintu/rankICC. 

\section{Population Parameters}
\label{s:parameters}
\subsection{Two Hierarchies}

Consider a two-level hierarchical distribution. A random variable from the distribution is denoted as $X_{ij}$, where $i$ represents the cluster it belongs to and $j$ is the index within cluster $i$. Fisher defined the ICC as the correlation between a random pair from the same cluster; that is, $\rho_I = corr(X_{ij}, X_{ij'})$, where $j\neq j'$, indicating that two different observations are drawn from cluster $i$. For a continuous hierarchical distribution, the ICC has also been expressed as the ratio of the between-cluster variance to the total variance \citep{filler1951}; $\rho_{Ir} = \sigma^2_b/(\sigma^2_b + \sigma^2_w)$, where $\sigma^2_b$ is the between-cluster variance (i.e., the variance of cluster means), and $\sigma^2_w$ is the within-cluster variance (i.e., the mean of within-cluster variances). These two definitions are equivalent only when cluster sizes are infinite. In general, the relationship between these two definitions is 
\begin{align*}
    \rho_I & = cov(X_{ij}, X_{ij'}) \big/ \sqrt{var(X_{ij})var(X_{ij'})}\\
    & = \{cov[E(X_{ij}|\mu_i), E(X_{ij'}|\mu_i)] + E[cov(X_{ij},X_{ij'}|\mu_i)]\}\big/(\sigma^2_b + \sigma^2_w)\\
    & = \{cov(\mu_i, \mu_i) + E[cov(X_{ij},X_{ij'}|\mu_i)]\}\big/(\sigma^2_b + \sigma^2_w)\\
    & = \rho_{Ir} + E[cov(X_{ij},X_{ij'}|\mu_i)]\big/(\sigma^2_b + \sigma^2_w),\numberthis \label{rhoI}
\end{align*} where $\mu_i$ is a random variable representing the mean of cluster $i$. If cluster sizes are finite, $\rho_{Ir} > \rho_I$ because $E[cov(X_{ij},X_{ij'}|\mu_i)]$ in (\ref{rhoI}) is negative. With equal cluster sizes of $m$, the value of $\rho_I$ is constrained between $-1/(m-1)$ and 1 \citep{fisher1925}. Note that $\rho_I$ can be negative when cluster sizes are finite, whereas $\rho_{Ir}$ is always non-negative. While $\rho_I$ is a correlation measure, $\rho_{Ir}$ is a measure of the fraction of total variance attributable to cluster means. Hence, $\rho_I$ is a more general measure of the intraclass correlation.

The rank ICC, to be defined below, is the rank-based version of Fisher's ICC, similar to Spearman's rank correlation which is the rank-based version of Pearson's correlation \citep{kruskal1958}. The relationship between the population parameters of Fisher's ICC and the rank ICC is identical to the relationship between those of Pearson's correlation and Spearman's rank correlation. The population parameters of Fisher's ICC and Pearson's correlation are correlations on the original scale of the variables, while the population parameter of Spearman's rank correlation is the grade correlation (i.e., the correlation between CDFs) for continuous variables \citep{kruskal1958} or more generally, the correlation of the population versions of midranks or ridits \citep{Bross1958, kendall1970}.

Let $F$ be the CDF of the two-level hierarchical distribution. Let $F(x-)=\lim_{t \uparrow x} F(t)$ and $F^*(x) = \{F(x) + F(x-)\}/2$. The population version of the rank ICC is defined as 
\begin{equation}
\gamma_I = corr[F^*(X_{ij}), F^*(X_{ij'})],  
\label{gammaI}
\end{equation}
where $(X_{ij}, X_{ij'})$ is a random pair drawn from a random cluster and $j\neq j'$. If $X$ is continuous, $\gamma_I = 12cov[F(X_{ij}), F(X_{ij'})]$, because $F^*(X) = F(X) \sim \text{Unif}(0,1)$ and its variance is $1/12$. If $X$ has a discrete or mixture distribution, $F^*(X_{ij})$ corresponds to the population version of ridits \citep{Bross1958}. The rank ICC $\gamma_I$ given in (\ref{gammaI}) is therefore Fisher's ICC on the cumulative probability scale. The rank ICC has the same boundaries as Fisher's ICC and can be negative with finite cluster sizes. 

\subsection{Multiple Hierarchies}

We extend the definition of the rank ICC to multiple hierarchies. For ease of understanding, we begin with three hierarchies. Starting from the innermost level, the three levels are named level 1, level 2, and level 3. One example is a population of schools, in which there are different classrooms and different students within each classroom. Here level 1 is the student, level 2 is the classroom, and level 3 is the school. Correlation may exist within both level-2 and level-3 units. A random variable drawn from a three-level hierarchical distribution is denoted as $X_{ijk}$, where $i$, $j$, and $k$ are indices for levels 3, 2, and 1, respectively. Let $F$ be the CDF of the three-level hierarchical distribution and $F^*(x) = \{F(x) + F(x-)\}/2$. The rank ICC at level 2, denoted as $\gamma_{I2}$, measures the correlation between a random pair of level-1 observations from the same level-2 unit. It is defined as  
\begin{equation}
\gamma_{I2} = corr[F^*(X_{ijk}), F^*(X_{ijk'})], 
\label{gammaI2}
\end{equation}
where $k \neq k'$. The rank ICC at level 3, denoted as $\gamma_{I3}$, measures the correlation between a random pair of level-1 observations from the same level-3 unit but different level-2 units. It is defined as 
\begin{equation}
\gamma_{I3} = corr[F^*(X_{ijk}), F^*(X_{ij'l})], 
\label{gammaI3}
\end{equation}
where $j \neq j'$ but $k$ and $l$ can be equal or different. At level 3, there are two potential sources of within-cluster correlation: one due to different level-2 units within the same level-3 unit and the other due to different level-1 units within the same level-2 unit. Our definition of $\gamma_{I3}$ captures the former; the latter has already been captured by $\gamma_{I2}$. If we were to ignore the second level and consider the rank correlation between two random level-1 units from the same level-3 unit irrespective of their level-2 information, the resulting definition would reflect both sources of correlation, which is not ideal; it could be quite different from $\gamma_{I3}$ with a small number of level-2 units within each level-3 unit. Our rank ICC definitions given by (\ref{gammaI2}) and (\ref{gammaI3}) have comparable interpretations to previously proposed definitions of ICC for 3 hierarchies on the original scale \citep{Siddiqui1996}.

The general definition of the rank ICC for a multiple-level hierarchical distribution can be similarly defined. Let $Q$ be the number of hierarchies and $X_{I_{Q}I_{Q-1}...I_{1}}$ denote a random variable from a $Q$-level hierarchical distribution, where $I_Q$, $I_{Q-1}$, ..., $I_1$ are indices for levels $Q$, $Q-1$, ..., $1$, respectively. The CDF of the $Q$-level hierarchical distribution is denoted as $F$, and $F^*(x) = \{F(x) + F(x-)\}/2$. The rank ICC at level $j$ ($j \in \{2,3,...,Q\}$) measures the correlation between level-1 observations from the same level-$j$ unit and different level-$(j-1)$ unit:
\begin{equation}
\gamma_{Ij} = corr[F^*(X_{I_{Q}I_{Q-1}...I_{j}I_{j-1}...I_{1}}),F^*(X_{I_{Q}I_{Q-1}...I_{j} I_{j-1}'...I_{1}'})],
\label{gammaIk}
\end{equation}
where $I_{j-1} \neq I_{j-1}'$, and for $l<j-1$, $I_{l}$ and $I_{l}'$ can be the same or different.

\section{Estimation and Inference}
\label{s:estimation}

\subsection{Estimation}
Since the rank ICC can be viewed as a function of the underlying distribution $\gamma_I(F)$, then our estimator of $\gamma_I$ is $\hat \gamma_I =\gamma_I(\hat F)$. Given two-level data $\{x_{ij}: i=1,2,...,n, j=1,2,...,k_i\}$ with a total number of observations of $N = \sum_{i=1}^n k_i$, a nonparametric estimator of the CDF is $\hat F(x) = \sum_{i=1}^n \sum_{j=1}^{k_i} w_{ij}I(x_{ij} \leq x)$, where $w_{ij}$ is the weight of observation $x_{ij}$ and $\sum_{i=1}^n \sum_{j=1}^{k_i} w_{ij} = 1$. The weight $w_{ij}$ depends on how we believe the data reflect the composition of the underlying hierarchical distribution; for example, $w_{ij}=1/(nk_i)$ corresponds to equal weights for clusters and $w_{ij}=1/N$ corresponds to equal weights for observations. Other weighting options will be described later in this section. The weight of cluster $i$ is denoted as $w_{i.} = \sum_{j=1}^{k_i} w_{ij}$. Similarly, we estimate $\hat F(x-) = \sum_{i=1}^n \sum_{j=1}^{k_i} w_{ij}I(x_{ij} < x)$, and define $\hat F^*(x) = \{\hat F(x) + \hat F(x-)\}/2$. Then our estimator of $\gamma_I$ is $\hat \gamma_I = corr\{\hat F^*(X_{ij}), \hat F^*(X_{ij'})\} $, where $(X_{ij}, X_{ij'})$ is a random pair drawn from a random cluster and $j\neq j'$. 

Because the rank ICC measures the correlation of a random pair from the same cluster, we could consider Monte Carlo estimation. That is, we first randomly select clusters with replacement and then randomly draw pairs of observations from the selected clusters. Then $\gamma_{I}$ could be estimated as the sample correlation of $\hat F^*(x)$ between the sampled pairs of observations. As the number of sampled pairs increases, the estimate of this approach will converge to a limit, which is our estimator:
\begin{equation}
\hat \gamma_I = \frac{\sum_{i=1}^n w_{i.} \sum_{1\le j<j'\le k_i} \frac{2}{k_i(k_i-1)} [\hat F^*(x_{ij}) - \bar{\hat{F^*}}][\hat F^*(x_{ij'}) - \bar{\hat{F^*}}]}{\sum_{i=1}^n \sum_{j=1}^{k_i} w_{ij} [\hat F^*(x_{ij}) - \bar{\hat{F^*}}]^2},  
\label{hatgammaI}
\end{equation}
where $\bar{\hat{F^*}} =  \sum_{i=1}^n \sum_{j=1}^{k_i} w_{ij} \hat F^* (x_{ij})$, and $k_i(k_i-1)/2$ is the number of possible unordered pairs in cluster $i$. 

The estimator $\hat \gamma_I$ given by (\ref{hatgammaI}) is consistent for $\gamma_I$ and is asymptotically normal. The proof of consistency and
asymptotic normality and the variance estimation of $\hat \gamma_I$ are in the Supporting Information. The results allow us to compute standard errors (SEs) of $\hat \gamma_I$ and to construct confidence intervals (CIs) for $\gamma_I$. 

The selection of weights, $w_{ij}$, warrants additional discussion. For populations with finite and unequal cluster sizes, if there is ambiguity in the relative contributions of clusters in a hierarchical distribution, then the rank ICC can have some ambiguity. One could assume all clusters have an equal contribution regardless of their cluster sizes, in which case it would be sensible to set $w_{ij}=1/(nk_i)$. Or one could assume the relative contributions are proportional to cluster sizes, in which case it would be sensible to set $w_{ij}=1/N$. The choice of $w_{ij}$ should be driven by subject matter knowledge. For example, if one is measuring the repeatability of an assay by collecting specimens (one per person) and measuring them multiple, unequal numbers of times, then it seems sensible to assume the clusters (people) contribute equally in the population. In contrast, if one is interested in the correlation of a trait between individuals within the same family, then it may (or may not) be sensible to assume each family contributes proportionally to the family size. These two weighting approaches have been applied to estimating the ICC on the original scale under variable cluster sizes \citep{Karlin1981}. For perfectly balanced data, $\hat \gamma_I$ is the same regardless of the weighting approach used.

However, in practice, there is often uncertainty in how we should assume clusters contribute to the underlying distribution and we may want to consider different weighting schemes. In fact, there may be bias-variance considerations that might suggest using weights that do not exactly match the true cluster contributions. For example, consider a population with equal cluster contribution. When $\gamma_I$ is close to zero, observations in the same cluster are almost independent, so treating all observations equally regardless of the cluster size can be more efficient than weighting observations inversely proportional to the size of their cluster. In contrast, when $\gamma_I$ is close to one, observations in the same clusters are almost redundant, favoring equal weight per cluster. But whether $\gamma_I$ is close to zero or one is often unknown before analysis. Therefore, one might use an iterative procedure to identify a more efficient weighting scheme. One approach is to use a linear combination of the two weights above, where the combination depends on the value of $\gamma_I$; that is, $w_{ij}(\gamma_I)=(1-\gamma_I)/N+\gamma_I/(nk_i)$. We call this the combination approach. Another approach is to compute the effective sample size (ESS) \citep{kish1965} for the clusters (i.e., $n_{i}^{(e)} = k_i/(1+(k_i-1)\gamma_I)$) and weight clusters and their observations in clusters accordingly; that is, $w_{i.}(\gamma_I)=n_{i}^{(e)} /\sum_{l=1}^n n_{l}^{(e)}$ and $w_{ij}(\gamma_I) = w_{i.}(\gamma_I)/k_i$. We call this second approach the ESS approach. These approaches require a working value of $\gamma_I$. We implement iterative procedures in which we (a) start with an initial value of $\gamma_I$, (b) update the weights, and (c) compute a new estimate of $\gamma_I$. We repeat steps (b) and (c) multiple times until the estimate of $\gamma_I$ converges.  Our simulations suggest that the choice of the initial value has no effect on the final estimate. 

With three or more hierarchies, the estimation of the rank ICC is similar to that described above for two hierarchies. Given three-level nested data $\{x_{ijk};i=1,2,...,n, j=1,2,...,n_i, k=1,2,...,m_{ij}\}$, the nonparametric estimator for the CDF is $\hat F(x) = \sum_{i=1}^n \sum_{j=1}^{n_i} \sum_{k=1}^{m_{ij}} w_{ijk}I(x_{ijk} \leq x)$, where $  \sum_{i=1}^n \sum_{j=1}^{n_i} \sum_{k=1}^{m_{ij}} w_{ijk} = 1$. Similarly, $\hat F(x-) = \sum_{i=1}^n \sum_{j=1}^{n_i} \sum_{k=1}^{m_{ij}} w_{ijk}I(x_{ijk} < x)$. Let $\hat F^*(x) = \{\hat F(x) + \hat F(x-)\}/2$. The general form of the estimator of $\gamma_{I2}$ is
\begin{equation}
\hat \gamma_{I2} = \frac{\sum_{i=1}^n \sum_{j=1}^{n_i} w_{ij.}  \sum_{1\le k<k'\le m_{ij}}
\frac{2}{m_{ij}(m_{ij}-1)} [\hat F^*(x_{ijk}) - \bar{\hat{F^*}}][\hat F^*(x_{ijk'}) - \bar{\hat{F^*}}]}{\sum_{i=1}^n \sum_{j=1}^{n_i} \sum_{k=1}^{m_{ij}} w_{ijk} [\hat F^*(x_{ijk}) - \bar{\hat{F^*}}]^2}, 
\label{hatgammaI2}
\end{equation}
where $w_{ij.} = \sum_{k=1}^{m_{ij}} w_{ijk}$ and $\bar{\hat{F^*}} = \sum_{i=1}^n \sum_{j=1}^{n_i} \sum_{k=1}^{m_{ij}} w_{ijk} \hat F^* (x_{ijk}) $. The general form of the estimator of $\gamma_{I3}$ is
\begin{equation}
\hat \gamma_{I3} = \frac{\sum_{i=1}^n w_{i..}\sum_{1\le j<j'\le n_i} \sum_{k=1}^{m_{ij}} \sum_{l=1}^{m_{ij'}} \frac{1}{c_i}  [\hat F^*(x_{ijk}) - \bar{\hat{F^*}}][\hat F^*(x_{ij'l}) - \bar{\hat{F^*}}]}{\sum_{i=1}^n \sum_{j=1}^{n_i} \sum_{k=1}^{m_{ij}} w_{ijk} [\hat F^*(x_{ijk}) - \bar{\hat{F^*}}]^2},  
\label{hatgammaI3}
\end{equation}
where  $w_{i..} =  \sum_{j=1}^{n_i}\sum_{k=1}^{m_{ij}} w_{ijk}$, and $c_{i}$ is the total number of possible unordered pairs in a level-3 unit; $c_{i} = \{(\sum_{j=1}^{n_i} m_{ij})^2-(\sum_{j=1}^{n_i}  m_{ij}^2)\}/2$. We show the asymptotic normality and consistency of $\hat \gamma_{I2}$ and $\hat \gamma_{I3}$ in the Supporting Information. There are several options for $w_{ijk}$ with three-level data, such as assigning equal weights to all level-1 units (i.e., $w_{ijk}=1/(\sum_{i=1}^n\sum_{j=1}^{n_i}m_{ij})$), assigning equal weights to all level-2 units (i.e., $w_{ijk}=1/(m_{ij}\sum_{i=1}^n n_{i})$), or assigning equal weights to all level-3 units (i.e., $w_{ijk}=1/(nn_{i}m_{ij})$). 

\subsection{Inference}

The distribution of $\hat \gamma_I$ can be approximated using asymptotics. The asymptotic standard error (SE) of $\hat \gamma_I$, presented in the Supporting Information, can be used to construct confidence intervals for $\gamma_I$ under normality. Because $\gamma_I$ is bounded, one might also consider estimating the large sample distribution of the Fisher transformed value (i.e., $\log\{(1+\hat \gamma_I)/(1-\hat \gamma_I)\}/2$) by the delta method to obtain confidence intervals \citep{fisher1915}. 

An alternative approach for estimating the distribution of $\hat \gamma_I$ is bootstrapping. There are two general ways to implement bootstrapping in clustered data; the cluster bootstrap and the two-stage bootstrap \citep{davison1997, Field2007}. In the cluster bootstrap, clusters are randomly selected with replacement. The two-stage bootstrap has an extra step, where in the selected clusters the observations are randomly drawn with replacement. In our setting, this intracluster sampling in the two-stage bootstrap can cause positive bias in estimating $\gamma_I$, because the same observation may be sampled twice in a two-stage bootstrap sample, thus inflating the estimated ICC, particularly in settings with smaller cluster sizes. Hence, we recommend using the cluster bootstrap for bootstrapping.

The standard errors of $\hat \gamma_{I2}$ and $\hat \gamma_{I3}$ with three hierarchies can be similarly computed. We have derived analytic formulas for asymptotic SEs of $\hat \gamma_{I2}$ and $\hat \gamma_{I3}$ given by (\ref{hatgammaI2}) and (\ref{hatgammaI3}) in the Supporting Information. In addition, one could bootstrap. Considering computational efficiency and the bias caused by intracluster sampling with replacement, we suggest a one-stage bootstrap for bootstrapping with three hierarchies; i.e., only sampling level-3 units with replacement. 

\section{Simulations}
\label{s:simulations}

\begin{figure}
     \centering
     \includegraphics[width=0.9\textwidth]{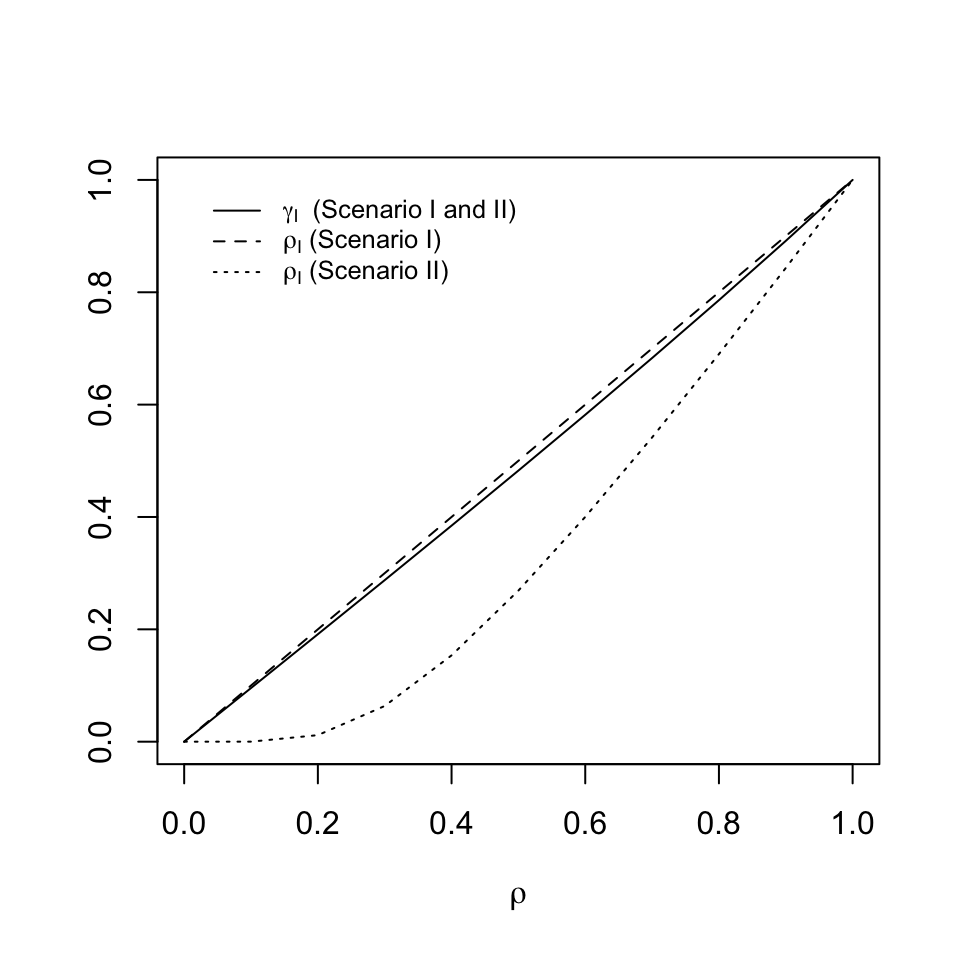}
     \caption[]{Parameters of rank ICC ($\gamma_I$) and Fisher's ICC ($\rho_I$) as a function of the within-cluster correlation ($\rho$) of $X_{ij}$ under normality (Scenario I) and after exponentiating the data (Scenario II).}
    \label{fig:fig7}
\end{figure}

 A simple additive model was used to generate two-level data: $X_{ij} = U_{i} + R_{ij}$, where $U_i\stackrel{i.i.d}{\sim}N(1,1)$ and $R_{ij} \stackrel{i.i.d}{\sim} N(0, (1-\rho)/\rho)$ with $\rho$ varying in $[0,1]$. Let $Y_{ij}$ be the observation of the $j$th individual in the $i$th cluster, where $i=1,2,....,n$; $j=1,2,...,k_i$; and $k_i$ is the cluster size of the $i$th cluster. We considered three scenarios: (I) $Y_{ij}=X_{ij}$; (II) $Y_{ij} = \exp(X_{ij})$; (III) $Y_{ij} = U^{'}_{i} + R_{ij}$, where $U^{'}_i$'s are i.i.d. following a log-normal distribution such that $var(U_i')=1$ and $\log(U_i') \stackrel{i.i.d}{\sim}N(1, \log(1/2+\sqrt{\exp(-2)+1/4}))$. In Scenarios I and II, since $X_{ij}$ is normally distributed, the rank ICC is $\gamma_I = 6\arcsin(\rho/2)/\pi$ \citep{Pearson1907}. The rank ICC is identical in Scenarios I and II while Fisher's ICC, $\rho_I$, is sensitive to skewness and depends on the scale of interest (Figure \ref{fig:fig7}). When the variable of interest is normal (Scenario I), $\gamma_I$ is close to $\rho_I$. In Scenario III, $Y_{ij}$ is not normally distributed so we empirically computed $\gamma_I$ by generating a million clusters each with 2 observations, and then computing Spearman's rank correlation.

We first evaluated the performance of our estimator of $\gamma_I$ for two-level data. The simulations were conducted at different sample sizes $n=$ 25, 50, 100, 200, 500, and 1000 with an equal cluster size ($k_i$=30). Furthermore, we also performed simulations with various configurations of cluster size at $n=200$: $k_i=2$; $k_i=30$; $k_i$ uniformly ranging from 2 to 50; and $k_i=2$ for half of the clusters and $k_i=30$ for the other half. Unless stated otherwise, for estimation, we assigned equal weights to clusters (i.e., $w_{ij}=1/(nk_i)$), which corresponds with the underlying equal cluster contribution in the simulated hierarchical distribution. We computed 95\% confidence intervals for $\gamma_I$ using the asymptotic SE and bootstrapping.  

\begin{figure}
     \centering
     \begin{subfigure}[b]{0.5\textwidth}
         \centering
         \includegraphics[width=0.9\textwidth]{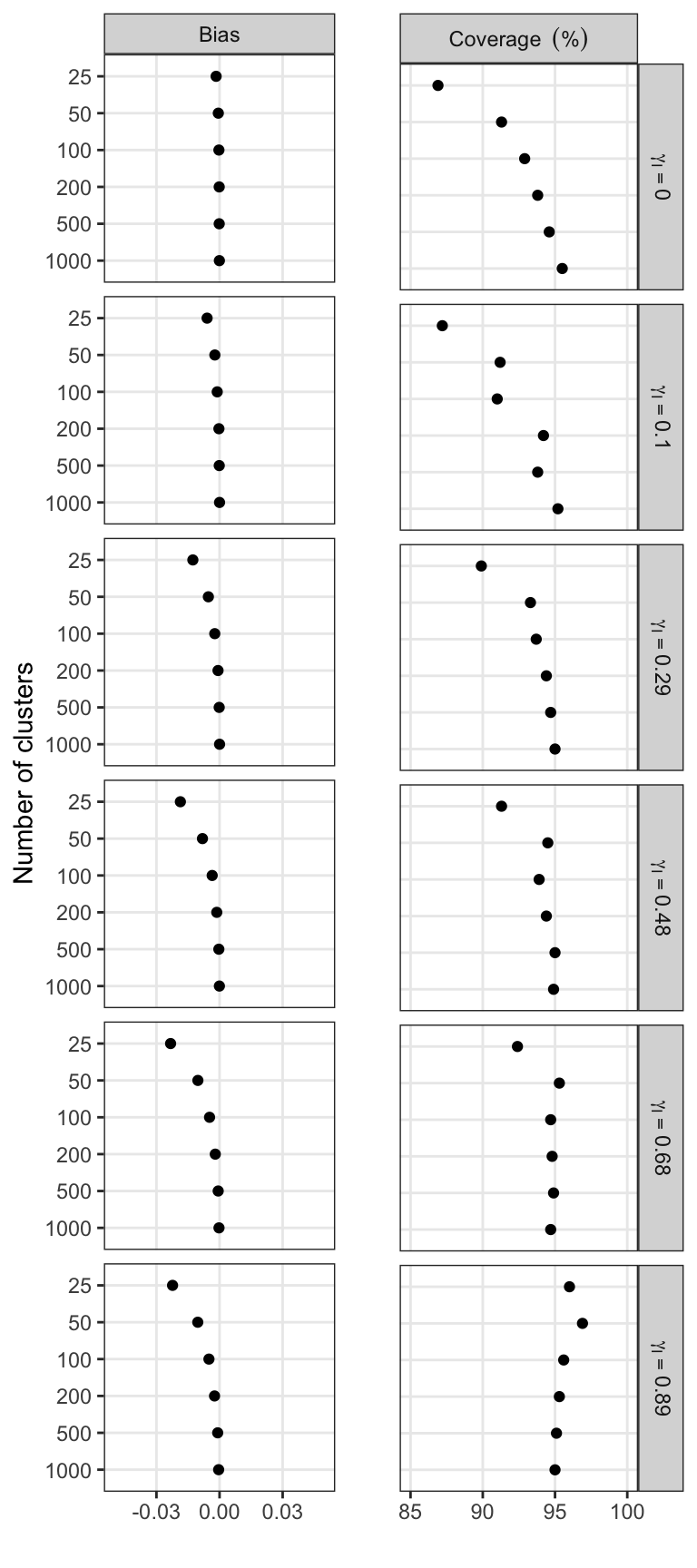}
         \caption{Scenario I and II}
         \label{fig:s1.1}
     \end{subfigure}\hfill
     \begin{subfigure}[b]{0.5\textwidth}
         \centering
         \includegraphics[width=0.9\textwidth]{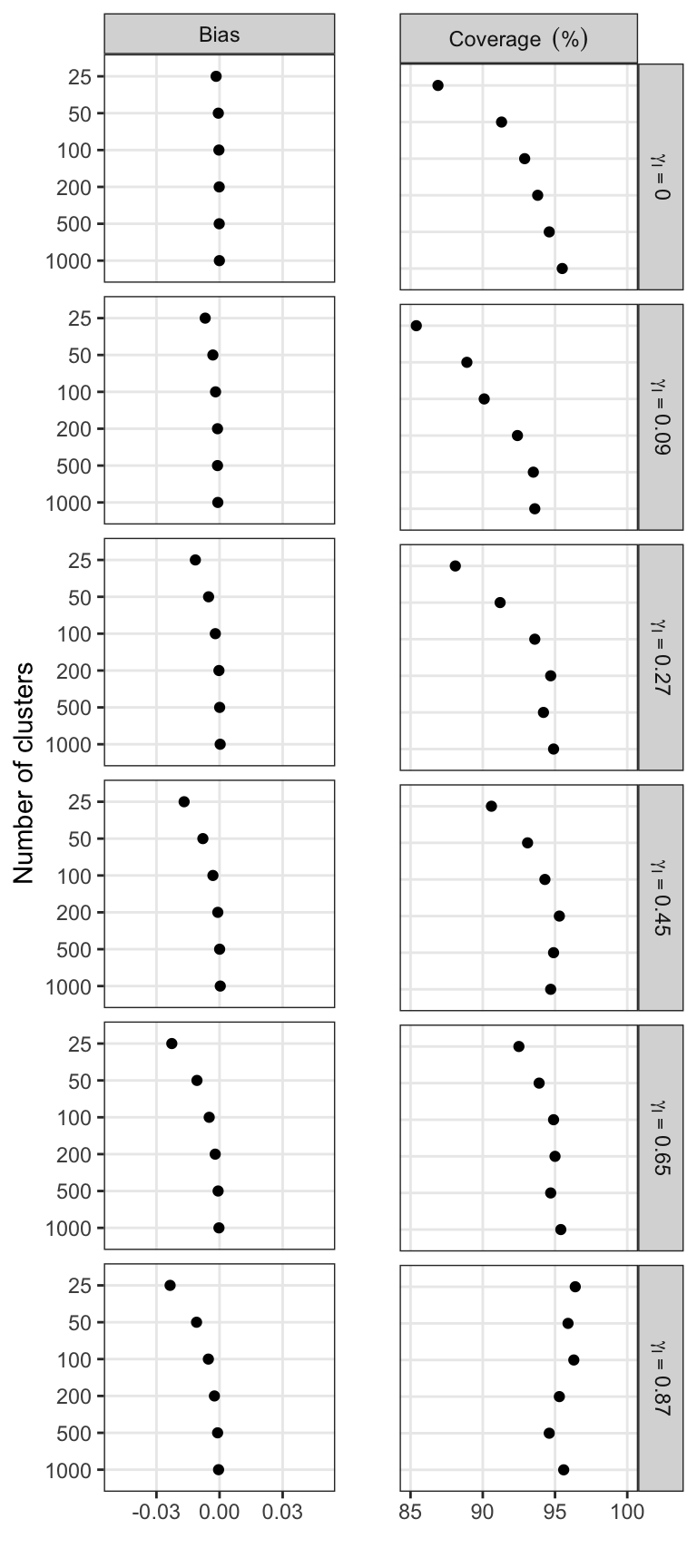}
         \caption{Scenario III}
         \label{fig:s3.1}
     \end{subfigure}
     \caption[]{Bias and coverage of 95\% confidence intervals for our estimator of $\gamma_I$ at different true values of $\gamma_I$ and different numbers of clusters under Scenarios I (normality), II (exponentiated outcomes), and III (exponentiated cluster means). The number of observations per cluster was set at 30.}
        \label{fig:fig1}
\end{figure}

\begin{figure}
     \centering
     \begin{subfigure}[b]{0.5\textwidth}
         \centering
         \includegraphics[width=0.9\textwidth]{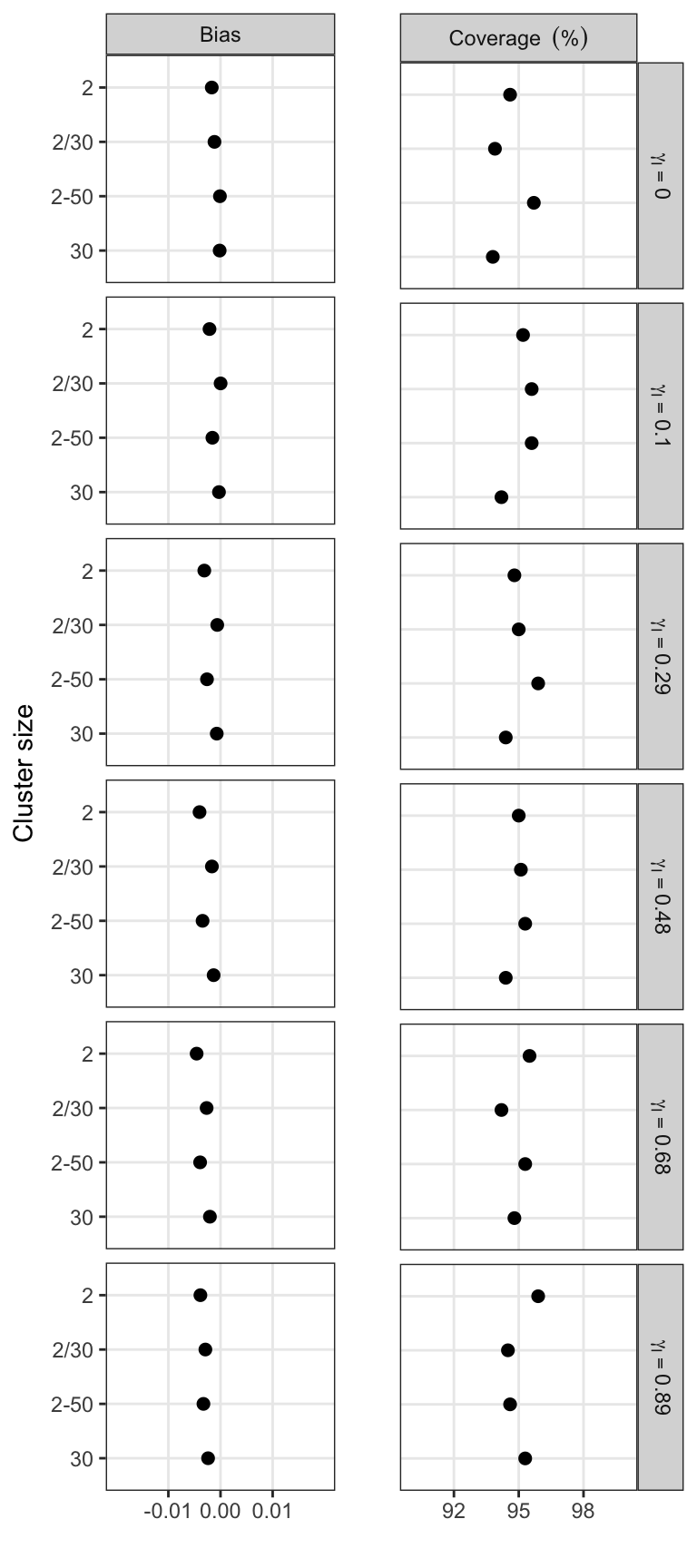}
         \caption{Scenario I and II}
         \label{fig:s1.2}
     \end{subfigure}\hfill
     \begin{subfigure}[b]{0.5\textwidth}
         \centering
         \includegraphics[width=0.9\textwidth]{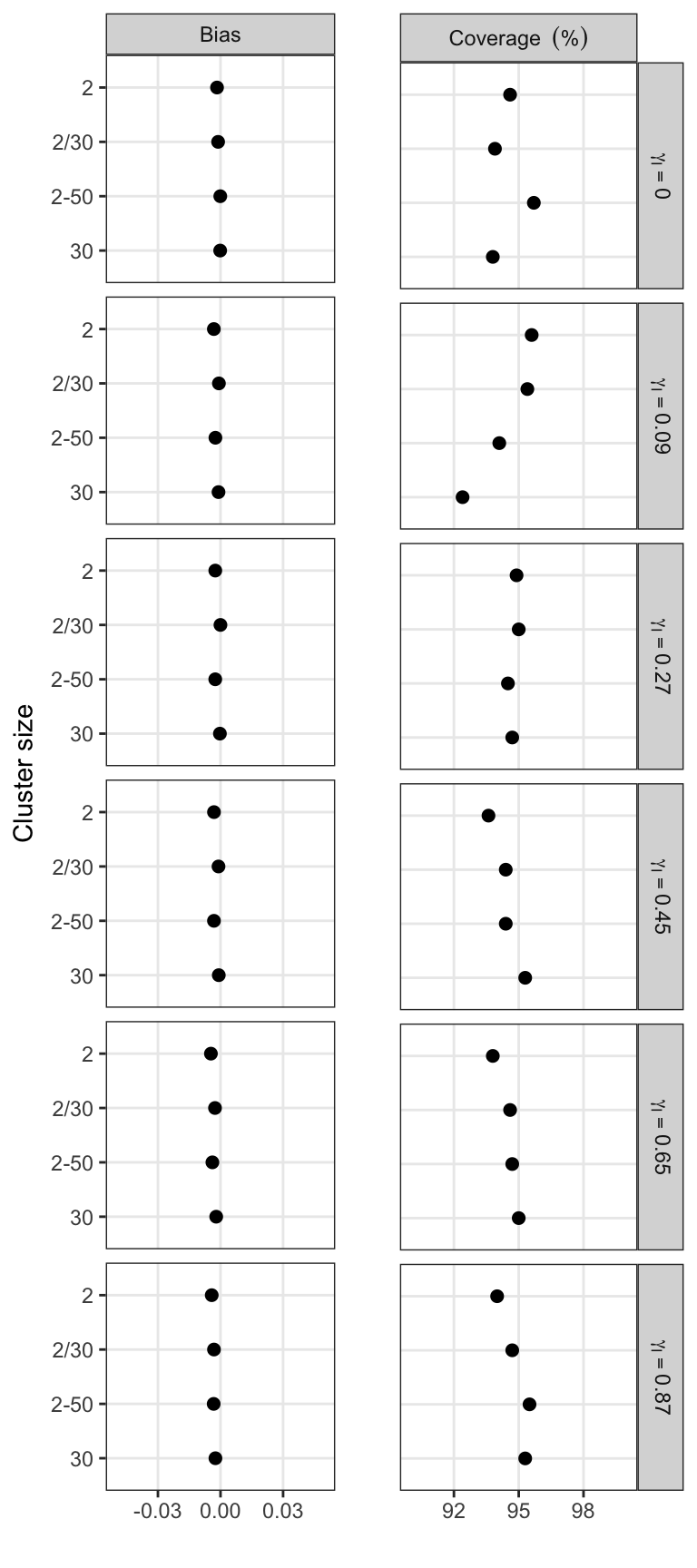}
         \caption{Scenario III}
         \label{fig:s3.2}
     \end{subfigure}
     \caption[]{Bias and coverage of 95\% confidence intervals for our estimator of $\gamma_I$ at different true values of $\gamma_I$ and different cluster sizes under Scenarios I (normality), II (exponentiated outcomes), and III (exponentiated cluster means). The number of clusters was set at 200. ``2-50'' means the cluster size follows a uniform distribution from 2 to 50, ``2/30'' means half of the clusters have size 2 and half have 30.}
     \label{fig:fig2}
\end{figure}

The bias of our estimator and the coverage of 95\% CIs based on the asymptotic SE under the different scenarios described above are shown in Figures \ref{fig:fig1} and \ref{fig:fig2}. In summary, our estimator of $\gamma_I$ had very low bias and good coverage with modest numbers of clusters across all scenarios we considered. It was also robust to the skewed data in Scenarios II and III. Although our estimator had slightly negative bias with a small number of clusters, this bias decreased as the number of clusters increased. Confidence intervals for $\gamma_I$ based on the asymptotic SE approximately covered at their nominal 0.95 level with $\geq$ 200 clusters for all true values of $\gamma_I$. For smaller values of $\gamma_I$, coverage could be low for $\leq$ 100 clusters. Fisher transformation did not appear to improve coverage (Web Figure $1$). The performance of estimators was fairly similar regardless of the size of clusters (Figure \ref{fig:fig2}). Additional simulations reported in Web Tables $1$-$5$ show that confidence intervals based on both the cluster bootstrap SE and percentiles had good coverage.

\begin{figure}
\centering
\includegraphics[width=0.9\textwidth]{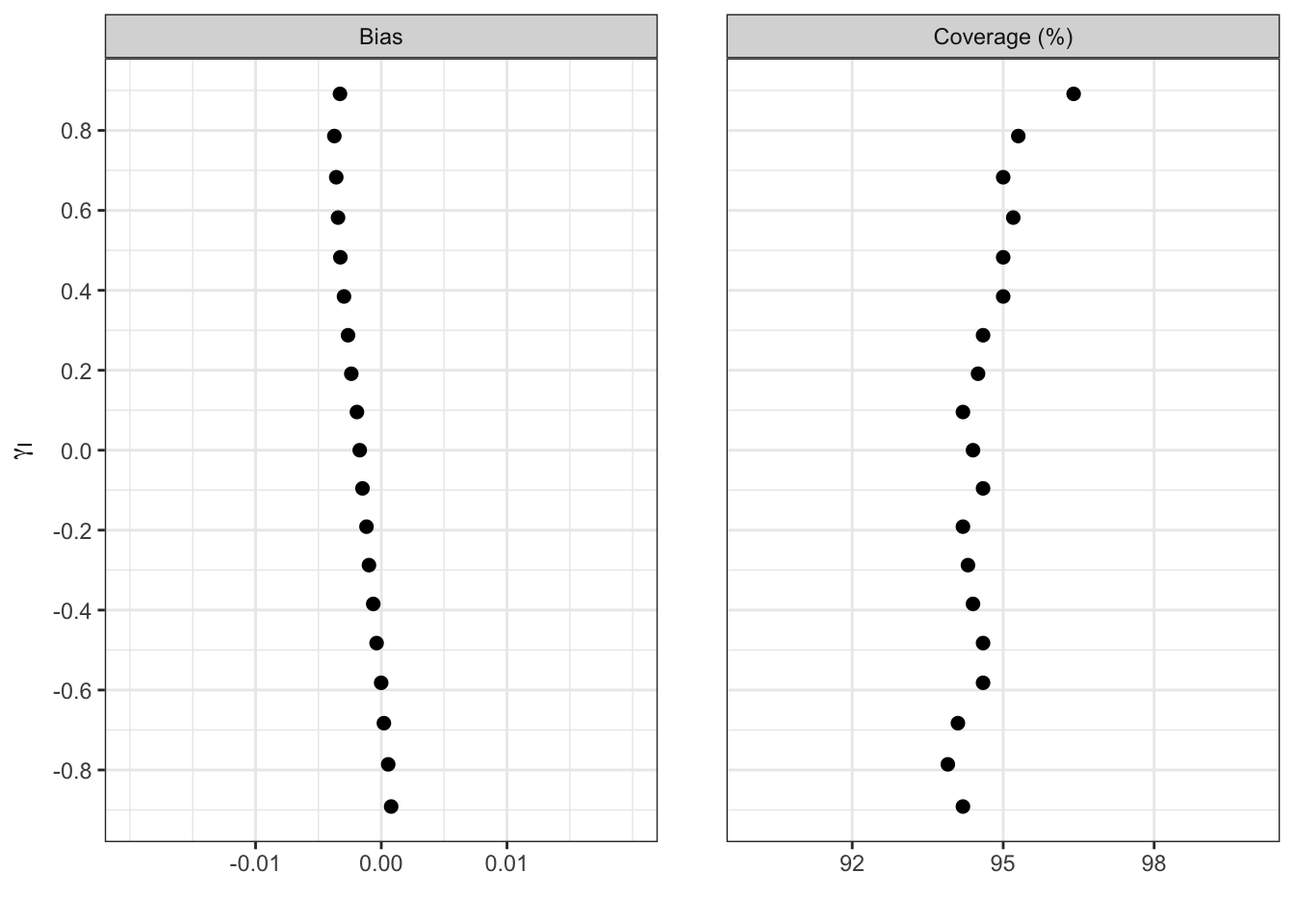}
\caption[]{Bias and coverage of 95\% confidence intervals for our estimator of $\gamma_I$ at different true positive and negative values of $\gamma_I$ when the cluster size in the population was 2. The number of clusters was set at 200.}
\label{fig:fig5}
\end{figure}

We then evaluated the performance of our estimator of $\gamma_I$ when the cluster size is 2 in the population and the rank ICC varies between $-1$ and $1$. Let $X_{i1}$ and $X_{i2}$ be the two observations in cluster $i$. We generated the two observations in cluster $i$ as follows: $X_{i1} = U_{i} + R_{i}$ and $X_{i2} = U_{i} - R_{i}$, where $U_i \stackrel{i.i.d}{\sim} N(1,1)$, $R_i \stackrel{i.i.d}{\sim} N(0, (1-\rho)/(1+\rho))$, and $\rho$ varies over $[-1, 1]$ (we set $var(U_i)=0$ and $var(R_i)=20$ when $\rho=-1$). We conducted 1000 simulations at $n=200$. Our estimator of $\gamma_I$ had low bias and good coverage (Figure \ref{fig:fig5}).

\begin{figure}
    \centering
    \includegraphics[width=0.9\textwidth]{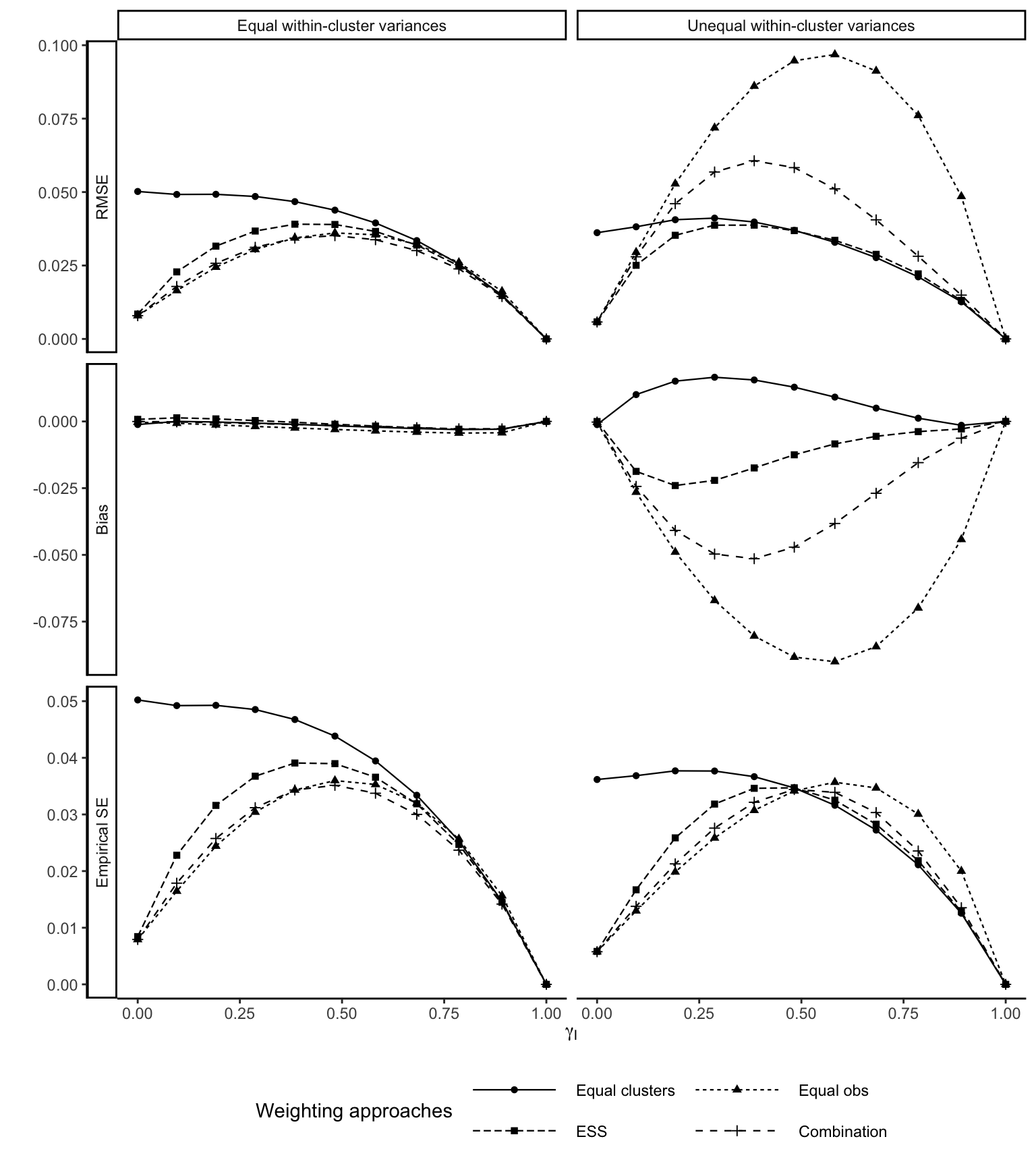}
    \caption{Root mean squared error (RMSE), bias, and empirical SE of estimates obtained by the four weighting approaches for our estimator of $\gamma_I$. ``Equal clusters'' refers to assigning equal weights to clusters, ``Equal obs'' refers to assigning equal weights to observations, ``ESS'' refers to the iterative weighting approach based on the effective sample size, and ``Combination'' refers to the iterative weighting approach based on the linear combination of equal weights for clusters and equal weights for observations. We set the tolerance of the two iterative approaches to 0.00001.}
    \label{fig:fig4}
\end{figure}

We next compared the performance of the four weighting approaches with 1) equal within-cluster variances and equal or unequal cluster sizes, and 2) within-cluster variances varying by cluster size. For 1), we used the same simulations described in the first paragraph of this section. For 2), we supposed the numbers of small clusters of size 2 and large clusters of size 30 are equal in the population, and simulated the data as $X_{ij} = U_{i} + R_{ij}$, with $R_{ij} \stackrel{i.i.d}{\sim} N(0, c(1-\rho)/\rho)$, where $c=0.5$ for small clusters and $c=1.5$ for large clusters and $\rho$ varying over $[0,1]$. We conducted 1000 simulations at $n=200$. Results of the two sets of simulations are shown in Figure \ref{fig:fig4} and Web Figure $2$. When cluster sizes were equal and within-cluster variances were equal, the estimates of the four weighting approaches were identical. When cluster sizes were unequal and within-cluster variances were equal, the four methods all had low bias and their mean squared errors were dominated by their variances. As hypothesized, assigning equal weights to clusters had the lowest efficiency when the rank ICC was close to zero, because treating large and small clusters equally resulted in lost information, even though the data were simulated in a manner such that equal cluster weighting matched the cluster contribution in the population. In contrast, when within-cluster variances varied by cluster sizes, assigning equal weights to observations contrary to the underlying distribution led to bias. The two iterative approaches had lower mean squared errors than assigning equal weights to clusters or to observations when the rank ICC was close to zero.

\begin{figure}
     \centering
     \includegraphics[width=0.9\textwidth]{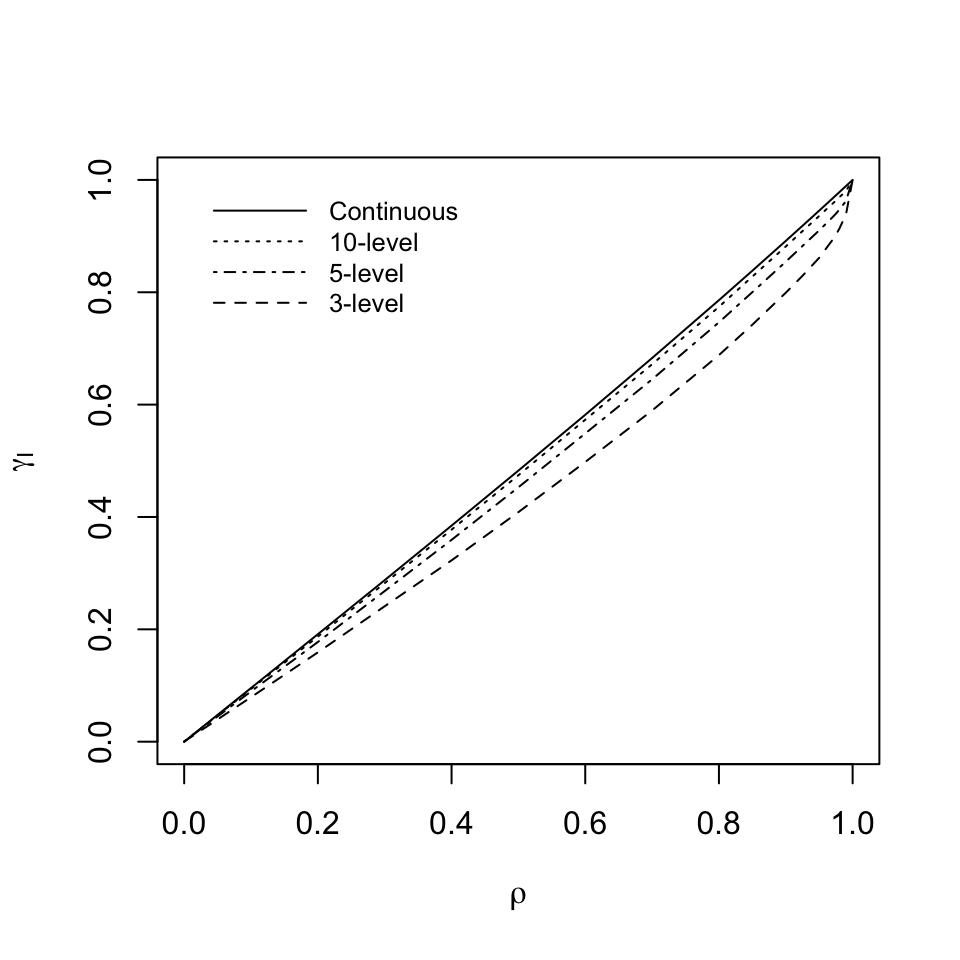}
     \caption[]{Parameters of rank ICC ($\gamma_I$) as a function of the within-cluster correlation ($\rho$) of $X_{ij}$ when data are continuous or discretized into ordered categorical variables with 3, 5, or 10 levels.}
    \label{fig:fig8}
\end{figure}

We also evaluated the performance of our estimator of $\gamma_I$ for ordered categorical variables. We simulated data of 3-level, 5-level, and 10-level ordered categorical variables by discretizing $X_{ij}$ in Scenario I with cut-offs at quantiles (i.e., using the 1/3 and 2/3 quantiles for 3 levels; the 0.2, 0.4, 0.6, 0.8 quantiles for 5 levels; and the 0.1, 0.2, ..., 0.8, 0.9 quantiles for 10 levels). Similar to Scenario III, we empirically computed $\gamma_I$ for the ordered categorical variables (Figure \ref{fig:fig8}). The rank ICCs of the 5-level and 10-level variables are close to the rank ICC of the continuous variable, while the rank ICC of the 3-level variable is slightly smaller. We conducted simulations for 3-level and 10-level ordered categorical variables at different sample sizes $n=$ 25, 50, 100, 200, 500, and 1000 with an equal cluster size ($k_i$=30). Our estimator of $\gamma_I$ of the ordered categorical variables generally had low bias and good coverage (Web Figure 3).

\begin{figure}
            \centering
            \includegraphics[width=0.8\textwidth]{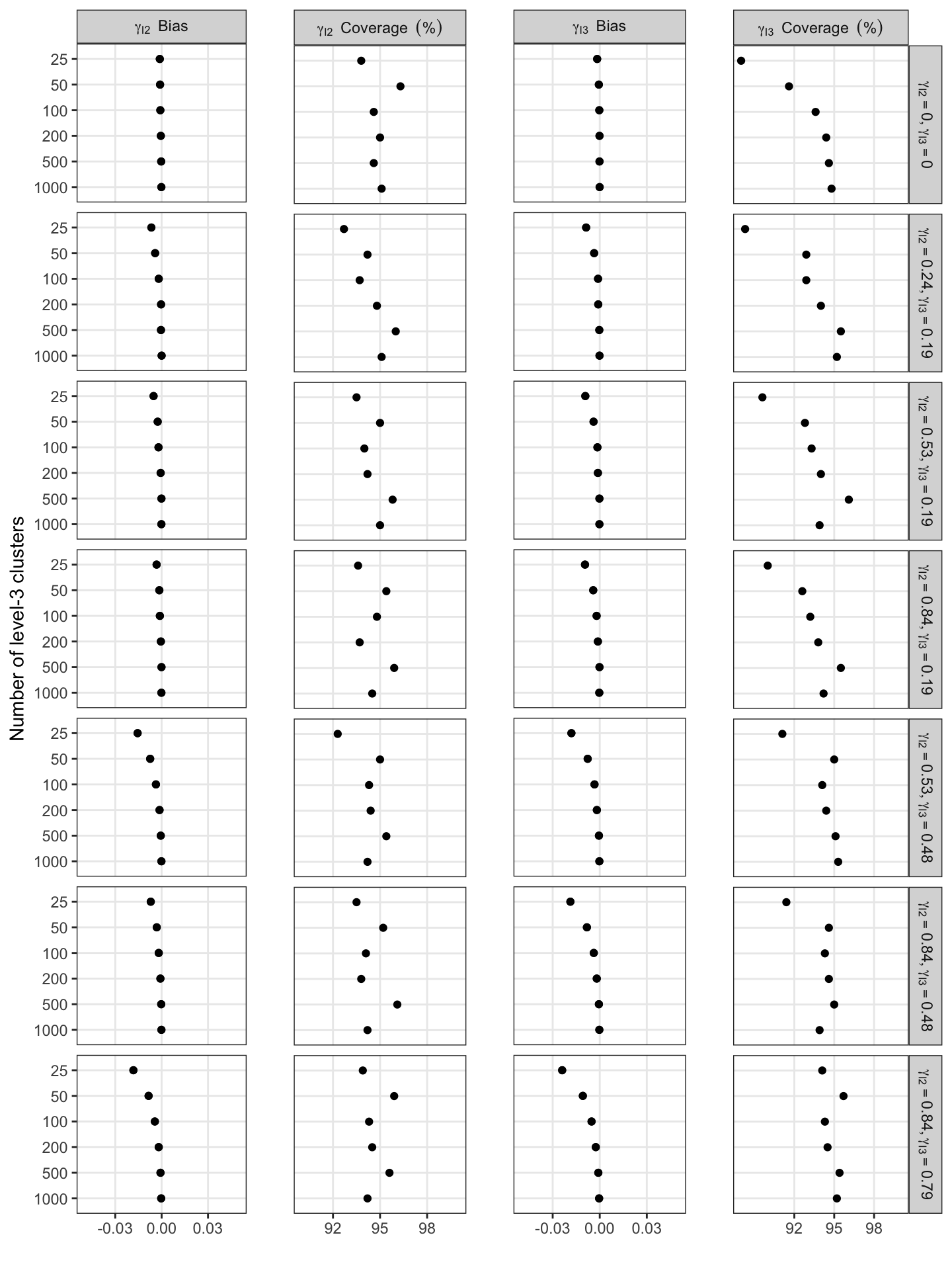}
            \caption[]{Bias and coverage of 95\% confidence intervals for our estimators of $\gamma_{I2}$ and $\gamma_{I3}$ at different true values of $\gamma_{I2}$ and $\gamma_{I3}$ and different numbers of level-3 units. The number of level-2 units in a level-3 unit was set at 15. The number of level-1 units in a level-2 unit was set at 2.} 
        \label{fig:fig3}
\end{figure}

We also investigated the performance of our estimators of $\gamma_{I2}$ and $\gamma_{I3}$ for data with three hierarchies. Let $X_{ijk}$ be the observation of the $k$th level-1 unit in the $j$th level-2 unit and the $i$th level-3 unit, where $i=1,2,...,n$,\; $j=1,2,...,n_i$; $k=1,2,...,m_{ij}$. We generated three-level data as follows: $X_{ijk} = U_i + V_{ij} + R_{ijk}$, where $U_i \stackrel{i.i.d}{\sim} N(1, 20\rho_{I3})$, $V_{ij} \stackrel{i.i.d}{\sim} N(0, 20(\rho_{I2}-\rho_{I3}))$, $R_{ijk} \stackrel{i.i.d}{\sim} N(0,20(1-\rho_{I2}))$, and $(\rho_{I2}, \rho_{I3}) \in \{(0,0),(0.25,0.20),(0.55,0.20),(0.85,0.20),(0.55,\\0.5),(0.85,0.5),(0.85,0.8)\}$. Because of normality, the true rank ICCs are $\gamma_{I2} = 6\arcsin(\rho_{I2}/2)/\pi$ and $\gamma_{I3} = 6\arcsin(\rho_{I3}/2)/\pi$. We conducted 1000 simulations for different sample sizes $n=$ 25, 50, 100, 200, 500, and 1000 under equal cluster sizes (i.e., $n_i=15$ and $m_{ij}=2$). Moreover, we also performed simulations under various patterns of cluster sizes: $(n_i, m_{ij}) \in \{(15, 2),(2, 15),(4, 2), (2$-$15, 2), (2$/$15, 2), (2, 2$-$15), (2, 2$-$15), (2$/$15, 2$/$15)\}$, where ``$2$-$15$'' means the cluster size follows a uniform distribution from 2 to 15, ``$2$/$15$'' means half of the clusters have size 2 and a half have 15. The results for $n_i=15$ and $m_{ij}=2$ are shown in Figure \ref{fig:fig3}, and the other results are in Web Figures $3$ and $4$ and Web Tables $6$ and $7$. Our estimators of $\gamma_{I2}$ and $\gamma_{I3}$ had very low bias and good coverage in all cases we considered. The asymptotic SE and the one-stage bootstrap had good performance in constructing confidence intervals, and the former was computationally efficient.

\section{Applications}
\label{s:applications}
%remove all brackets 
\subsection{Albumin-Creatinine Ratio}
\begin{figure}
            \centering
            \includegraphics[width=0.8\textwidth]{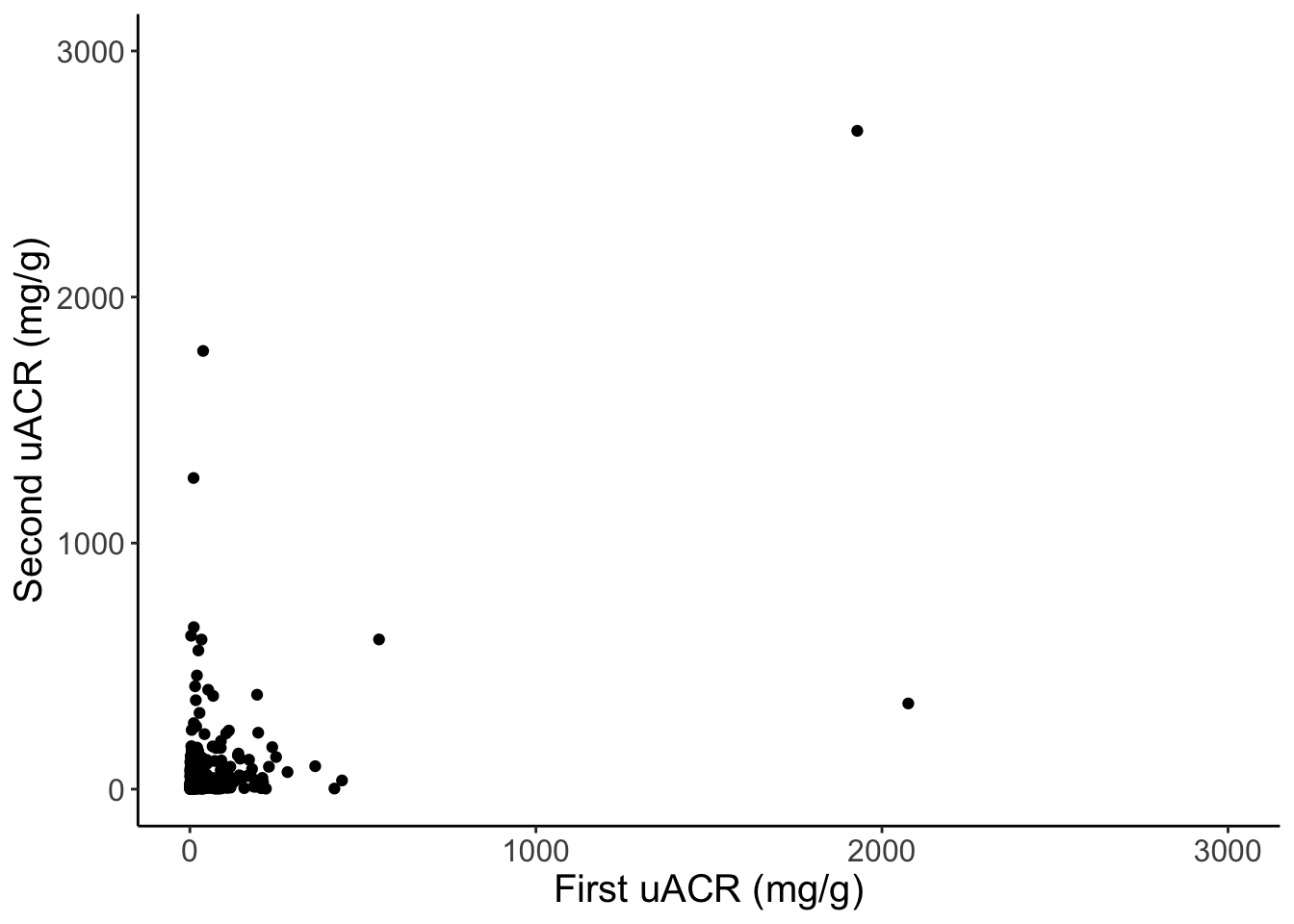}
            \caption[]{Scatter plot of the first and second uACR measurements of each person in the example of albumin-creatinine ratio.} 
        \label{fig:scatteracr}
\end{figure}

In a cross-sectional study, 598 people living with HIV in Nigeria on stable dolutegravir-based antiretroviral therapy provided first-morning void urine specimens at two visits 4 to 8 weeks apart \citep{wudil2021}. The collected urine specimens were used to calculate the urine albumin-creatinine ratio (uACR). There is interest in estimating the intraclass correlation of uACR. Each patient is considered a cluster, and each cluster has two observations. The uACR measurements are right-skewed, and the empirical distributions of the first and second uACR measurements were comparable (Figure \ref{fig:scatteracr}). The rank ICC estimate was 0.217 (95\% CI: 0.140-0.295, Table \ref{tab:acr}). The traditional ICC estimate on the original scale obtained from a random effects model was 0.493, which was driven by a single pair of measurements with extreme values. After removing that pair, the rank ICC estimate was almost unchanged (0.213, 95\% CI: 0.136-0.291) while the traditional ICC on the original scale dropped dramatically to 0.160, illustrating the robustness of the rank ICC compared to the traditional ICC. Instead of removing extreme observations, one could consider transforming the data. The traditional ICC estimate was 0.254 after log transformation and 0.345 after square root transformation, illustrating the sensitivity of the traditional ICC to the choice of scale. 

\subsection{Status Epilepticus}
\label{s:seizures}
\begin{figure}
            \centering
            \includegraphics[width=0.8\textwidth]{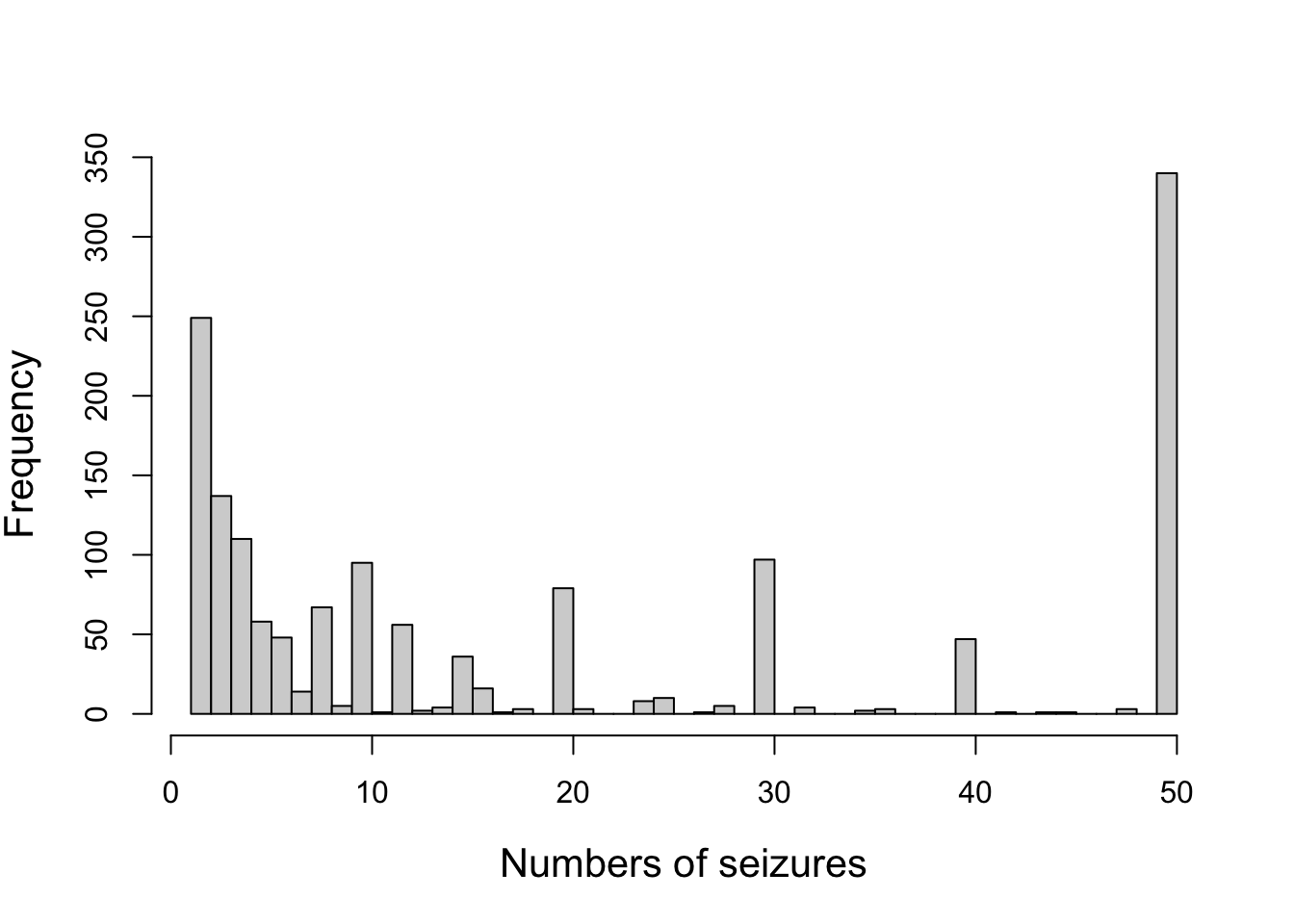}
            \caption[]{Histogram of numbers of seizures of children with untreated epilepsy from the 60 primary healthcare centers in the example of status epilepticus.} 
        \label{fig:seizures}
\end{figure}

The Bridging the Childhood Epilepsy Treatment Gap in Africa (BRIDGE) study is a non-inferiority randomized clinical trial of childhood epilepsy care at 60 randomly selected primary healthcare centers (PHCs) in northern Nigeria \citep{aliyu2019}. The trial is designed to understand if task-shifting childhood epilepsy treatment by trained community health workers can be as effective at reducing seizures as treatment by trained physicians. The study recruited 1507 children with untreated epilepsy from the participating PHCs. Each child's number of seizures in the six months prior to randomization was collected (Figure \ref{fig:seizures}), with a median of 10 (range 1-50). There is interest in estimating the ICC for the number of seizures across PHCs. Cluster size ranged from 19 to 31 children per PHC. Since the PHCs were the units of randomization in this study, it seems reasonable to treat them equally. The rank ICC based on assigning equal weights to clusters was estimated as 0.0482 (95\% CI: 0.023-0.073), which suggested low association between the number of seizures in children within a PHC (Table \ref{tab:seizures}). Other methods of weight assignment yielded similar estimates: assigning equal weights to children resulted in an estimate of 0.0514 (95\% CI: 0.025-0.078), the ESS weighting approach yielded 0.0496 (95\% CI: 0.024-0.075), and the combination weighting approach resulted in 0.0512 (95\% CI: 0.025-0.078). For comparison, the ICC estimated using a linear random effects model was 0.0426, and the ICCs estimated using generalized linear random effects models were 0.0268 (quasi-Poisson) and 0.0168 (negative binomial) \citep{nakagawa2017}.

\subsection{Patient Health Questionaire-9 Score}
\begin{figure}
            \centering
            \includegraphics[width=0.8\textwidth]{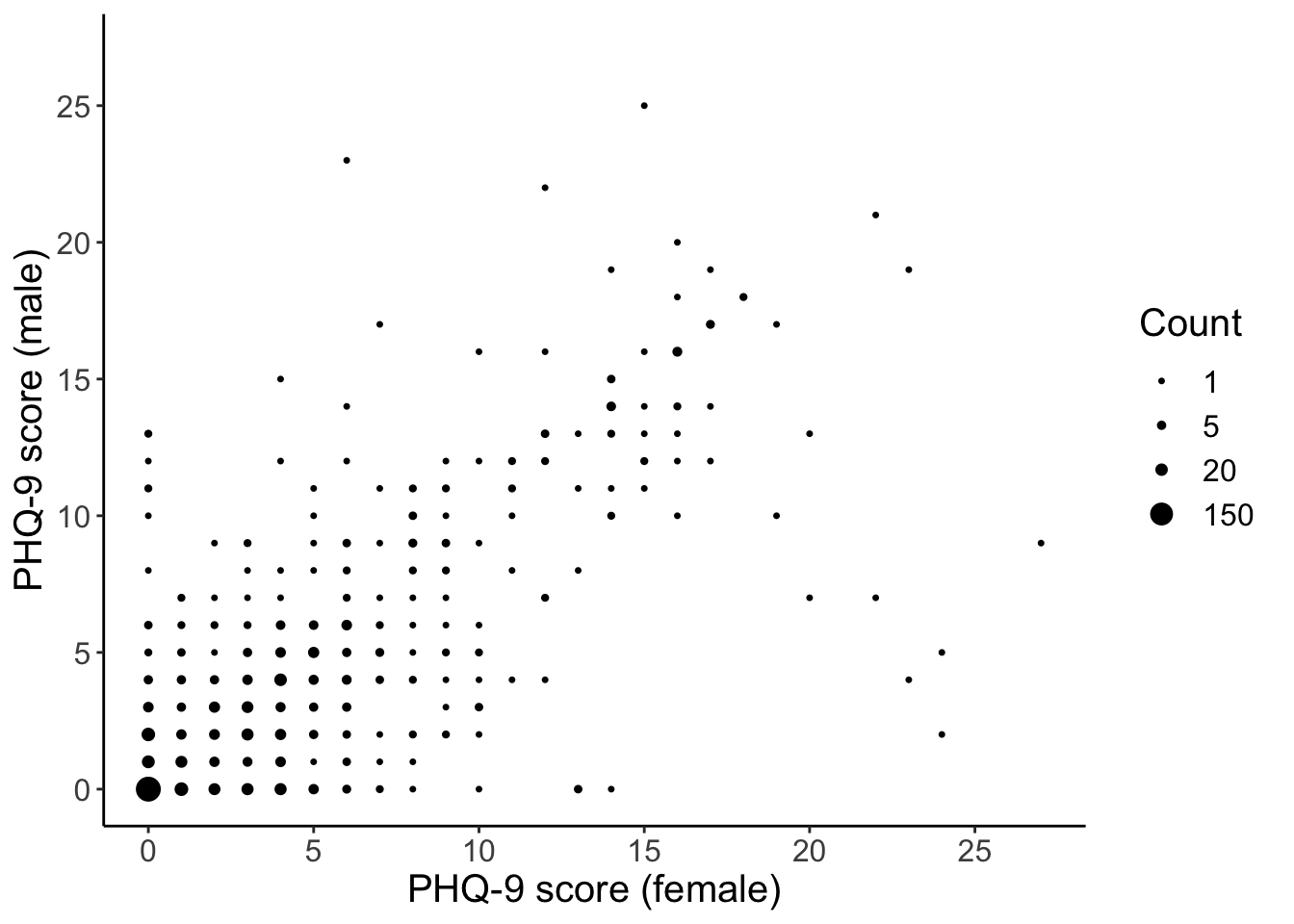}
            \caption[]{Scatter plot of PHQ-9 scores of male and female partners enrolled in the clustered randomized clinical trial in the example of Patient Health Questionaire-9 score.} 
        \label{fig:scatterphq}
\end{figure}

In a third example, we used baseline data from the Homens para $\text{Sa}\acute{\text{u}}\text{de}$ Mais (HoPS+) study, a clustered randomized controlled trial in $\text{Zamb}\acute{\text{e}}\text{zia}$ Province, Mozambique \citep{audet2018}. The trial aimed to measure the impact of incorporating male partners with HIV into prenatal care for pregnant women living with HIV on retention in care, adherence to treatment, and mother-to-child HIV transmission. The trial enrolled 813 couples living with HIV (with a pregnant female) at 24 clinical sites. Depressive symptoms at the time of study enrollment were evaluated with the Patient Health Questionaire-9 (PHQ-9), a nine-item scale that measures depressive symptoms over the previous two weeks. The ordinal PHQ-9 score had a median of 2 (interquartile range 0-5), ranging from 0 to 27 (Figure \ref{fig:scatterphq}). The data have three levels: the innermost level is the person, the middle level is the couple, and the outer level is the clinical site. The number of couples at a clinical site ranged from 2 to 68. Our estimates assigned equal weights to couples. The estimated rank ICC at the couple level, $\hat \gamma_{I2}$, was 0.678 (95\% CI: 0.518-0.838), suggesting substantial clustering of PHQ-9 scores within couples (Table \ref{tab:phq}). The estimated rank ICC at the clinical site level, $\hat \gamma_{I3}$, was 0.397 (95\% CI: 0.242-0.552), which was higher than expected, suggesting a fairly high correlation within clinics. This was confirmed by the estimated rank ICC among females at the clinical level (0.418, 95\% CI: 0.260-0.576) and among males (0.395, 95\% CI: 0.243-0.548). For comparison, the ICC estimates obtained from a linear random effects model were 0.792 at the couple level and 0.474 at the clinical site level, both larger than their rank ICC counterparts. 

\section{Discussion}
\label{s:disucssion}
In this paper, we defined the rank ICC as a natural extension of Fisher’s ICC to the rank scale, and described its population parameter. Our approach maintains the spirit of Fisher’s ICC while creating a nonparametric rank ICC measure analogous to Spearman’s rank correlation. The rank ICC is simply interpreted as the rank correlation between a pair of observations from the same cluster. We also extended the rank ICC for distributions with more than two hierarchies (i.e., equation (\ref{gammaIk})). Our estimator of the rank ICC is insensitive to extreme values and skewed distributions, and does not depend on the scale of the data. It is also consistent and asymptotically normal, with low bias and good coverage in our simulations. Our framework is general, and applicable to any orderable variables with estimable distributions. 

We also discussed assigning weights to clusters and observations under different cases when estimating the rank ICC for two-level data with heterogeneous cluster sizes. In general, 
the selection of weights should be driven by subject matter knowledge. However, in practice, there may be uncertainty in how clusters contribute to the underlying distribution, and efficiency considerations might guide the choice of weights.

There is a relationship between the rank ICC and Spearman's rank correlation when the cluster size is two. With two ordered observations per cluster following the same marginal distribution, the population parameter of the rank ICC is mathematically equal to that of Spearman's rank correlation between the first and second observations. However, their estimation procedures differ; in estimating Spearman's rank correlation, we separately estimate the variances and means of the first and second observations, but in estimating $\gamma_I$, we pool the data to estimate their overall variance and mean. For example, in the albumin-creatinine ratio study, the estimate of Spearman's rank correlation between the first and second uACR measurements was 0.236, close but not equal to the rank ICC estimate, 0.217.

Our rank ICC fills an important gap in the analysis of clustered data. Given Fisher’s introduction of the ICC nearly 100 years ago, we are surprised that a rank-based ICC has not been developed until now. We suspect that some researchers may have simply ranked their data and then used the ratio of the between-cluster and total variances on the rank scale as a rank-based ICC measure, as suggested by others for ordered categorical data \citep{hallgren2012, denham2016}. Although not completely unreasonable, such an approach is ad hoc and does not correspond with a sensible population parameter. Alternatively, some researchers may prefer estimating the similarity within clusters via constructing models for continuous and ordered categorical clustered data, in particular random effects models \citep{agresti2001,skrondal2004,koo2016}. With linear mixed models, the ICC is calculated using estimates of the variance of the random effects and the residuals. These model-derived ICC estimates may be sensitive to the choice of the model: e.g., the form of the linear predictor, potential response variable transformation, non-normality of residuals, and/or non-normality of random effects. With generalized random effects models (e.g., for count or ordinal response variables), the ICC is evaluated on the continuous latent variable scale after a link transformation, which complicates interpretation and remains sensitive to model choice \citep{skrondal2004}. These models may also be sensitive to the method used to derive the within-cluster variance \citep{nakagawa2017}. In contrast, our rank ICC does not require fitting a model and provides a simple and interpretable one-number summary of within-cluster similarity across many types of variables. 
 
Our rank ICC has some limitations. It does not adjust for the effect of other variables on within-cluster similarity. For example, in the status epilepticus study (Section \ref{s:seizures}), there may be interest in measuring the intraclass correlation after adjusting for child age. Our rank ICC cannot do this, whereas a model-derived ICC estimate can. Future work could consider extensions to develop covariate-adjusted conditional and partial rank ICCs. An additional limitation is that our rank ICC appears to have slightly negative bias with small numbers of clusters when $\gamma_I$ is large; this problem goes away as the number of clusters increases. Furthermore, our rank ICC can be time-consuming to calculate with very large sample sizes. In such settings, analysts may consider empirically estimating the rank ICC by randomly sampling clusters with replacement, then sampling pairs of observations from the selected clusters, and finally estimating Spearman’s correlation across many sampled pairs. 

Future work could consider the use of the rank ICC in designing clustered randomized controlled trials for skewed or ordered categorical outcomes.

\section*{Acknowledgments}
We would like to thank the study investigators for providing data used in our example applications. This study was supported in part by funding from the National Institutes of Health (R01AI093234; U01DK112271 for Nigerian uACR study; R01NS113171 for BRIDGE; and R01MH113478 for HoPS+). 

\bibliographystyle{biom} 
\bibliography{reference.bib}

\begin{thebibliography}{}

\bibitem[\protect\citeauthoryear{Agresti and Natarajan}{Agresti and
  Natarajan}{2001}]{agresti2001}
Agresti, A. and Natarajan, R. (2001).
\newblock Modeling clustered ordered categorical data: A survey.
\newblock {\em Int Stat Rev} {\bf 69,} 345--371.

\bibitem[\protect\citeauthoryear{Aliyu, Abdullahi, Iliyasu, Salihu, Adamu,
  Sabo, et~al\mbox{.}}{Aliyu et~al.}{2019}]{aliyu2019}
Aliyu, M.~H., Abdullahi, A.~T., Iliyasu, Z., Salihu, A.~S., Adamu, H., Sabo,
  U., et~al. (2019).
\newblock Bridging the childhood epilepsy treatment gap in northern nigeria
  ({BRIDGE)}: Rationale and design of pre-clinical trial studies.
\newblock {\em Contemp Clin Trials Commun} {\bf 15,} 100362.

\bibitem[\protect\citeauthoryear{Audet, Graves, Barreto, De~Schacht, Gong,
  Shepherd, et~al\mbox{.}}{Audet et~al.}{2018}]{audet2018}
Audet, C.~M., Graves, E., Barreto, E., De~Schacht, C., Gong, W., Shepherd,
  B.~E., et~al. (2018).
\newblock Partners-based {HIV} treatment for seroconcordant couples attending
  antenatal and postnatal care in rural mozambique: A cluster randomized trial
  protocol.
\newblock {\em Contemp Clin Trials Commun} {\bf 71,} 63--69.

\bibitem[\protect\citeauthoryear{Bross}{Bross}{1958}]{Bross1958}
Bross, I. D.~J. (1958).
\newblock How to use ridit analysis.
\newblock {\em Biometrics} {\bf 14,} 18--38.

\bibitem[\protect\citeauthoryear{Chakraborty, Solomon, and Anstrom}{Chakraborty
  et~al.}{2021}]{rishi2021}
Chakraborty, H., Solomon, N., and Anstrom, K.~J. (2021).
\newblock A method to estimate intra-cluster correlation for clustered
  categorical data.
\newblock {\em Commun Stat Theory Methods} {\bf 52,} 429--444.

\bibitem[\protect\citeauthoryear{Davison and Hinkley}{Davison and
  Hinkley}{1997}]{davison1997}
Davison, A.~C. and Hinkley, D.~V. (1997).
\newblock {\em Bootstrap Methods and Their Application}.
\newblock Cambridge Series in Statistical and Probabilistic Mathematics.
  Cambridge University Press.

\bibitem[\protect\citeauthoryear{Denham}{Denham}{2016}]{denham2016}
Denham, B. (2016).
\newblock {\em Categorical Statistics for Communication Research}.
\newblock John Wiley and Sons, Ltd.

\bibitem[\protect\citeauthoryear{Donner}{Donner}{1986}]{Donner1986}
Donner, A. (1986).
\newblock A review of inference procedures for the intraclass correlation
  coefficient in the one-way random effects model.
\newblock {\em Int Stat Rev} {\bf 54,} 67--82.

\bibitem[\protect\citeauthoryear{Field and Welsh}{Field and
  Welsh}{2007}]{Field2007}
Field, C.~A. and Welsh, A.~H. (2007).
\newblock Bootstrapping clustered data.
\newblock {\em J R Stat Soc Series B Stat Methodol} {\bf 69,} 369--390.

\bibitem[\protect\citeauthoryear{Fieller and Smith}{Fieller and
  Smith}{1951}]{filler1951}
Fieller, E. and Smith, C. (1951).
\newblock Note on the analysis of variance and intraclass correlation.
\newblock {\em Ann Eugen} {\bf 16,} 97--104.

\bibitem[\protect\citeauthoryear{Fisher}{Fisher}{1925}]{fisher1925}
Fisher, R. (1925).
\newblock {\em Statistical Methods for Research Workers}.
\newblock Oliver \& Boyd: Edinburgh.

\bibitem[\protect\citeauthoryear{Fisher}{Fisher}{1915}]{fisher1915}
Fisher, R.~A. (1915).
\newblock Frequency distribution of the values of the correlation coefficient
  in samples from an indefinitely large population.
\newblock {\em Biometrika} {\bf 10,} 507--521.

\bibitem[\protect\citeauthoryear{Hallgren}{Hallgren}{2012}]{hallgren2012}
Hallgren, K. (2012).
\newblock Computing inter-rater reliability for observational data: an overview
  and tutorial.
\newblock {\em Tutor Quant Methods Psychol} {\bf 8,} 23--34.

\bibitem[\protect\citeauthoryear{Harris}{Harris}{1913}]{Harris1913}
Harris, J.~A. (1913).
\newblock On the calculation of intra-class and inter-class coefficients of
  correlation from class moments when the number of possible combinations is
  large.
\newblock {\em Biometrika} {\bf 9,} 446--472.

\bibitem[\protect\citeauthoryear{Hedges and Hedberg}{Hedges and
  Hedberg}{2007}]{hedges2007}
Hedges, L.~V. and Hedberg, E.~C. (2007).
\newblock Intraclass correlation values for planning group-randomized trials in
  education.
\newblock {\em Educ Eval Policy Anal} {\bf 29,} 60--87.

\bibitem[\protect\citeauthoryear{Karlin, Cameron, and Williams}{Karlin
  et~al.}{1981}]{Karlin1981}
Karlin, S., Cameron, E.~C., and Williams, P.~T. (1981).
\newblock Sibling and parent--offspring correlation estimation with variable
  family size.
\newblock {\em Proc Natl Acad Sci U S A} {\bf 78,} 2664--2668.

\bibitem[\protect\citeauthoryear{Kendall}{Kendall}{1970}]{kendall1970}
Kendall, M. (1970).
\newblock {\em Rank Correlation Methods}.
\newblock Charles Griffin.

\bibitem[\protect\citeauthoryear{Kish}{Kish}{1965}]{kish1965}
Kish, L. (1965).
\newblock {\em Survey Sampling}.
\newblock Wiley.

\bibitem[\protect\citeauthoryear{Koo and Li}{Koo and Li}{2016}]{koo2016}
Koo, T.~K. and Li, M.~Y. (2016).
\newblock A guideline of selecting and reporting intraclass correlation
  coefficients for reliability research.
\newblock {\em J Chiropr Med} {\bf 15,} 155--163.

\bibitem[\protect\citeauthoryear{Kruskal}{Kruskal}{1958}]{kruskal1958}
Kruskal, W.~H. (1958).
\newblock Ordinal measures of association.
\newblock {\em J Am Stat Assoc} {\bf 53,} 814--861.

\bibitem[\protect\citeauthoryear{Murray, Varnell, and Blitstein}{Murray
  et~al.}{2004}]{murray2004}
Murray, D.~M., Varnell, S.~P., and Blitstein, J.~L. (2004).
\newblock Design and analysis of group-randomized trials: a review of recent
  methodological developments.
\newblock {\em Am J Public Health} {\bf 94,} 423--432.

\bibitem[\protect\citeauthoryear{Nakagawa, Johnson, and Schielzeth}{Nakagawa
  et~al.}{2017}]{nakagawa2017}
Nakagawa, S., Johnson, P. C.~D., and Schielzeth, H. (2017).
\newblock The coefficient of determination r2 and intra-class correlation
  coefficient from generalized linear mixed-effects models revisited and
  expanded.
\newblock {\em J R Soc Interface} {\bf 14,} 20170213.

\bibitem[\protect\citeauthoryear{Pearson}{Pearson}{1907}]{Pearson1907}
Pearson, K.~G. (1907).
\newblock {\em On Further Methods of Determining Correlation}.
\newblock Cambridge University Press.

\bibitem[\protect\citeauthoryear{Rothery}{Rothery}{1979}]{Rothery1979}
Rothery, P. (1979).
\newblock A nonparametric measure of intraclass correlation.
\newblock {\em Biometrika} {\bf 66,} 629--639.

\bibitem[\protect\citeauthoryear{Shirahata}{Shirahata}{1981}]{Shirahata1981}
Shirahata, S. (1981).
\newblock Intraclass rank tests for independence.
\newblock {\em Biometrika} {\bf 68,} 451--456.

\bibitem[\protect\citeauthoryear{Shirahata}{Shirahata}{1982}]{Shirahata1982}
Shirahata, S. (1982).
\newblock Nonparametric measures of intraclass correlation.
\newblock {\em Commun Stat Theory Methods} {\bf 11,} 1707--1721.

\bibitem[\protect\citeauthoryear{Shrout and Fleiss}{Shrout and
  Fleiss}{1979}]{Shrout1979}
Shrout, P. and Fleiss, J. (1979).
\newblock Intraclass correlations: uses in assessing rater reliability.
\newblock {\em Psychol Bull} {\bf 86,} 420—428.

\bibitem[\protect\citeauthoryear{Siddiqui, Hedeker, Flay, and Hu}{Siddiqui
  et~al.}{1996}]{Siddiqui1996}
Siddiqui, O., Hedeker, D., Flay, B.~R., and Hu, F.~B. (1996).
\newblock Intraclass correlation estimates in a school-based smoking prevention
  study. outcome and mediating variables, by sex and ethnicity.
\newblock {\em Am J Epidemiol} {\bf 144 4,} 425--33.

\bibitem[\protect\citeauthoryear{Skrondal and Rabe-Hesketh}{Skrondal and
  Rabe-Hesketh}{2004}]{skrondal2004}
Skrondal, A. and Rabe-Hesketh, S. (2004).
\newblock {\em Generalized Latent Variable Modeling: Multilevel, Longitudinal
  and Structural Equation Models}.
\newblock Chapman \& Hall/CRC.

\bibitem[\protect\citeauthoryear{Wudil, Aliyu, Prigmore, Ingles, Ahonkhai,
  Musa, et~al\mbox{.}}{Wudil et~al.}{2021}]{wudil2021}
Wudil, U., Aliyu, M., Prigmore, H., Ingles, D., Ahonkhai, A., Musa, B., et~al.
  (2021).
\newblock Apolipoprotein-1 risk variants and associated kidney phenotypes in an
  adult hiv cohort in nigeria.
\newblock {\em Kidney Int} {\bf 100,} 146--154.

\end{thebibliography}

\section*{Supporting Information}
Web Appendices, Tables, and Figures referenced in Sections \ref{s:intro}, \ref{s:estimation}, and \ref{s:simulations} are available with this paper on arXiv.

\newpage
\begin{table}
    \centering
    \caption{Estimates of rank ICC and traditional ICC of uACR in the example of albumin-creatinine ratio.}
    \begin{tabular}{>{\RaggedRight}p{2.5cm} >{\RaggedRight}p{3cm} >{\RaggedRight}p{3cm} >{\RaggedRight}p{3cm} >{\RaggedRight}p{3cm}}
        \toprule & \textbf{Original data} & \textbf{Extreme values removed} & \textbf{Log transformation} & \textbf{Square root transformation} \\
        \midrule 
        \textbf{Rank ICC [95\% CI]} & 0.217 [0.140, 0.295] & 0.213 [0.136, 0.291] & 0.217 [0.140, 0.295] & 0.217 [0.140, 0.295] \\
        \textbf{ICC} & 0.493 & 0.160 & 0.254 & 0.345 \\
        \bottomrule
        \hspace{3em}
    \end{tabular}
    \label{tab:acr}
\end{table}
       
\begin{table}
    \centering
    \caption{Estimates of rank ICC and traditional ICC for the number of seizures across the primary healthcare centers in the sample of status epilepticus.}
    \begin{tabular}{>{\RaggedRight}p{3cm}>{\RaggedRight}p{7.5cm}>{\RaggedRight}p{4cm}}
        \toprule
         & Equal weights for clusters & 0.0482   [0.023, 0.073]\\
        \textbf{Rank ICC [95\% CI]}& Equal weights for observations & 0.0514  [0.025, 0.078] \\
         & Iterative weighting based on the effective sample size & 0.0496 [0.024, 0.075]\\
        & Iterative weighting based on the combination  & 0.0512 [0.025, 0.078]\\
        \midrule
          &Linear & 0.0426  \\
        \textbf{ICC} & Quasi-Poisson link  &  0.0268 \\
          &  Negative binomial link &  0.0168 \\
        \bottomrule
        \hspace{3em}
    \end{tabular}
    \label{tab:seizures}
\end{table}
        
\begin{table}
      \centering
      \caption{Estimates of rank ICC and traditional ICC of PHQ-9 score at the couple level and the clinical site level in the example of Patient Health Questionaire-9 score.}
      \begin{tabular}{>{\RaggedRight}p{2.5cm} >{\RaggedRight}p{3cm} >{\RaggedRight}p{3cm} >{\RaggedRight}p{3cm} >{\RaggedRight}p{3cm}}
        \toprule
         & \textbf{The couple level} & \textbf{The clinical site level} & \textbf{Females at the clinical level} & \textbf{Males at the clinical level} \\
        \midrule
        \textbf{Rank ICC [95\% CI]} & 0.678 [0.518, 0.838] & 0.397 [0.242, 0.552] & 0.418 [0.260, 0.576] & 0.395 [0.243, 0.548] \\
        \textbf{ICC} & 0.792 & 0.474 & 0.452 & 0.497 \\
        \bottomrule
      \end{tabular}
      \label{tab:phq}
    \end{table}
    
\clearpage

\newpage 

\newpage 

\newpage 

\newpage 

\newpage 

\newpage 
\clearpage
\newpage 

\clearpage
\newpage 

\end{document}

\label{firstpage}
\maketitle
\section{Web Appendix A}
\label{s:appendixA}
\subsection{Proof of asymptotic properties of $\hat \gamma_I$}

Let $g_n(\textbf{x})$ and $ h_n(\textbf{x})$ be two functions such that $\hat \gamma_I = [\frac{1}{n}\sum_{i=1}^n g_n(\textbf{x}_i)] /[\frac{1}{n} \sum_{i=1}^n h_n(\textbf{x}_i)]$, where $\textbf{x}_i = (x_{i1},...,x_{ik_i})$ denotes a vector of observations in a cluster. Let $\textbf{P}_n$ denotes an empirical measure and $\textbf{P}$ denotes the underlying probability measure such that $\textbf{P}_n \{g(\textbf{x})\} = \frac{1}{n} \sum_{i=1}^n g(\textbf{x}_i)$ and $E[\textbf{P}_n \{g(\textbf{x})\}] = \textbf{P} \{g(\textbf{x})\}$, and same for $h(\textbf{x})$. We then have $\hat \gamma_I =\frac{\textbf{P}_n\{g_n(\textbf{x})\}}{\textbf{P}_n\{h_n(\textbf{x})\}}$ and $\gamma_I = \frac{\textbf{P}\{g(\textbf{x})\}}{\textbf{P}\{h(\textbf{x})\}}$. Here $\hat F^*$ belongs to a Donsker class which contains all distribution functions in $X$'s support, where a Donsker class is a sets of functions with the useful property that empirical processes indexed by the class converge weakly to a certain Gaussian process.\\\\
ASSUMPTION 1 \;\; For any function $\Tilde{f}$ in a Donsker class, there exits a corresponding function $f$ in another Donsker class such that $\sum_{i}\sum_{j} w_{ij}\Tilde{f}(x_{ij}) - E[\Tilde{f}(x_{ij})] = (\textbf{P}_n - \textbf{P})\{f(\textbf{x})\} + O_p(1/\sqrt{n})$. \\\\
REMARK 1 \;\; Assumption 1 trivially holds if $w_{ij}=\frac{1}{nk_i}$, where $k_i$ is a bounded variable and the two Donsker classes are the same. The asymptotic properties of $\hat \gamma_I$ shown under $w_{ij}=\frac{1}{nk_i}$ in the followings are also valid for any $w_{ij}$ (e.g., $\frac{1}{\sum_i ki}$) that satisfies Assumption 1. \\\\
We consider $w_{ij}=\frac{1}{nk_i}$ in the followings. Because $\hat F^*$ belongs to a Donsker class which contains all distribution functions in $X$'s support and $\hat F^*$ uniformly converges to $F^*$ with probability one, $g_n$ and $h_n$ also belong to some Donsker classes based on Assumption 1 and converge in $L_2(P)$ distance to their limits $g(\textbf{x})$ and $h(\textbf{x})$, respectively, where $g(\textbf{x})=E\big[\sum_{j'>j}\frac{2w_{i.}}{k_i(k_i-1)} [ F^*(x_{ij}) - E(F^*)][F^*(x_{ij'}) - E(F^*)]\big]$ and $h(\textbf{x})=E\big[\sum_{j=1}^{k_i} w_{ij} [ F^*(x_{ij}) -E(F^*)]^2 \big]$. Also, $|g_n - g| \rightarrow 0$, $|h_n - h| \rightarrow 0$.\\\\
Then $\sqrt{n}(\hat \gamma_I  - \gamma_I)$ can be expressed in terms of the empirical process $\sqrt{n}(\textbf{P}_n-\textbf{P})$,
 \begin{align*}
& \sqrt{n}(\hat \gamma_I  - \gamma_I) = \sqrt{n} \Big(\frac{\textbf{P}_n\{g_n(\textbf{x})\}}{\textbf{P}_n\{h_n(\textbf{x})\}} - \gamma_I \Big)\\
& = \sqrt{n} \Big(\frac{\textbf{P}_n\{g_n(\textbf{x})\}}{\textbf{P}_n\{h_n(\textbf{x})\}} - \frac{\textbf{P}\{g_n(\textbf{x})\}}{\textbf{P}\{h_n(\textbf{x})\}}  + \frac{\textbf{P}\{g_n(\textbf{x})\}}{\textbf{P}\{h_n(\textbf{x})\}} -\gamma_I \Big)\\
& = \sqrt{n} \Big(\frac{\textbf{P}_n\{g_n(\textbf{x})\}}{\textbf{P}_n\{h_n(\textbf{x})\}} - \frac{\textbf{P}\{g_n(\textbf{x})\}}{\textbf{P}_n\{h_n(\textbf{x})\}} + \frac{\textbf{P}\{g_n(\textbf{x})\}}{\textbf{P}_n\{h_n(\textbf{x})\}}  - \frac{\textbf{P}\{g_n(\textbf{x})\}}{\textbf{P}\{h_n(\textbf{x})\}}  + \frac{\textbf{P}\{g_n(\textbf{x})\}}{\textbf{P}\{h_n(\textbf{x})\}} -\gamma_I \Big)\\
 & =\sqrt{n}(\textbf{P}_n-\textbf{P})\big\{\frac{g_n(\textbf{x})}{\textbf{P}_n\{h_n(\textbf{x})\}}  - \frac{ h_n(\textbf{x})\textbf{P}\{g_n(\textbf{x})\}}{\textbf{P}_n\{h_n(\textbf{x})\}\textbf{P}\{h_n(\textbf{x})\}}\} + \sqrt{n}\Big[\frac{\textbf{P}\{g_n(\textbf{x})\}}{\textbf{P}\{h_n(\textbf{x})\}} - \gamma_I \Big]\\
 & = \sqrt{n}(\textbf{P}_n-\textbf{P}) \big\{\frac{g(\textbf{x})}{E[h(\textbf{x})]} - \frac{ h(\textbf{x})E[g(\textbf{x})]}{(E[h(\textbf{x})])^2}\big\} + O_p(1) + \sqrt{n}\Big[\frac{\textbf{P}\{g_n(\textbf{x})\}}{\textbf{P}\{h_n(\textbf{x})\}} - \gamma_I \Big]
 \end{align*}
We perform the linearization for $\sqrt{n}\Big[\frac{\textbf{P}\{g_n(\textbf{x})\}}{\textbf{P}\{h_n(\textbf{x})\}} - \gamma_I \Big]$ around $(F^*, \bar F^*)$, where $\bar F^*$ denotes the functional component of $E(F^*)$.
\begin{align*}
\sqrt{n}\Big[\frac{\textbf{P}\{g_n(\textbf{x})\}}{\textbf{P}\{h_n(\textbf{x})\}} - \gamma_I \Big]    & = \sqrt{n}\frac{\textbf{P}\big\{\nabla_{F^*} g_n(\textbf{x})(\hat F^* - F^*)\big\}}{\textbf{P}\{h_n(\textbf{x})\}}\\
& \:\:\:\:\:  - \sqrt{n}\frac{\textbf{P}\{g_n(\textbf{x})\}}{(\textbf{P}\{h_n(\textbf{x})\})^2} \textbf{P}\{\nabla_{F^*} h_n(\textbf{x}) (\hat F^* - F^*)\} \\
& \:\:\:\:\:  +\sqrt{n}\frac{\textbf{P}\big\{\nabla_{\bar F^*} g_n(\textbf{x}) (\bar{\hat{F^*}} - E[F^*]) \big\}}{\textbf{P}\{h_n(\textbf{x})\}}\\
& \:\:\:\:\:  -\sqrt{n}\frac{\textbf{P}\{g_n(\textbf{x})\}}{(\textbf{P}\{h_n(\textbf{x})\})^2}\textbf{P}\{\nabla_{\bar F^*} h_n(\textbf{x})   (\bar{\hat{F^*}} - E[F^*])\}  +  O_p(1)\\
& = (1) + (2) + (3) + (4) +  O_p(1)
\end{align*}
We then transform $(1)-(4)$ into expressions of $\sqrt{n}(\textbf{P}_n-\textbf{P})$.
%%%%%%%%%%%%%%%%%%%%%%%%%%%%%%%%%
\allowdisplaybreaks
\begin{align*} (1) &   = \sqrt{n}\frac{\textbf{P}\big\{\nabla_{F^*} g_n(\textbf{x})(\hat F^* - F^*)\big\}}{\textbf{P}\{h_n(\textbf{x})\}} \\
& \text{Obtain derivative of each $F^*(x_{ij})$}\\
(1) & =  \sqrt{n} \textbf{P} \Big\{\frac{2w_{i.}}{k_i(k_i-1)}  \sum\limits_{j'>j} (\hat F^*(x_{ij'}) -\bar{\hat{F^*}})(\hat F^*(x_{ij}) - F^*(x_{ij}))\Big\} \big/ \textbf{P}\{h_n(\textbf{x})\}  \\
& \:\:\:\:\:  +  \sqrt{n} \textbf{P} \Big\{\frac{2w_{i.}}{k_i(k_i-1)}  \sum\limits_{j'>j} (\hat F^*(x_{ij}) -\bar{\hat{F^*}})(\hat F^*(x_{ij'}) - F^*(x_{ij'}))\Big\} \big/ \textbf{P}\{h_n(\textbf{x})\}  \\
& =  \sqrt{n}  E \Big[\frac{2w_{i.}}{k_i(k_i-1)}  \sum\limits_{j'>j} (\hat F^*(x_{ij'}) -\bar{\hat{F^*}})(\hat F^*(x_{ij}) - F^*(x_{ij}))\Big] \big/ E[h_n(\textbf{x})]  \\
& \:\:\:\:\:  + \sqrt{n}  E \Big[\frac{2w_{i.}}{k_i(k_i-1)}  \sum\limits_{j'>j} (\hat F^*(x_{ij}) -\bar{\hat{F^*}})(\hat F^*(x_{ij'}) - F^*(x_{ij'}))\Big]  \big/ E[h_n(\textbf{x})]  \\
& \text{Plug in the expression of $\hat F^*(x_{ij})$ and $\hat F^*(x_{ij'})$} \\
(1) & =\frac{\sqrt{n} }{E[h_n(\textbf{x})] } \Big\{ E \Big[\frac{2w_{i.}}{k_i(k_i-1)}  \sum\limits_{j'>j} \Big (\hat F^*(x_{ij'}) -\bar{\hat{F^*}} \Big) \Big( \sum\limits_{i^{'}}\sum\limits_{j^{''}} w_{i^{'}  j^{''}} [I(x_{i^{'} j^{''}} < x_{ij}) \\
& \:\:\:\:\: +I(x_{i^{'}j^{''}} \leq x_{ij})]/2 - F^*(x_{ij})+ \frac{2w_{i.}}{k_i(k_i-1)}  \sum\limits_{j'>j} \Big (\hat F^*(x_{ij}) -\bar{\hat{F^*}} \Big)\\
& \:\:\:\:\: \times \Big( \sum\limits_{i^{'}}\sum\limits_{j^{''}} w_{i^{'}  j^{''}} [I(x_{i^{'} j^{''}} < x_{ij'})+ I(x_{i^{'}j^{''}} \leq x_{ij'})]/2 - F^*(x_{ij'}) \Big)\Big] \Big\} \\
& \text{Take $i'$ and $j^{''}$ outside the expectation}\\
(1) & =\frac{\sqrt{n} }{E[h_n(\textbf{x})] } \sum\limits_{i^{'}}\sum\limits_{j^{''}} w_{i^{'}  j^{''}} E  \Big\{\frac{2w_{i.}}{k_i(k_i-1)}  \sum\limits_{j'>j} \Big (\hat F^*(x_{ij'}) -\bar{\hat{F^*}} \Big) \Big(  I(x_{i^{'} j^{''}} < x_{ij})\\
& \:\:\:\:\:  +I(x_{i^{'}j^{''}} \leq x_{ij})\Big)/2 + \frac{2w_{i.}}{k_i(k_i-1)}  \sum\limits_{j'>j} \Big (\hat F^*(x_{ij}) -\bar{\hat{F^*}} \Big) \Big(  I(x_{i^{'} j^{''}} < x_{ij'})\\
& \:\:\:\:\:  +I(x_{i^{'}j^{''}} \leq x_{ij'})\Big)/2\Big\} -  \frac{\sqrt{n} }{E[h_n(\textbf{x})] } E \Big\{\frac{2w_{i.}}{k_i(k_i-1)}  \sum\limits_{j'>j} \Big [(\hat F^*(x_{ij'}) -\bar{\hat{F^*}}) F^*(x_{ij}) \\
& \:\:\:\:\: + (\hat F^*(x_{ij}) -\bar{\hat{F^*}})  F^*(x_{ij'})\Big] \Big\}\\
&\text{The expectation in the first component is a function of $x_{i'j^{''}}$,}\\
&\text{we denote it as $\Tilde{a}_1(x_{i'j^{''}})$}\\
(1) & = \frac{\sqrt{n} }{E[h(\textbf{x})] } \{\sum\limits_{i^{'}}\sum\limits_{j^{''}} w_{i^{'}  j^{''}} \Tilde{a}_1(x_{i'j^{''}}) - E[\Tilde{a}_1(x_{i'j^{''}})]\} + O_p(1) \\
& \text{Because $\Tilde{a}_1(x_{ij})$ belongs to a Donsker class}\\
(1) & =  \frac{\sqrt{n} }{E[h(\textbf{x})] } (\textbf{P}_n - \textbf{P})\{ a_1(\textbf{x})\} + O_p(1) \text{ (under Assumption 1)}\\ 
\end{align*}
The derivation of $(2)-(4)$ is similar to that of $(1)$ and under Assumption 1.
 \begin{align*}
 %%%%%%%%%%%%%%%%%%%%%%%%%%%%%%%%%
(2)&  =  - \sqrt{n}\frac{\textbf{P}\{g_n(\textbf{x})\}}{(\textbf{P}\{h_n(\textbf{x})\})^2} \textbf{P}\{\nabla_{F^*} h_n(\textbf{x}) (\hat F^* - F^*)\} \\
& - \sqrt{n}\frac{\textbf{P}\{g_n(\textbf{x})\}}{(\textbf{P}\{h_n(\textbf{x})\})^2} \textbf{P}\Big\{ 2\sum\limits_{j} w_{ij} (\hat F^*(x_{ij}) -\bar{\hat{F^*}}) ( \hat F^*(x_{ij}) - F^*(x_{ij}))\Big\}  \\
 & = - \sqrt{n}\frac{ E[g_n(\textbf{x})]}{(E[h_n(\textbf{x})])^2}  E \Big\{2\sum\limits_{j} w_{ij} (\hat F^*(x_{ij}) -\bar{\hat{F^*}}) ( \hat F^*(x_{ij}) - F^*(x_{ij}))  \Big\}\\
 & = - \sqrt{n}\frac{ E[g_n(\textbf{x})]}{(E[h_n(\textbf{x})])^2} E \Big\{2\sum\limits_{j} w_{ij} (\hat F^*(x_{ij}) -\bar{\hat{F^*}}) \Big( \sum\limits_{i^{'}}\sum\limits_{j^{'}} w_{i^{'}  j^{'}} [I(x_{i^{'} j^{'}} < x_{ij})\\
 & \:\:\:\:\:  + I(x_{i^{'}j^{'}} \leq x_{ij})]/2 - F^*(x_{ij}) \Big) \Big\}\\
 & = - \sqrt{n}\frac{ E[g_n(\textbf{x})]}{(E[h_n(\textbf{x})])^2}  \sum\limits_{i^{'}}\sum\limits_{j^{'}}  w_{i^{'}  j^{'}}  E\Big\{2 \sum\limits_{j} w_{ij} (\hat F^*(x_{ij}) -\bar{\hat{F^*}}) [I(x_{i^{'} j^{'}} < x_{ij})\\
 & \:\:\:\:\:  +I(x_{i^{'}j^{'}} \leq x_{ij})]/2\Big\} + \sqrt{n}\frac{ E[g_n(\textbf{x})]}{(E[h_n(\textbf{x})])^2}  E\Big\{2 \sum\limits_{j} w_{ij}  (\hat F^*(x_{ij}) -\bar{\hat{F^*}}) F^*(x_{ij}) \Big\}\\
 & \text{Let $\Tilde{a}_2(x_{i'j'})$ denote the expectation in the first component}\\
(2) & = - \sqrt{n}\frac{ E[g_n(\textbf{x})]}{(E[h_n(\textbf{x})])^2}  \{ \sum\limits_{i^{'}}\sum\limits_{j^{'}}  w_{i^{'}  j^{'}} \Tilde{a}_2(x_{i'j'})  - E[ \Tilde{a}_2(x_{i'j'}) ]\} + O_p(1) \\
& = - \frac{\sqrt{n} E[g(\textbf{x})]}{(E[h(\textbf{x})])^2} (\textbf{P}_n - \textbf{P})\{a_2(\textbf{x}) \} + O_p(1)\\\\
%%%%%%%%%%%%%%%%%%%%%%%%%%%%%%%%%
(3) & = \sqrt{n}\frac{\textbf{P}\big\{\nabla_{\bar F^*} g_n(\textbf{x}) (\bar{\hat{F^*}} - E[F^*]) \big\}}{\textbf{P}\{h_n(\textbf{x})\}}\\
& =  \frac{-\sqrt{n}}{\textbf{P}\{h_n(\textbf{x})\}}\textbf{P}\Big\{\frac{2w_{i.}}{k_i(k_i-1)}  \big[\sum\limits_{j'>j} (\hat F^*(x_{ij}) -\bar{\hat{F^*}} + \hat F^*(x_{ij'}) -\bar{\hat{F^*}})\big] (\bar{\hat{F^*}} - E[F^*])\Big\}  \\
& = \frac{-\sqrt{n}}{E[h_n(\textbf{x})]} E \Big\{ \sum\limits_{j'>j} \frac{2w_{i.}}{k_i(k_i-1)} [\hat F^*(x_{ij'}) - \bar{\hat{F^*}} + \hat F^*(x_{ij}) - \bar{\hat{F^*}}] (\bar{\hat{F^*}} - E[F^*])\} \\
& = -  \frac{\sqrt{n}}{E[h_n(\textbf{x})]}  E \Big\{    \Big( \sum\limits_{j'>j} \frac{2w_{i.}}{k_i(k_i-1)} [\hat F^*(x_{ij'}) - \bar{\hat{F^*}} + \hat F^*(x_{ij}) - \bar{\hat{F^*}}]\Big) \Big(\sum\limits_{i^{'}}\sum\limits_{j^{'}} w_{i'j'} \\
& \:\:\:\:\:  \big[ \hat F^*(x_{i'j'}) + \frac{1}{n}  \sum\limits_{i}\sum\limits_{j}  w_{ij}  [I(x_{i'j'} < x_{ij})+I(x_{i'j'} \leq x_{ij})]/2 \big] - E[F^*] \Big)\Big\}\\
& = -  \frac{\sqrt{n}}{E[h_n(\textbf{x})]}  \sum\limits_{i^{'}}\sum\limits_{j^{'}} w_{i' j^{'}}  E \Big\{   \Big( \sum\limits_{j'>j} \frac{2w_{i.}}{k_i(k_i-1)} [\hat F^*(x_{ij'}) - \bar{\hat{F^*}} + \hat F^*(x_{ij}) - \bar{\hat{F^*}}]\Big) \\
& \:\:\:\:\: \times \Big( \hat F^*(x_{i'j'}) + \frac{1}{n}  \sum\limits_{i}\sum\limits_{j}  w_{ij}  [I(x_{i'j'} < x_{ij}) +I(x_{i'j'} \leq x_{ij})]/2 \Big)\Big\}\\
& \:\:\:\:\:  + \frac{\sqrt{n}}{E[h_n(\textbf{x})]}E \Big\{   \sum\limits_{j'>j} \frac{2w_{i.}}{k_i(k_i-1)} [\hat F^*(x_{ij'}) - \bar{\hat{F^*}} + \hat F^*(x_{ij}) - \bar{\hat{F^*}}] E[F^*]\Big\}\\
 & \text{We denote the expectation in the first component as $\Tilde{a}_3(x_{i'j'})$}\\
(3) & = -  \frac{\sqrt{n}}{E[h_n(\textbf{x})]} \{\sum\limits_{i^{'}}\sum\limits_{j^{'}} w_{i' j^{'}} \Tilde{a}_3(x_{i'j'}) - E[\Tilde{a}_3(x_{i'j'})]\}\\
& =  -  \frac{\sqrt{n}}{E[h(\textbf{x})]} (\textbf{P}_n - \textbf{P})\{a_3(\textbf{x})\} + O_p(1)\\\\
%%%%%%%%%%%%%%%%%%%%%%%%%%%%%%%%%
(4) & =-\sqrt{n}\frac{\textbf{P}\{g_n(\textbf{x})\}}{(\textbf{P}\{h_n(\textbf{x})\})^2}\textbf{P}\{\nabla_{\bar F^*} h_n(\textbf{x})   (\bar{\hat{F^*}} - E[F^*])\}\\ 
& = \sqrt{n}\frac{E[g_n(\textbf{x})]}{(E[h_n(\textbf{x})])^2} \textbf{P}\Big\{ 2\sum\limits_{j} w_{ij} (\hat F^*(x_{ij}) - \bar{\hat{F^*}})  (\bar{\hat{F^*}}- E[F^*])\Big\}  \\
& =  \sqrt{n}\frac{E[g_n(\textbf{x})]}{(E[h_n(\textbf{x})])^2}  \sum\limits_{i^{'}}\sum\limits_{j^{'}} w_{i' j^{'}}  E \Big\{  \sum\limits_{j} 2w_{ij} (\hat F^*(x_{ij}) - \bar{\hat{F^*}}) \\
& \:\:\:\:\: \:\:\:\:\:  \times \big( \hat F^*(x_{i'j'}) + \frac{1}{n}  \sum\limits_{i}\sum\limits_{j}  w_{ij}  [I(x_{i'j'} < x_{ij})+I(x_{i'j'} \leq x_{ij})]/2\big)\Big\}\\
& \:\:\:\:\:  - \sqrt{n}\frac{E[g(\textbf{x})]}{(E[h(\textbf{x})])^2} E \Big\{    2\sum\limits_{j} w_{ij} (\hat F^*(x_{ij}) - \bar{\hat{F^*}}) E[F^*]\Big\}\\
 & \text{The expectation in the first component is denoted as $\Tilde{a}_4(x_{i'j'})$}\\
(4) & = \sqrt{n}\frac{E[g_n(\textbf{x})]}{(E[h_n(\textbf{x})])^2}  \{\sum\limits_{i^{'}}\sum\limits_{j^{'}} w_{i' j^{'}} \Tilde{a}_4(x_{i'j'}) - E[\Tilde{a}_4(x_{i'j'})]\}\\
& =  \sqrt{n}\frac{E[g(\textbf{x})]}{(E[h(\textbf{x})])^2}  (\textbf{P}_n - \textbf{P})\{a_4(\textbf{x})\} + O_p(1)
\end{align*}
%%%%%%%%%%%%%%%%%%%%%%%%%%%%%%%%%
With the derivation of $(1)-(4)$ above, we can express $\sqrt{n}\Big[\frac{\textbf{P}\{g_n(\textbf{x})\}}{\textbf{P}\{h_n(\textbf{x})\}} - \gamma_I \Big]$ by the empirical process. Let $a(\textbf{x}) =\frac{1}{E[h(\textbf{x})]}[a_1(\textbf{x})-a_3(\textbf{x})] - \frac{E[g(\textbf{x})]}{(E[h(\textbf{x})])^2}[a_2(\textbf{x})-a_4(\textbf{x})]$, note $a(\textbf{x})$ is a deterministic function of $\textbf{x}$. We then have $\sqrt{n}\Big[\frac{\textbf{P}\{g_n(\textbf{x})\}}{\textbf{P}\{h_n(\textbf{x})\}} - \gamma_I \Big] = \sqrt{n}(\textbf{P}_n-\textbf{P}) \{a(\textbf{x})\} + O_p(1)$. \\\\
Accordingly, we have the expression for $\sqrt{n}(\hat \gamma_I - \gamma_I)$ in terms of the empirical process, $$\sqrt{n}(\hat \gamma_I - \gamma_I) = \sqrt{n}(\textbf{P}_n-\textbf{P}) \{t(\textbf{x})\} + O_p(1)$$ where $ t(\textbf{x}) = \frac{g(\textbf{x})}{E[h(\textbf{x})]} - \frac{ h(\textbf{x})E[g(\textbf{x})]}{(E[h(\textbf{x})])^2} + a(\textbf{x})$. By the central limit theorem, $$\sqrt{n} (\textbf{P}_n -  \textbf{P}) \{ t(\textbf{x})\} \xrightarrow[]{d} N(0, \sigma^2_{t})$$ $\sigma^2_{t}$ is the variance of $t$. Thus, we can say that as $n \rightarrow \infty$,  $$\sqrt{n}(\hat \gamma_I - \gamma_I) \xrightarrow[]{d} N(0, \sigma^2_{t})$$ The large sampling distribution of $\hat \gamma_I$ is $N(\gamma_I, \sigma^2_{t}/n)$. Since $\frac{1}{\sqrt{n}} \rightarrow 0$ as $n \rightarrow \infty$, by the Slutsky's theorem, we have $$\hat \gamma_I - \gamma_I = \frac{1}{\sqrt{n}} \sqrt{n}(\hat \gamma_I - \gamma_I)   \xrightarrow[]{d} 0 \Rightarrow \hat \gamma_I - \gamma_I  \xrightarrow[]{p} 0 $$ That is, $\hat \gamma_I$ converges to $\gamma_I$ in probability. Therefore, $\hat \gamma_I$ is a consistent estimator of $\gamma_I$. 
\subsection{Variance estimation of $\hat \gamma_I$}
Given two-level data $\{x_{ij}, i=1,...,n, j=1,...,k_i\}$, we can estimate the variance of $\hat \gamma_I$ using the sample variance of $t(\textbf{x})$. We first obtain the estimate of $t(\textbf{x})$ for each cluster. Let $A_n =  \frac{1}{n} \sum_i^n g_n(\textbf{x}_i)$ and $B_n =  \frac{1}{n} \sum_i^n h_n(\textbf{x}_i)$. For each cluster, we compute 
$$\hat t(\textbf{x}_i) = \frac{g_n(\textbf{x}_i)}{B_n} - \frac{h_n(\textbf{x}_i) A_n}{B_n^2} + \hat a(\textbf{x}_i)$$
where $\hat a(\textbf{x}_i) = I_i + II_i + III_i + IV_i$, and the four items are counterparts of components in $(1)-(4)$. We describe $I_i-IV_i$ in the followings. 
\begin{align*}
& I_i  = \frac{1}{B_n} \sum\limits_{j^{''}=1}^{k_i} w_{i j^{''}} \sum\limits_{i'=1}^n \frac{2w_{i'.} }{k_{i'}(k_{i'}-1)} \sum\limits_{j'>j} \Big(\{\hat F^*(x_{i'j'}) - \bar{\hat{F^*}}\}\{[ I(x_{ij^{''}} \leq x_{i'j})+\\
& \:\:\:\:\: I(x_{i j^{''}} < x_{i'j})]/2 \} + \sum\limits_{j'>j} \{\hat F^*(x_{i'j}) - \bar{\hat{F^*}}\}\{[ I(x_{ij^{''}} \leq x_{i'j'})+I(x_{ij^{''}} < x_{i'j'})]/2 \}\Big)\\
& II_i = -\frac{A_n}{B_n^2} \sum\limits_{j=1}^{k_i} w_{ij} \sum\limits_{i'=1}^n \sum\limits_{j'=1}^{k_{i'}} 2w_{i'j'}[\hat F^*(x_{i'j'}) - \bar{\hat{F^*}}][I(x_{ij} \leq x_{i'j'})+I(x_{ij} < x_{i'j'})]/2\\
& III_i = - \frac{C_{ni}}{B_n} \sum\limits_{i'=1}^n \frac{2 w_{i'.}}{k_{i'}(k_{i'}-1)} \sum\limits_{j'>j}[\hat F^*(x_{i'j}) - \bar{\hat{F^*}}+\hat F^*(x_{i'j'}) -\bar{\hat{F^*}}]\\
& IV_i = \frac{A_nC_{ni}}{B_n^2}  \sum\limits_{i'=1}^n \sum\limits_{j'=1}^{k_{i'}} 2w_{i'j'}[\hat F^*(x_{i'j'}) - \bar{\hat{F^*}}]
\end{align*}
where $C_{ni} = \sum_{j=1}^{k_i} w_{ij} \hat F^*(x_{ij}) + \frac{1}{n} \sum\limits_{i'=1}^n \sum\limits_{j'=1}^{k_{i'}} w_{i'j'}  \big\{\sum\limits_{j=1}^{k_i} w_{ij} [I(x_{ij} \leq x_{i'j'})+I(x_{ij} < x_{i'j'})]/2\big\}$. \\
Then we can obtain the sample variance of $t(\textbf{x})$ with $\{\hat t(\textbf{x}_i), i=1,...,n\}$, denoted as $\hat \sigma_t^2$. The variance of $\hat \gamma_I$ is estimated by $\hat \sigma_t^2/n$.
%%%%%%%%%%%%%%%%%%%%%%%%%%%%%%%%%%%%%%%%%%%%%%%%%%%%%%%%%%%%%%%%%%%%%%%%%gamma2 and gamma3
\subsection{Proof of asymptotic properties of $\hat \gamma_{I2}$ and $\hat \gamma_{I3}$}
Let $g_n^{(2)}(\textbf{x})$, $g_n^{(3)}(\textbf{x})$ and $ h_n(\textbf{x})$ be three functions such that $\hat \gamma_{I2} = [\frac{1}{n}\sum_{i=1}^n g_n^{(2)}(\textbf{x}_i)] /[\frac{1}{n} \sum_{i=1}^n h_n(\textbf{x}_i)]$ and $\hat \gamma_{I3} = [\frac{1}{n}\sum_{i=1}^n g_n^{(3)}(\textbf{x}_i)] /[\frac{1}{n} \sum_{i=1}^n h_n(\textbf{x}_i)]$, where $\textbf{x}_i = (x_{i11},...,x_{i1m_{i1}},..., x_{in_i1},...,x_{in_im_{ij}})$ denotes a vector of observations in a level-3 unit. We then have $\hat \gamma_{I2} =\frac{\textbf{P}_n\{g_n^{(2)}(\textbf{x})\}}{\textbf{P}_n\{h_n(\textbf{x})\}}$, $\gamma_{I2} = \frac{\textbf{P}\{g^{(2)}(\textbf{x})\}}{\textbf{P}\{h(\textbf{x})\}}$, $\hat \gamma_{I3} =\frac{\textbf{P}_n\{g_n^{(3)}(\textbf{x})\}}{\textbf{P}_n\{h_n(\textbf{x})\}}$, and $\gamma_{I3} = \frac{\textbf{P}\{g^{(3)}(\textbf{x})\}}{\textbf{P}\{h(\textbf{x})\}}$. \\\\
ASSUMPTION 2 \;\; For any function $\Tilde{f}$ in a Donsker class, there exits a corresponding function $f$ in another Donsker class such that $\sum_{i}\sum_{j}\sum_{k} w_{ijk}\Tilde{f}(x_{ijk}) - E[\Tilde{f}(x_{ijk})] = (\textbf{P}_n - \textbf{P})\{f(\textbf{x})\} + O_p(1/\sqrt{n})$. \\\\
REMARK 2\;\; Assumption 2 trivially holds if $w_{ijk}=\frac{1}{nn_im_{ij}}$, where $n_i$ and $m_{ij}$ are bounded variables and the two Donsker classes are the same. The asymptotic properties of $\hat \gamma_{I2}$ and $\hat \gamma_{I3}$ shown under $w_{ijk}=\frac{1}{nn_im_{ij}}$ in the followings are also valid for any $w_{ijk}$ (e.g., $w_{ijk} = \frac{1}{\sum_i \sum_j m_{ij}}$ or $\frac{1}{m_{ij}\sum_{i=1}^n n_i}$) that satisfies Assumption 2. \\\\
We consider $w_{ijk}=\frac{1}{nn_im_{ij}}$ in the following proofs. Because $\hat F^*$ belongs to a Donsker class which contains all distribution functions in $X$'s support and $\hat F^*$ uniformly converges to $F^*$ with probability one, $g^{(2)}_n$, $g^{(3)}_n$, and $h_n$ also belong to some Donsker classes based on Assumption 2 and converge in $L_2(P)$ distance to their limits $g^{(2)}(\textbf{x})$, $g^{(3)}(\textbf{x})$ and $h(\textbf{x})$, respectively, where $g^{(2)}(\textbf{x})=E\big[\sum_{j}\sum_{k'>k}\frac{2w_{ij.}}{m_{ij}(m_{ij}-1)} [ F^*(x_{ijk}) - E(F^*)][F^*(x_{ijk'}) - E(F^*)]\big]$, $g^{(3)}(\textbf{x})=E\big[\sum_{j'>j}\sum_{k}\sum_{l}\frac{w_{i..}}{c_i}[ F^*(x_{ijk}) - E(F^*)][F^*(x_{ij'l}) - E(F^*)]\big]$, and $h(\textbf{x})=E\big[\sum_{j=1}^{n_i}\sum_{k=1}^{m_{ij}} w_{ijk} [ F^*(x_{ijk}) -E(F^*)]^2 \big]$. Also, $|g^{(2)}_n - g^{(2)}| \rightarrow 0$, $|g^{(3)}_n - g^{(3)}| \rightarrow 0$, and $|h_n - h| \rightarrow 0$. \\\\ 
Then we express $\sqrt{n}(\hat \gamma_{I2}  - \gamma_{I2})$ and $\sqrt{n}(\hat \gamma_{I3}  - \gamma_{I3})$ in terms of the empirical process.
 \begin{align*}
& \sqrt{n}(\hat \gamma_{I2}  - \gamma_{I2})  = \sqrt{n} \Big(\frac{\textbf{P}_n\{g_n^{(2)}(\textbf{x})\}}{\textbf{P}_n\{h_n(\textbf{x})\}} - \gamma_{I2} \Big)\\
& = \sqrt{n} \Big(\frac{\textbf{P}_n\{g_n^{(2)}(\textbf{x})\}}{\textbf{P}_n\{h_n(\textbf{x})\}} - \frac{\textbf{P}\{g_n^{(2)}(\textbf{x})\}}{\textbf{P}\{h_n(\textbf{x})\}}  + \frac{\textbf{P}\{g_n^{(2)}(\textbf{x})\}}{\textbf{P}\{h_n(\textbf{x})\}} -\gamma_{I2} \Big)\\
& = \sqrt{n} \Big(\frac{\textbf{P}_n\{g_n^{(2)}(\textbf{x})\}}{\textbf{P}_n\{h_n(\textbf{x})\}} - \frac{\textbf{P}\{g_n^{(2)}(\textbf{x})\}}{\textbf{P}_n\{h_n(\textbf{x})\}} + \frac{\textbf{P}\{g_n^{(2)}(\textbf{x})\}}{\textbf{P}_n\{h_n(\textbf{x})\}}  - \frac{\textbf{P}\{g_n^{(2)}(\textbf{x})\}}{\textbf{P}\{h_n(\textbf{x})\}}  + \frac{\textbf{P}\{g_n^{(2)}(\textbf{x})\}}{\textbf{P}\{h_n(\textbf{x})\}} -\gamma_{I2} \Big)\\
 & =\sqrt{n}(\textbf{P}_n-\textbf{P})\big\{\frac{g_n^{(2)}(\textbf{x})}{\textbf{P}_n\{h_n(\textbf{x})\}}  - \frac{ h_n(\textbf{x})\textbf{P}\{g_n^{(2)}(\textbf{x})\}}{\textbf{P}_n\{h_n(\textbf{x})\}\textbf{P}\{h_n(\textbf{x})\}}\} + \sqrt{n}\Big[\frac{\textbf{P}\{g_n^{(2)}(\textbf{x})\}}{\textbf{P}\{h_n(\textbf{x})\}} - \gamma_{I2}\Big]\\
 & = \sqrt{n}(\textbf{P}_n-\textbf{P}) \big\{\frac{g^{(2)}(\textbf{x})}{E[h(\textbf{x})]} - \frac{ h(\textbf{x})E[g^{(2)}(\textbf{x})]}{(E[h(\textbf{x})])^2}\big\} + O_p(1) + \sqrt{n}\Big[\frac{\textbf{P}\{g_n^{(2)}(\textbf{x})\}}{\textbf{P}\{h_n(\textbf{x})\}} - \gamma_{I2}\Big]
 \end{align*}
 Similarly,  
  \begin{align*}
& \sqrt{n}(\hat \gamma_{I3}  - \gamma_{I3})  = \sqrt{n} \Big(\frac{\textbf{P}_n\{g_n^{(3)}(\textbf{x})\}}{\textbf{P}_n\{h_n(\textbf{x})\}} - \gamma_{I3} \Big)\\
 & =\sqrt{n}(\textbf{P}_n-\textbf{P})\big\{\frac{g_n^{(3)}(\textbf{x})}{\textbf{P}_n\{h_n(\textbf{x})\}}  - \frac{ h_n(\textbf{x})\textbf{P}\{g_n^{(3)}(\textbf{x})\}}{\textbf{P}_n\{h_n(\textbf{x})\}\textbf{P}\{h_n(\textbf{x})\}}\} + \sqrt{n}\Big[\frac{\textbf{P}\{g_n^{(3)}(\textbf{x})\}}{\textbf{P}\{h_n(\textbf{x})\}} - \gamma_{I3} \Big]\\
 & = \sqrt{n}(\textbf{P}_n-\textbf{P}) \big\{\frac{g^{(3)}(\textbf{x})}{E[h(\textbf{x})]} - \frac{ h(\textbf{x})E[g^{(3)}(\textbf{x})]}{(E[h(\textbf{x})])^2}\big\} + O_p(1) + \sqrt{n}\Big[\frac{\textbf{P}\{g_n^{(3)}(\textbf{x})\}}{\textbf{P}\{h_n(\textbf{x})\}} - \gamma_{I3} \Big]
\end{align*}
We perform the linearization for $\sqrt{n}\Big[\frac{\textbf{P}\{g_n^{(2)}(\textbf{x})\}}{\textbf{P}\{h_n(\textbf{x})\}} - \gamma_{I2} \Big]$ and $\sqrt{n}\Big[\frac{\textbf{P}\{g_n^{(3)}(\textbf{x})\}}{\textbf{P}\{h_n(\textbf{x})\}} - \gamma_{I3} \Big]$ separately around $(F^*, \bar F^*)$, where $\bar F^*$ denotes the functional component of $E(F^*)$.
\begin{align*}
\sqrt{n}\Big[\frac{\textbf{P}\{g_n^{(2)}(\textbf{x})\}}{\textbf{P}\{h_n(\textbf{x})\}} - \gamma_{I2} \Big]    & = \sqrt{n}\frac{\textbf{P}\big\{\nabla_{F^*} g_n^{(2)}(\textbf{x})(\hat F^* - F^*)\big\}}{\textbf{P}\{h_n(\textbf{x})\}} \\
& \:\:\:\:\: - \sqrt{n}\frac{\textbf{P}\{g_n^{(2)}(\textbf{x})\}}{(\textbf{P}\{h_n(\textbf{x})\})^2} \textbf{P}\{\nabla_{F^*} h_n(\textbf{x}) (\hat F^* - F^*)\} \\
& \:\:\:\:\:  +\sqrt{n}\frac{\textbf{P}\big\{\nabla_{\bar F^*} g_n^{(2)}(\textbf{x}) (\bar{\hat{F^*}} - E[F^*]) \big\}}{\textbf{P}\{h_n(\textbf{x})\}}\\
& \:\:\:\:\:  -\sqrt{n}\frac{\textbf{P}\{g_n^{(2)}(\textbf{x})\}}{(\textbf{P}\{h_n(\textbf{x})\})^2}\textbf{P}\{\nabla_{\bar F^*} h_n(\textbf{x})   (\bar{\hat{F^*}} - E[F^*])\}  +  O_p(1)\\
& = (*1) + (*2) + (*3) + (*4) +  O_p(1)
\end{align*}
\begin{align*}
\sqrt{n}\Big[\frac{\textbf{P}\{g_n^{(3)}(\textbf{x})\}}{\textbf{P}\{h_n(\textbf{x})\}} - \gamma_{I3} \Big]    & = \sqrt{n}\frac{\textbf{P}\big\{\nabla_{F^*} g_n^{(3)}(\textbf{x})(\hat F^* - F^*)\big\}}{\textbf{P}\{h_n(\textbf{x})\}}\\
& \:\:\:\:\: - \sqrt{n}\frac{\textbf{P}\{g_n^{(3)}(\textbf{x})\}}{(\textbf{P}\{h_n(\textbf{x})\})^2} \textbf{P}\{\nabla_{F^*} h_n(\textbf{x}) (\hat F^* - F^*)\} \\
& \:\:\:\:\:  +\sqrt{n}\frac{\textbf{P}\big\{\nabla_{\bar F^*} g_n^{(3)}(\textbf{x}) (\bar{\hat{F^*}} - E[F^*]) \big\}}{\textbf{P}\{h_n(\textbf{x})\}}\\
& \:\:\:\:\:  -\sqrt{n}\frac{\textbf{P}\{g_n^{(3)}(\textbf{x})\}}{(\textbf{P}\{h_n(\textbf{x})\})^2}\textbf{P}\{\nabla_{\bar F^*} h_n(\textbf{x})   (\bar{\hat{F^*}} - E[F^*])\}  +  O_p(1)\\
& = (*5) + (*6) + (*7) + (*8) +  O_p(1)
\end{align*}
Then transform $(*1)-(*8)$ into expressions of $\sqrt{n}(\textbf{P}_n-\textbf{P})$.
%%%%%%%%%%%%%%%%%%%%%%%%%%%%%%%%
\begin{align*} 
(*1) &   = \sqrt{n}\frac{\textbf{P}\big\{\nabla_{F^*} g_n^{(2)}(\textbf{x})(\hat F^* - F^*)\big\}}{\textbf{P}\{h_n(\textbf{x})\}} \\
& =  \frac{\sqrt{n}}{\textbf{P}\{h_n(\textbf{x})\}}\textbf{P} \Big\{\sum\limits_j\frac{2w_{ij.}}{m_{ij}(m_{ij}-1)}  \sum\limits_{k'>k} (\hat F^*(x_{ijk'}) -\bar{\hat{F^*}})(\hat F^*(x_{ijk}) - F^*(x_{ijk}))\Big\} \\
& \:\:\:\:\:  +  \frac{\sqrt{n}}{\textbf{P}\{h_n(\textbf{x})\}} \textbf{P} \Big\{\sum\limits_j\frac{2w_{ij.}}{m_{ij}(m_{ij}-1)}  \sum\limits_{k'>k} (\hat F^*(x_{ijk}) -\bar{\hat{F^*}})(\hat F^*(x_{ijk'}) - F^*(x_{ijk'}))\Big\} \\
& =  \frac{\sqrt{n} }{E[h_n(\textbf{x})]}  E \Big[\sum\limits_j\frac{2w_{ij.}}{m_{ij}(m_{ij}-1)}  \sum\limits_{k'>k} (\hat F^*(x_{ijk'}) -\bar{\hat{F^*}})(\hat F^*(x_{ijk}) - F^*(x_{ijk}))\Big]\\
& \:\:\:\:\:  + \frac{\sqrt{n} }{E[h_n(\textbf{x})]}  E \Big[\sum\limits_j\frac{2w_{ij.}}{m_{ij}(m_{ij}-1)}  \sum\limits_{k'>k} (\hat F^*(x_{ijk}) -\bar{\hat{F^*}})(\hat F^*(x_{ijk'}) - F^*(x_{ijk'}))\Big]  \\
& =\frac{\sqrt{n} }{E[h_n(\textbf{x})]} \Big\{ E \Big[\sum\limits_j\frac{2w_{ij.}}{m_{ij}(m_{ij}-1)}  \sum\limits_{k'>k} \Big(\hat F^*(x_{ijk'}) -\bar{\hat{F^*}} \Big)\\
& \:\:\:\:\: \times \Big( \sum\limits_{i^{'}}\sum\limits_{j^{''},\;k^{''}} w_{i^{'}  j^{''}k^{''}} [I(x_{i^{'} j^{''}k^{''}} < x_{ijk}) + I(x_{i^{'}j^{''}k^{''}} \leq x_{ijk})]/2 - F^*(x_{ijk}) \Big)\Big]  \\
& \:\:\:\:\: + E \Big[\sum\limits_j\frac{2w_{ij.}}{m_{ij}(m_{ij}-1)} \sum\limits_{k'>k} \Big(\hat F^*(x_{ijk}) -\bar{\hat{F^*}} \Big)\\
& \:\:\:\:\: \times  \Big( \sum\limits_{i^{'}}\sum\limits_{j^{''},\;k^{''}} w_{i^{'} j^{''}k^{''}} \times [I(x_{i^{'} j^{''}k^{''}} < x_{ijk'})+ I(x_{i^{'}j^{''}k^{''}} \leq x_{ijk'})]/2 - F^*(x_{ijk'}) \Big)\Big] \Big\} \\
& =\frac{\sqrt{n} }{E[h_n(\textbf{x})] } \sum\limits_{i^{'}}\sum\limits_{j^{''},k^{''}} w_{i^{'}  j^{''}k^{''}} E  \Big\{\sum\limits_j\frac{2w_{ij.}}{m_{ij}(m_{ij}-1)} \sum\limits_{k'>k} \Big[ \Big(\hat F^*(x_{ijk'}) -\bar{\hat{F^*}}\Big)\\
& \:\:\:\:\: \times \Big(I(x_{i^{'} j^{''}k^{''}} < x_{ijk}) + I(x_{i^{'}j^{''}k^{''}} \leq x_{ijk})\Big)/2  + \Big(\hat F^*(x_{ijk}) -\bar{\hat{F^*}}\Big)\\
& \:\:\:\:\:  \times \Big(I(x_{i^{'} j^{''}k^{''}} < x_{ijk'}) + I(x_{i^{'}j^{''}k^{''}} \leq x_{ijk'})\Big)/2 \Big]\Big\}\\
& \:\:\:\:\:-  \frac{\sqrt{n} }{E[h_n(\textbf{x})] } E \Big\{\sum\limits_j\frac{2w_{ij.}}{m_{ij}(m_{ij}-1)} \sum\limits_{k'>k} \Big [(\hat F^*(x_{ijk'}) -\bar{\hat{F^*}}) F^*(x_{ijk})  + \\
& \:\:\:\:\:  (\hat F^*(x_{ijk}) -\bar{\hat{F^*}})  F^*(x_{ijk'})\Big] \Big\}\\
& \text{The expectation in the first component is a function of $x_{i'j^{''}k^{''}}$,}\\
& \text{we denote it as $\Tilde{a}^*_1(x_{i'j^{''}k{''}})$}\\
& = \frac{\sqrt{n} }{E[h(\textbf{x})] } \{\sum\limits_{i^{'}}\sum\limits_{j^{''}}\sum\limits_{k^{''}} w_{i^{'}  j^{''}k^{''}} \Tilde{a}^*_1(x_{i'j^{''}k^{''}}) - E[\Tilde{a}^*_1(x_{i'j^{''}k^{''}})]\} + O_p(1) \\
& \text{Because $\Tilde{a}^*_1(x_{ijk})$ belongs to a Donsker class}\\
(*1) & =  \frac{\sqrt{n} }{E[h(\textbf{x})] } (\textbf{P}_n - \textbf{P})\{ a^*_1(\textbf{x})\} + O_p(1) \text{ (under Assumption 2)}
\end{align*}
The derivation of $(*2)-(*8)$ is similar to that of $(*1)$ and under Assumption 2.
%%%%%%%%%%%%%%%%%%%%%%%%
 \begin{align*}
(*2)&  =  - \sqrt{n}\frac{\textbf{P}\{g_n^{(2)}(\textbf{x})\}}{(\textbf{P}\{h_n(\textbf{x})\})^2} \textbf{P}\{\nabla_{F^*} h_n(\textbf{x}) (\hat F^* - F^*)\} \\
& - \sqrt{n}\frac{\textbf{P}\{g_n^{(2)}(\textbf{x})\}}{(\textbf{P}\{h_n(\textbf{x})\})^2} \textbf{P}\Big\{ 2\sum\limits_{j} \sum\limits_{k} w_{ijk} (\hat F^*(x_{ijk}) -\bar{\hat{F^*}}) ( \hat F^*(x_{ijk}) - F^*(x_{ijk}))\Big\}  \\
 & = - \sqrt{n}\frac{ E[g_n^{(2)}(\textbf{x})]}{(E[h_n(\textbf{x})])^2}  E \Big\{2\sum\limits_{j} \sum\limits_{k} w_{ijk} (\hat F^*(x_{ijk}) -\bar{\hat{F^*}}) ( \hat F^*(x_{ijk}) - F^*(x_{ijk}))  \Big\}\\
 & = - \sqrt{n}\frac{ E[g_n^{(2)}(\textbf{x})]}{(E[h_n(\textbf{x})])^2} E \Big\{2\sum\limits_{j}\sum\limits_{k} w_{ijk} (\hat F^*(x_{ijk}) -\bar{\hat{F^*}}) \Big( \sum\limits_{i^{'}}\sum\limits_{j^{'}\;k^{'}} w_{i^{'}  j^{'}k^{'}}\\
  & \:\:\:\:\: \times [I(x_{i^{'} j^{'} k^{'}} < x_{ijk}) +I(x_{i^{'}j^{'}k^{'}} \leq x_{ijk})]/2 - F^*(x_{ijk}) \Big) \Big\}\\
 & = - \sqrt{n}\frac{ E[g_n^{(2)}(\textbf{x})]}{(E[h_n(\textbf{x})])^2}  \sum\limits_{i^{'}}\sum\limits_{j^{'},\;k^{'}}  w_{i^{'}  j^{'}k^{'}}  E\Big\{2 \sum\limits_{j}\sum\limits_{k} w_{ijk} (\hat F^*(x_{ijk}) -\bar{\hat{F^*}}) \\
 & \:\:\:\:\: \times [I(x_{i^{'} j^{'}k^{'}} < x_{ijk}) +I(x_{i^{'}j^{'}k^{'}} \leq x_{ijk})]/2\Big\}  \\
 & \:\:\:\:\: + \sqrt{n}\frac{ E[g_n^{(2)}(\textbf{x})]}{(E[h_n(\textbf{x})])^2}  E\Big\{2 \sum\limits_{j} \sum\limits_{k}w_{ijk}  (\hat F^*(x_{ijk}) -\bar{\hat{F^*}}) F^*(x_{ijk}) \Big\}\\
 & \text{Let $\Tilde{a}^*_2(x_{i'j'k'})$ denote the expectation in the first component}\\
(*2) & = - \sqrt{n}\frac{ E[g_n^{(2)}(\textbf{x})]}{(E[h_n(\textbf{x})])^2}  \{ \sum\limits_{i^{'}}\sum\limits_{j^{'}}\sum\limits_{k^{'}}  w_{i^{'}  j^{'}k^{'}} \Tilde{a}^*_2(x_{i'j'k'})  - E[ \Tilde{a}^*_2(x_{i'j'k'}) ]\} + O_p(1) \\
& = - \frac{\sqrt{n} E[g^{(2)}(\textbf{x})]}{(E[h(\textbf{x})])^2} (\textbf{P}_n - \textbf{P})\{a_2^*(\textbf{x}) \} + O_p(1)\\\\
%%%%%%%%%%%%%%%%%%%%%%%%%%%%%
(*3) & = \sqrt{n}\frac{\textbf{P}\big\{\nabla_{\bar F^*} g_n^{(2)}(\textbf{x}) (\bar{\hat{F^*}} - E[F^*]) \big\}}{\textbf{P}\{h_n(\textbf{x})\}}\\
& =  \sqrt{n}\textbf{P}\Big\{ - \sum\limits_{j}\frac{2w_{ij.}}{m_{ij}(m_{ij}-1)}  \big[\sum\limits_{k'>k} (\hat F^*(x_{ijk}) -\bar{\hat{F^*}} + \hat F^*(x_{ijk'}) -\bar{\hat{F^*}})\big] \\
& \:\:\:\:\: \times (\bar{\hat{F^*}} - E[F^*])\Big\} / \textbf{P}\{h_n(\textbf{x})\}  \\
& = -\sqrt{n} E \Big\{ \sum\limits_{j}\frac{2w_{ij.}}{m_{ij}(m_{ij}-1)} \sum\limits_{k'>k} [\hat F^*(x_{ijk'}) - \bar{\hat{F^*}} + \hat F^*(x_{ijk}) - \bar{\hat{F^*}}] \\
& \:\:\:\:\: \times (\bar{\hat{F^*}} - E[F^*])\}  / E[h_n(\textbf{x})] \\
& = -  \frac{\sqrt{n}}{E[h_n(\textbf{x})]}  E \Big\{ \Big( \sum\limits_{j}\frac{2w_{ij.}}{m_{ij}(m_{ij}-1)} \sum\limits_{k'>k} [\hat F^*(x_{ijk'}) - \bar{\hat{F^*}} + \hat F^*(x_{ijk}) - \bar{\hat{F^*}}] \Big) \\
& \:\:\:\:\: \times\Big(\sum\limits_{i^{'}}\sum\limits_{j^{'},\;k^{'}} w_{i'j'k'} \big[ \hat F^*(x_{i'j'k'}) + \frac{1}{n}  \sum\limits_{i,\;j,\;k} w_{ijk}  [I(x_{i'j'k'} < x_{ijk}) \\
& \:\:\:\:\: + I(x_{i'j'k'} \leq x_{ijk})]/2 \big] - E[F^*] \Big)\Big\}\\
& = -  \frac{\sqrt{n}}{E[h_n(\textbf{x})]}  \sum\limits_{i^{'}}\sum\limits_{j^{'},\;k'} w_{i' j^{'}k'}  E \Big\{ \Big(\sum\limits_{j}\frac{2w_{ij.}}{m_{ij}(m_{ij}-1)} \sum\limits_{k'>k} [\hat F^*(x_{ijk'}) - \bar{\hat{F^*}}+ \hat F^*(x_{ijk}) \\
& \:\:\:\:\:  - \bar{\hat{F^*}}]\Big) \Big( \hat F^*(x_{i'j'k'}) + \frac{1}{n}  \sum\limits_{i,\;j,\;k}  w_{ij}  [I(x_{i'j'k'} < x_{ijk})+I(x_{i'j'k'} \leq x_{ijk})]/2 \Big)\Big\}\\
& \:\:\:\:\:  + \frac{\sqrt{n}}{E[h_n(\textbf{x})]}E \Big\{   \sum\limits_{j}\frac{2w_{ij.}}{m_{ij}(m_{ij}-1)} \sum\limits_{k'>k} [\hat F^*(x_{ijk'}) - \bar{\hat{F^*}} + \hat F^*(x_{ijk}) - \bar{\hat{F^*}}] E[F^*]\Big\}\\
 & \text{We denote the expectation in the first component as $\Tilde{a}^*_3(x_{i'j'k'})$}\\
(*3) & = -  \frac{\sqrt{n}}{E[h_n(\textbf{x})]} \{\sum\limits_{i^{'}}\sum\limits_{j^{'}}\sum\limits_{k^{'}} w_{i' j^{'}k'} \Tilde{a}^*_3(x_{i'j'k'}) - E[\Tilde{a}^*_3(x_{i'j'k'})]\}\\
& =  -  \frac{\sqrt{n}}{E[h(\textbf{x})]} (\textbf{P}_n - \textbf{P})\{a_3^*(\textbf{x})\} + O_p(1)\\\\
%%%%%%%%%%%%%%%%%%%%%%%%%%%%%
(*4) & =-\sqrt{n}\frac{\textbf{P}\{g_n^{(2)}(\textbf{x})\}}{(\textbf{P}\{h_n(\textbf{x})\})^2}\textbf{P}\{\nabla_{\bar F^*} h_n(\textbf{x})   (\bar{\hat{F^*}} - E[F^*])\}\\ 
& = \sqrt{n}\frac{E[g^{(2)}_n(\textbf{x})]}{(E[h_n(\textbf{x})])^2} \textbf{P}\Big\{ 2\sum\limits_{j} \sum\limits_{k} w_{ijk} (\hat F^*(x_{ijk}) - \bar{\hat{F^*}})  (\bar{\hat{F^*}}- E[F^*])\Big\}  \\
& =  \sqrt{n}\frac{E[g_n(\textbf{x})]}{(E[h_n(\textbf{x})])^2}  \sum\limits_{i^{'}}\sum\limits_{j^{'}} \sum\limits_{k'} w_{i' j^{'}k'}  E \Big\{  \big(\sum\limits_{j} \sum\limits_{k} 2w_{ijk} (\hat F^*(x_{ijk}) - \bar{\hat{F^*}})\big) \\
& \:\:\:\:\: \:\:\:\:\:  \times \big( \hat F^*(x_{i'j'k'}) + \frac{1}{n}  \sum\limits_{i}\sum\limits_{j}\sum\limits_{k}  w_{ijk}  [I(x_{i'j'k'} < x_{ijk})+I(x_{i'j'k'} \leq x_{ijk})]/2\big)\Big\}\\
& \:\:\:\:\:  - \sqrt{n}\frac{E[g(\textbf{x})]}{(E[h(\textbf{x})])^2} E \Big\{    2\sum\limits_{j}\sum\limits_{k}w_{ijk} (\hat F^*(x_{ijk}) - \bar{\hat{F^*}}) E[F^*]\Big\}\\
 & \text{The expectation in the first component is denoted as $\Tilde{a}^*_4(x_{i'j'k'})$}\\
(*4) & = \sqrt{n}\frac{E[g_n^{(2)}(\textbf{x})]}{(E[h_n(\textbf{x})])^2}  \{\sum\limits_{i^{'}}\sum\limits_{j^{'}}\sum\limits_{k'} w_{i' j^{'}k'} \Tilde{a}^*_4(x_{i'j'k'}) - E[\Tilde{a}_4(x_{i'j'k'})]\}\\
& =  \sqrt{n}\frac{E[g^{(2)}(\textbf{x})]}{(E[h(\textbf{x})])^2}  (\textbf{P}_n - \textbf{P})\{a^*_4(\textbf{x})\} + O_p(1)
\end{align*}
%%%%%%%%%%%%%%%%%%%%%%%%%%%%%%%%
\begin{align*} 
(*5) &   = \sqrt{n}\frac{\textbf{P}\big\{\nabla_{F^*} g_n^{(3)}(\textbf{x})(\hat F^* - F^*)\big\}}{\textbf{P}\{h_n(\textbf{x})\}} \\
& =  \sqrt{n} \textbf{P} \Big\{\frac{w_{i..}}{c_i}  \sum\limits_{j'>j} \sum\limits_{k,\;l} (\hat F^*(x_{ijk}) -\bar{\hat{F^*}})(\hat F^*(x_{ij'l}) - F^*(x_{ij'l}))\Big\} \big/ \textbf{P}\{h_n(\textbf{x})\} \\
& \:\:\:\:\:  + \sqrt{n} \textbf{P} \Big\{\frac{w_{i..}}{c_i}  \sum\limits_{j'>j} \sum\limits_{k,\;l} (\hat F^*(x_{ij'l}) -\bar{\hat{F^*}})(\hat F^*(x_{ijk}) - F^*(x_{ijk}))\Big\} \big/ \textbf{P}\{h_n(\textbf{x})\} \\
& =  \sqrt{n}  E \Big[\frac{w_{i..}}{c_i}  \sum\limits_{j'>j} \sum\limits_{k,\;l} (\hat F^*(x_{ijk}) -\bar{\hat{F^*}})(\hat F^*(x_{ij'l}) - F^*(x_{ij'l}))\Big] \big/ E[h_n(\textbf{x})] \\
& \:\:\:\:\:  + \sqrt{n}  E \Big[\frac{w_{i..}}{c_i}  \sum\limits_{j'>j} \sum\limits_{k,\;l}(\hat F^*(x_{ij'l}) -\bar{\hat{F^*}})(\hat F^*(x_{ijk}) - F^*(x_{ijk}))\Big]  \big/ E[h_n(\textbf{x})] \\
& =\frac{\sqrt{n} }{E[h_n(\textbf{x})]} \Big\{ E \Big[\frac{w_{i..}}{c_i}  \sum\limits_{j'>j} \sum\limits_{k,\;l} \Big(\hat F^*(x_{ijk}) -\bar{\hat{F^*}} \Big)\\ 
& \:\:\:\:\:\ \times \Big( \sum\limits_{i^{'}}\sum\limits_{j^{''},\;k^{''}} w_{i^{'}  j^{''}k^{''}} [I(x_{i^{'} j^{''}k^{''}} < x_{ij'l})  + I(x_{i^{'}j^{''}k^{''}} \leq x_{ij'l})]/2 - F^*(x_{ij'l}) \Big)\Big]\\
& \:\:\:\:\: + E \Big[\frac{w_{i..}}{c_i}  \sum\limits_{j'>j} \sum\limits_{k,\;l} \Big(\hat F^*(x_{ij'l}) -\bar{\hat{F^*}} \Big) \\
& \:\:\:\:\: \times \Big( \sum\limits_{i^{'}}\sum\limits_{j^{''},\;k^{''}} w_{i^{'} j^{''}k^{''}} [I(x_{i^{'} j^{''}k^{''}} < x_{ijk})+I(x_{i^{'}j^{''}k^{''}} \leq x_{ijk})]/2 - F^*(x_{ijk}) \Big)\Big] \Big\} \\
& =\frac{\sqrt{n} }{E[h_n(\textbf{x})] } \sum\limits_{i^{'}}\sum\limits_{j^{''},k^{''}} w_{i^{'}  j^{''}k^{''}} E  \Big\{\frac{w_{i..}}{c_i}  \sum\limits_{j'>j} \sum\limits_{k,\;l} \Big[ \Big(\hat F^*(x_{ijk}) -\bar{\hat{F^*}}\Big)\\
& \:\:\:\:\: \times \Big(I(x_{i^{'} j^{''}k^{''}} < x_{ij'l}) + I(x_{i^{'}j^{''}k^{''}} \leq x_{ij'l})\Big)/2 \\
& \:\:\:\:\: + \Big(\hat F^*(x_{ij'l}) -\bar{\hat{F^*}}\Big) \Big(I(x_{i^{'} j^{''}k^{''}} < x_{ijk})+ I(x_{i^{'}j^{''}k^{''}} \leq x_{ijk})\Big)/2 \Big]\Big\} \\
& \:\:\:\:\: -  \frac{\sqrt{n} }{E[h_n(\textbf{x})] } E \Big\{\frac{w_{i..}}{c_i}  \sum\limits_{j'>j} \sum\limits_{k,\;l} \Big [(\hat F^*(x_{ijk}) -\bar{\hat{F^*}}) F^*(x_{ij'l}) \\
& \:\:\:\:\:  + (\hat F^*(x_{ij'l}) -\bar{\hat{F^*}})  F^*(x_{ijk})\Big] \Big\}\\
& \text{We denote the expectation in the first component as $\Tilde{a}^*_5(x_{i'j^{''}k{''}})$}\\
& = \frac{\sqrt{n} }{E[h(\textbf{x})] } \{\sum\limits_{i^{'}}\sum\limits_{j^{''}}\sum\limits_{k^{''}} w_{i^{'}  j^{''}k^{''}} \Tilde{a}^*_5(x_{i'j^{''}k^{''}}) - E[\Tilde{a}^*_5(x_{i'j^{''}k^{''}})]\} + O_p(1) \\
& =  \frac{\sqrt{n} }{E[h(\textbf{x})] } (\textbf{P}_n - \textbf{P})\{ a^*_5(\textbf{x})\} + O_p(1) \text{ (under Assumption 2)}
\end{align*}
%%%%%%%%%%%%%%%%%%%%%%%%
 \begin{align*}
(*6)&  = - \sqrt{n}\frac{\textbf{P}\{g_n^{(3)}(\textbf{x})\}}{(\textbf{P}\{h_n(\textbf{x})\})^2} \textbf{P}\{\nabla_{F^*} h_n(\textbf{x}) (\hat F^* - F^*)\} \\
& - \sqrt{n}\frac{\textbf{P}\{g_n^{(3)}(\textbf{x})\}}{(\textbf{P}\{h_n(\textbf{x})\})^2} \textbf{P}\Big\{ 2\sum\limits_{j} \sum\limits_{k} w_{ijk} (\hat F^*(x_{ijk}) -\bar{\hat{F^*}}) ( \hat F^*(x_{ijk}) - F^*(x_{ijk}))\Big\}  \\
 & = - \sqrt{n}\frac{ E[g_n^{(3)}(\textbf{x})]}{(E[h_n(\textbf{x})])^2}  E \Big\{2\sum\limits_{j} \sum\limits_{k} w_{ijk} (\hat F^*(x_{ijk}) -\bar{\hat{F^*}}) ( \hat F^*(x_{ijk}) - F^*(x_{ijk}))  \Big\}\\
 & = - \sqrt{n}\frac{ E[g_n^{(3)}(\textbf{x})]}{(E[h_n(\textbf{x})])^2} E \Big\{2\sum\limits_{j}\sum\limits_{k} w_{ijk} (\hat F^*(x_{ijk}) -\bar{\hat{F^*}})\\
  & \:\:\:\:\: \times \Big( \sum\limits_{i^{'}}\sum\limits_{j^{'}\;k^{'}} w_{i^{'}  j^{'}k^{'}} [I(x_{i^{'} j^{'}k^{'}} < x_{ijk}) +I(x_{i^{'}j^{'}k^{'}} \leq x_{ijk})]/2 - F^*(x_{ijk}) \Big) \Big\}\\
 & = - \sqrt{n}\frac{ E[g_n^{(3)}(\textbf{x})]}{(E[h_n(\textbf{x})])^2}  \sum\limits_{i^{'}}\sum\limits_{j^{'},\;k^{'}}  w_{i^{'}  j^{'}k^{'}}  E\Big\{2 \sum\limits_{j}\sum\limits_{k} w_{ijk} (\hat F^*(x_{ijk}) -\bar{\hat{F^*}}) \\
  & \:\:\:\:\: \times [I(x_{i^{'} j^{'}k^{'}} < x_{ijk}) +I(x_{i^{'}j^{'}k^{'}} \leq x_{ijk})]/2\Big\} \\
   & \:\:\:\:\: + \sqrt{n}\frac{ E[g_n^{(3)}(\textbf{x})]}{(E[h_n(\textbf{x})])^2}  E\Big\{2 \sum\limits_{j} \sum\limits_{k}w_{ijk}  (\hat F^*(x_{ijk}) -\bar{\hat{F^*}}) F^*(x_{ijk}) \Big\}\\
 & \text{Let $\Tilde{a}^*_6(x_{i'j'k'})$ denote the expectation in the first component}\\
(*6) & = - \sqrt{n}\frac{ E[g_n^{(3)}(\textbf{x})]}{(E[h_n(\textbf{x})])^2}  \{ \sum\limits_{i^{'}}\sum\limits_{j^{'}}\sum\limits_{k^{'}}  w_{i^{'}  j^{'}k^{'}} \Tilde{a}^*_6(x_{i'j'k'})  - E[ \Tilde{a}^*_6(x_{i'j'k'}) ]\} + O_p(1) \\
& = - \frac{\sqrt{n} E[g^{(3)}(\textbf{x})]}{(E[h(\textbf{x})])^2} (\textbf{P}_n - \textbf{P})\{a_6^*(\textbf{x}) \} + O_p(1)\\\\
%%%%%%%%%%%%%%%%%%%%%%%%%%%%%
(*7) & = \sqrt{n}\frac{\textbf{P}\big\{\nabla_{\bar F^*} g_n^{(3)}(\textbf{x}) (\bar{\hat{F^*}} - E[F^*]) \big\}}{\textbf{P}\{h_n(\textbf{x})\}}\\
& =  \sqrt{n}\textbf{P}\Big\{ - \frac{w_{i..}}{c_i}  \sum\limits_{j'>j} \big[ \sum\limits_{k,\;l} (\hat F^*(x_{ijk}) -\bar{\hat{F^*}} + \hat F^*(x_{ij'l}) -\bar{\hat{F^*}})\big] \\
 & \:\:\:\:\: \times (\bar{\hat{F^*}} - E[F^*])\Big\} / \textbf{P}\{h_n(\textbf{x})\}  \\
& = - \frac{\sqrt{n}}{E[h_n(\textbf{x})]} E \Big\{ \frac{w_{i..}}{c_i}  \sum\limits_{j'>j} \big[ \sum\limits_{k,\;l}  [\hat F^*(x_{ijk}) - \bar{\hat{F^*}} + \hat F^*(x_{ij'l}) - \bar{\hat{F^*}}] (\bar{\hat{F^*}} - E[F^*])\} \\
& = -  \frac{\sqrt{n}}{E[h_n(\textbf{x})]}  E \Big\{   \frac{w_{i..}}{c_i}  \sum\limits_{j'>j} \big[ \sum\limits_{k,\;l} [\hat F^*(x_{ijk}) - \bar{\hat{F^*}} + \hat F^*(x_{ij'l}) - \bar{\hat{F^*}}] \\
& \:\:\:\:\: \times\Big(\sum\limits_{i^{'}}\sum\limits_{j^{''},\;k^{''}} w_{i'j^{''}k^{''}} \big[ \hat F^*(x_{i'j^{''}k^{''}}) + \frac{1}{n}  \sum\limits_{i,\;j,\;k} w_{ijk}  [I(x_{i'j^{''}k^{''}} < x_{ijk})\\
 & \:\:\:\:\: + I(x_{i'j^{''}k^{''}} \leq x_{ijk})]/2 \big] - E[F^*] \Big)\Big\}\\
& = -  \frac{\sqrt{n}}{E[h_n(\textbf{x})]}  \sum\limits_{i^{'}}\sum\limits_{j^{''},\;k^{''}} w_{i' j^{''}k^{''}}  E \Big\{ \frac{w_{i..}}{c_i}  \sum\limits_{j'>j} \sum\limits_{k,\;l} [\hat F^*(x_{ijk}) - \bar{\hat{F^*}} + \hat F^*(x_{ij'l}) - \bar{\hat{F^*}}] \\
& \:\:\:\:\: \times\Big( \hat F^*(x_{i'j^{''}k^{''}}) + \frac{1}{n}  \sum\limits_{i,\;j,\;k}  w_{ij}  [I(x_{i'j^{''}k^{''}} < x_{ijk})+I(x_{i'j^{''}k^{''}} \leq x_{ijk})]/2 \Big)\Big\}\\
& \:\:\:\:\:  + \frac{\sqrt{n}}{E[h_n(\textbf{x})]}E \Big\{   \frac{w_{i..}}{c_i}  \sum\limits_{j'>j} \big[ \sum\limits_{k,\;l}  [\hat F^*(x_{ijk}) - \bar{\hat{F^*}} + \hat F^*(x_{ij'l}) - \bar{\hat{F^*}}] E[F^*]\Big\}\\
 & \text{We denote the expectation in the first component as $\Tilde{a}^*_7(x_{i'j^{''}k^{''}})$}\\
(*7) & = -  \frac{\sqrt{n}}{E[h_n(\textbf{x})]} \{\sum\limits_{i^{'}}\sum\limits_{j^{''}}\sum\limits_{k^{''}} w_{i' j^{'}k^{''}} \Tilde{a}^*_7(x_{i'j^{''}k^{''}}) - E[\Tilde{a}^*_7(x_{i'j^{''}k^{''}})]\}\\
& =  -  \frac{\sqrt{n}}{E[h(\textbf{x})]} (\textbf{P}_n - \textbf{P})\{a_7^*(\textbf{x})\} + O_p(1)\\\\
%%%%%%%%%%%%%%%%%%%%%%%%%%%%%
(*8) & =-\sqrt{n}\frac{\textbf{P}\{g_n^{(3)}(\textbf{x})\}}{(\textbf{P}\{h_n(\textbf{x})\})^2}\textbf{P}\{\nabla_{\bar F^*} h_n(\textbf{x})   (\bar{\hat{F^*}} - E[F^*])\}\\ 
& = \sqrt{n}\frac{E[g^{(3)}_n(\textbf{x})]}{(E[h_n(\textbf{x})])^2} \textbf{P}\Big\{ 2\sum\limits_{j} \sum\limits_{k} w_{ijk} (\hat F^*(x_{ijk}) - \bar{\hat{F^*}})  (\bar{\hat{F^*}}- E[F^*])\Big\}  \\
& =  \sqrt{n}\frac{E[g_n^{(3)}(\textbf{x})]}{(E[h_n(\textbf{x})])^2}  \sum\limits_{i^{'}}\sum\limits_{j^{'}} \sum\limits_{k'} w_{i' j^{'}k'}  E \Big\{  \sum\limits_{j} \sum\limits_{k} 2w_{ijk} (\hat F^*(x_{ijk}) - \bar{\hat{F^*}}) \\
& \:\:\:\:\: \:\:\:\:\:  \times \big( \hat F^*(x_{i'j'k'}) + \frac{1}{n}  \sum\limits_{i}\sum\limits_{j}\sum\limits_{k}  w_{ijk}  [I(x_{i'j'k'} < x_{ijk})+I(x_{i'j'k'} \leq x_{ijk})]/2\big)\Big\}\\
& \:\:\:\:\:  - \sqrt{n}\frac{E[g^{(3)}(\textbf{x})]}{(E[h(\textbf{x})])^2} E \Big\{    2\sum\limits_{j}\sum\limits_{k}w_{ijk} (\hat F^*(x_{ijk}) - \bar{\hat{F^*}}) E[F^*]\Big\}\\
 & \text{The expectation in the first component is denoted as $\Tilde{a}^*_4(x_{i'j'k'})$}\\
(*8) & = \sqrt{n}\frac{E[g_n^{(3)}(\textbf{x})]}{(E[h_n(\textbf{x})])^2}  \{\sum\limits_{i^{'}}\sum\limits_{j^{'}}\sum\limits_{k'} w_{i' j^{'}k'} \Tilde{a}^*_8(x_{i'j'k'}) - E[\Tilde{a}_8(x_{i'j'k'})]\}\\
& =  \sqrt{n}\frac{E[g^{(3)}(\textbf{x})]}{(E[h(\textbf{x})])^2}  (\textbf{P}_n - \textbf{P})\{a^*_8(\textbf{x})\} + O_p(1)
\end{align*}
%%%%%%%%%%%%%%%%%%%%%%%%%%%%%%%%
Let $a^{(2)}(\textbf{x}) =\frac{1}{E[h(\textbf{x})]}[a_1^*(\textbf{x})-a_3^*(\textbf{x})] - \frac{E[g^{(2)}(\textbf{x})]}{(E[h(\textbf{x})])^2}[a_2^*(\textbf{x})-a_4^*(\textbf{x})]$ and $a^{(3)}(\textbf{x}) =\frac{1}{E[h(\textbf{x})]}[a_5^*(\textbf{x})-a_7^*(\textbf{x})] - \frac{E[g^{(3)}(\textbf{x})]}{(E[h(\textbf{x})])^2}[a_6^*(\textbf{x})-a_8^*(\textbf{x})]$. With the derivation of $(*1)-(*8)$ above, we have $\sqrt{n}\Big[\frac{\textbf{P}\{g_n^{(2)}(\textbf{x})\}}{\textbf{P}\{h_n(\textbf{x})\}} - \gamma_{I2} \Big] = \sqrt{n}(\textbf{P}_n-\textbf{P}) \{a^{(2)}(\textbf{x})\} + O_p(1)$ and $\sqrt{n}\Big[\frac{\textbf{P}\{g_n^{(3)}(\textbf{x})\}}{\textbf{P}\{h_n(\textbf{x})\}} - \gamma_{I3} \Big] = \sqrt{n}(\textbf{P}_n-\textbf{P}) \{a^{(3)}(\textbf{x})\} + O_p(1)$. \\\\
Accordingly, we have the expressions for $\sqrt{n}(\hat \gamma_{I2} - \gamma_{I2})$ and $\sqrt{n}(\hat \gamma_{I3} - \gamma_{I3})$ in terms of the empirical process, 
$$\sqrt{n}(\hat \gamma_{I2} - \gamma_{I2}) = \sqrt{n}(\textbf{P}_n-\textbf{P}) \{t_2(\textbf{x})\} + O_p(1)$$ 
$$\sqrt{n}(\hat \gamma_{I3} - \gamma_{I3}) = \sqrt{n}(\textbf{P}_n-\textbf{P}) \{t_3(\textbf{x})\} + O_p(1)$$ 
where $t_2(\textbf{x}) = \frac{g^{(2)}(\textbf{x})}{E[h(\textbf{x})]} - \frac{ h(\textbf{x})E[g^{(2)}(\textbf{x})]}{(E[h(\textbf{x})])^2} + a^{(2)}(\textbf{x})$ and $t_3(\textbf{x}) = \frac{g^{(3)}(\textbf{x})}{E[h(\textbf{x})]} - \frac{ h(\textbf{x})E[g^{(3)}(\textbf{x})]}{(E[h(\textbf{x})])^2} + a^{(3)}(\textbf{x})$. By the central limit theorem, 
$$\sqrt{n} (\textbf{P}_n -  \textbf{P}) \{ t_2(\textbf{x})\} \xrightarrow[]{d} N(0, \sigma^2_{t_2})$$ 
$$\sqrt{n} (\textbf{P}_n -  \textbf{P}) \{ t_3(\textbf{x})\} \xrightarrow[]{d} N(0, \sigma^2_{t_3})$$ 
$\sigma^2_{t_2}$ is the variance of $t_2$ and $\sigma^2_{t_3}$ is the variance of $t_3$. Thus, we can say that as $n \rightarrow \infty$,  $$\sqrt{n}(\hat \gamma_{I2} - \gamma_{I2}) \xrightarrow[]{d} N(0, \sigma^2_{t_2})$$ 
$$\sqrt{n}(\hat \gamma_{I3} - \gamma_{I3}) \xrightarrow[]{d} N(0, \sigma^2_{t_3})$$ 
The large sampling distribution of $\hat \gamma_{I2}$ is $N(\gamma_I, \sigma^2_{t_2}/n)$ and of $\hat \gamma_{I3}$ is $N(\gamma_I, \sigma^2_{t_3}/n)$. Since $\frac{1}{\sqrt{n}} \rightarrow 0$ as $n \rightarrow \infty$, by the Slutsky's theorem, we have 
$$\hat \gamma_{I2} - \gamma_{I2} = \frac{1}{\sqrt{n}} \sqrt{n}(\hat \gamma_{I2} - \gamma_{I2})   \xrightarrow[]{d} 0 \Rightarrow \hat \gamma_{I2} - \gamma_{I2}  \xrightarrow[]{p} 0 $$ 
Similarly, $\hat \gamma_{I3} - \gamma_{I3}  \xrightarrow[]{p} 0$. Therefore, $\hat \gamma_{I2}$ is a consistent estimator of $\gamma_{I2}$, and $\hat \gamma_{I3}$ is a consistent estimator of $\gamma_{I3}$. 
%%%%%%%%%%%%%%%%%%%%%%%%%%%%%%%%%
\subsection{Variance estimation of $\hat \gamma_{I2}$ and $\hat \gamma_{I3}$}
With a three-level dataset $\{x_{ijk}, i=1,...,n, j=1,...,n_i, k=1,...,m_{ij}\}$, we can use the sample variances of $t_2(\textbf{x})$ and $t_3(\textbf{x})$ to estimate the variances of $\hat \gamma_{I2}$ and $\hat \gamma_{I3}$. We first obtain the estimates of $t_2(\textbf{x})$ and $t_3(\textbf{x})$ for each cluster. Let $A_n^{(2)} =  \frac{1}{n} \sum_i^n g_n^{(2)}(\textbf{x}_i)$, $A_n^{(3)} =  \frac{1}{n} \sum_i^n g_n^{(3)}(\textbf{x}_i)$ and $B_n =  \frac{1}{n} \sum_i^n h_n(\textbf{x}_i)$. For each cluster, we compute 
$$\hat t_2(\textbf{x}_i) = \frac{g_n^{(2)}(\textbf{x}_i)}{B_n} - \frac{h_n(\textbf{x}_i) A_n^{(2)}}{B_n^2} + \hat a^{(2)}(\textbf{x}_i)$$
$$\hat t_3(\textbf{x}_i) = \frac{g_n^{(3)}(\textbf{x}_i)}{B_n} - \frac{h_n(\textbf{x}_i) A_n^{(3)}}{B_n^2} + \hat a^{(3)}(\textbf{x}_i)$$
where $\hat a^{(2)}(\textbf{x}_i) = I_i + II_i + III_i + IV_i$, $\hat a^{(3)}(\textbf{x}_i) = V_i + VI_i + VII_i + VIII_i$, and the eight items are counterparts of components in $(*1)-(*8)$. We describe $I_i-VIII_i$ in the followings. 
\begin{align*}
& I_i  = \frac{1}{B_n} \sum\limits_{j=1}^{n_{i}} \sum\limits_{k=1}^{m_{ij}}  w_{i jk} \sum\limits_{i'=1}^n \sum\limits_{j'=1}^{n_{i'}}  \frac{2w_{i'j'.}}{m_{i'j}(k_{i'}-1)} \sum\limits_{k'>k^{''}} \Big(\{\hat F^*(x_{i'j'k'}) - \bar{\hat{F^*}}\} \\
& \;\;\;\;   \times \{[ I(x_{ijk} \leq x_{i'j'k^{''}}) +I(x_{ijk} < x_{i'j'k^{''}})]/2 \} \\
& \;\;\;\;  + \{\hat F^*(x_{i'j'k^{''}}) - \bar{\hat{F^*}}\}\{[ I(x_{ijk} \leq x_{i'j'k'})+I(x_{ijk} < x_{i'j'k'})]/2 \}\Big) \\
& II_i = -\frac{A^{(2)}_n}{B_n^2} \sum\limits_{j=1}^{n_i}  \sum\limits_{k=1}^{m_{ij}}  w_{ijk} \sum\limits_{i'=1}^n \sum\limits_{j'=1}^{n_{i'}} \sum\limits_{k'=1}^{m_{i'j'}}  2w_{i'j'k'}[\hat F^*(x_{i'j'k'}) - \bar{\hat{F^*}}]\\
& \;\;\;\;  \times [I(x_{ijk} \leq x_{i'j'k'})+I(x_{ijk} < x_{i'j'k'})]/2\\
& III_i = - \frac{C_{ni}}{B_n} \sum\limits_{i'=1}^n \sum\limits_{j'=1}^{n_{i'}} \frac{2 w_{i'j'.}}{m_{i'j'}(m_{i'j'}-1)} \sum\limits_{k'>k}[\hat F^*(x_{i'j'k}) - \bar{\hat{F^*}}+\hat F^*(x_{i'j'k'}) -\bar{\hat{F^*}}]\\
& IV_i = \frac{A_n^{(2)}C_{ni}}{B_n^2}  \sum\limits_{i'=1}^n \sum\limits_{j'=1}^{n_{i'}} \sum\limits_{k'=1}^{m_{i'j'}}2w_{i'j'k'}[\hat F^*(x_{i'j'k'}) - \bar{\hat{F^*}}]\\
& V_i  = \frac{1}{B_n} \sum\limits_{j=1}^{n_{i}} \sum\limits_{k=1}^{m_{ij}}  w_{i jk} \sum\limits_{i'=1}^n \frac{w_{i'..}}{w_{i'}} \sum\limits_{j'>j^{''}} \sum\limits_{k',\;l} \Big(\{\hat F^*(x_{i'j'k'}) - \bar{\hat{F^*}}\} \\
& \;\;\;\;  \times \{ [ I(x_{ijk} \leq x_{i'j^{''}l}) +I(x_{ijk} < x_{i'j^{''}l})]/2 \} \\
& \;\;\;\;  + \{\hat F^*(x_{i'j^{''}l}) - \bar{\hat{F^*}}\}\{[ I(x_{ijk} \leq x_{i'j'k'})+I(x_{ijk} < x_{i'j'k'})]/2 \} \Big) \\
& VI_i = -\frac{A^{(3)}_n}{B_n^2} \sum\limits_{j=1}^{n_i}  \sum\limits_{k=1}^{m_{ij}}  w_{ijk} \sum\limits_{i'=1}^n \sum\limits_{j'=1}^{n_{i'}} \sum\limits_{k'=1}^{m_{i'j'}}  2w_{i'j'k'}[\hat F^*(x_{i'j'k'}) - \bar{\hat{F^*}}]\\
& \;\;\;\;  \times [I(x_{ijk} \leq x_{i'j'k'})+I(x_{ijk} < x_{i'j'k'})]/2\\
& VII_i = - \frac{C_{ni}}{B_n} \sum\limits_{i'=1}^n \frac{w_{i'..}}{w_{i'}} \sum\limits_{j'>j} \sum\limits_{k,\;l} [\hat F^*(x_{i'j'k}) - \bar{\hat{F^*}}+\hat F^*(x_{i'j^{''}l}) -\bar{\hat{F^*}}]\\
& VIII_i = \frac{A_n^{(3)}C_{ni}}{B_n^2}  \sum\limits_{i'=1}^n \sum\limits_{j'=1}^{n_{i'}} \sum\limits_{k'=1}^{m_{i'j'}}2w_{i'j'k'}[\hat F^*(x_{i'j'k'}) - \bar{\hat{F^*}}]
\end{align*}
where $C_{ni} = \sum_{j=1}^{n_i} \sum_{k=1}^{m_{ij}}  w_{ijk} \hat F^*(x_{ijk}) + \frac{1}{n} \sum\limits_{i'=1}^n \sum\limits_{j'=1}^{n_{i'}} \sum\limits_{k'=1}^{m_{i'j'}}  w_{i'j'k'}  \big\{\sum\limits_{j=1}^{n_i} \sum\limits_{k=1}^{m_{ij}} w_{ijk} [I(x_{ijk} \leq x_{i'j'k'})+I(x_{ijk} < x_{i'j'k'})]/2\big\}$. \\
Then we can obtain the sample variances of $t_2(\textbf{x})$ and $t_3(\textbf{x})$ with $\{\hat t_2(\textbf{x}_i), i=1,...,n\}$ and $\{\hat t_3(\textbf{x}_i), i=1,...,n\}$, denoted as $\hat \sigma_{t_2}^2$ and $\hat \sigma_{t_3}^2$. The asymptotic variances of $\hat \gamma_{I2}$ and $\hat \gamma_{I3}$ are estimated by $\hat \sigma_{t_2}^2/n$ and $\hat \sigma_{t_3}^2/n$, respectively. 

%%%%%%%%%%%%%%%%%%%%%%%%%%%%%%%%%%%%  
\beginsupplement
\section{Web Table 1}
\label{s:table1}
 \begin{table}[h!]
    \centering
    \caption{\small{Bias, standard error (SE) and coverage of 95\% confidence intervals based on the six possible inference approaches for our estimator of $\gamma_I$ at different number of clusters under Scenario I (normality). We set $\gamma_I$ to be 0.48 and the cluster size to be 30. There were $200$ replicates per bootstrap.}}
    \label{tab:g1extra}
     \begin{subtable}{\textwidth}
        \centering
        \caption{Coverage probabilities of 95\% CIs}
        \begin{tabular}{p{1cm}p{2cm}p{2cm}p{2cm}p{2cm}p{2cm}p{2cm}}
        \hline
        Size & Asymptotic SE & Fisher transformation & Cluster bootstrap percentile	& Two-stage bootstrap percentile & Cluster bootstrap SE	& Two-stage bootstrap SE  \\ 
        \hline
        25 & 0.913 & 0.911 & 0.860 & 0.911	& 0.916 & 0.923 \\
        50 & 0.945 & 0.935 & 0.915 & 0.965 & 0.949 & 0.950 \\
        100 & 0.939 & 0.938 & 0.922 & 0.953 & 0.935 & 0.941 \\
        200	& 0.944 & 0.945 & 0.921 & 0.923 & 0.943 & 0.948	\\	
        500	& 0.950 & 0.947 & 0.944 & 0.841 & 0.946 & 0.947 \\
        1000 & 0.949  & 0.950 &	0.934 &	0.679 &	0.949 &	0.954\\  
         \hline
        \end{tabular}
    \label{tab:covprobg1}
    \end{subtable}
    \begin{subtable}{\textwidth}
    \centering
        \caption{Bias and SE of $\hat \gamma_I$}
        \begin{tabular}{p{1.5cm} p{1.5cm}p{2cm}p{2.5cm}p{3cm}p{2.5cm}}
        \hline
        Size & Percent bias (\%) & Empirical SE	& Averaged asymptotic SE & Averaged cluster bootstrap SE & Averaged two-stage bootstrap SE \\ 
        \hline
        25 & 	-3.859& 	0.077	& 0.075	& 0.074& 	0.076\\
        50 & -1.684& 	0.053& 	0.054& 	0.054& 	0.054\\
        100 & 	-0.726	& 0.039	& 0.038& 	0.038& 	0.039\\
        200	& -0.274& 	0.027& 	0.027& 	0.027& 	0.027\\	
        500	& -0.069& 	0.017& 	0.017& 	0.017& 	0.017\\
        1000 & -0.015& 	0.012& 0.012& 	0.012& 	0.012 \\  
         \hline
        \end{tabular}
    \label{tab:seg1}
    \end{subtable}
    \label{tab:sumg1}
        \begin{subtable}{\textwidth}
        \centering
        \caption{Non-Coverage at the tails of 95\% CIs of $\gamma_I$}
        \begin{tabular}[t]{p{1cm}p{2cm}p{2cm}p{2cm}p{2cm}p{2cm}p{2cm}}
        \hline
        Size & Asymptotic SE & Fisher transformation & Cluster bootstrap percentile	& Two-stage bootstrap percentile & Cluster bootstrap SE	& Two-stage bootstrap SE \\
        \hline
        25 & R=0.07, L=0.017 & R=0.08, L=0.009 & R=0.073, L=0.011 & R=0.067, L=0.01 & R=0.133, L=0.007 & R=0.081, L=0.008\\
        \hline
        50 & R=0.04, L=0.015 & R=0.055, L=0.01 & R=0.039, L=0.012 & R=0.038, L=0.012 & R=0.081, L=0.004 & R=0.027, L=0.008\\
        \hline
        100 & R=0.035, L=0.026 & R=0.042, L=0.02 & R=0.042, L=0.023 & R=0.037, L=0.022 & R=0.061, L=0.017 & R=0.012, L=0.035\\
        \hline
        200 & R=0.034, L=0.022 & R=0.037, L=0.018 & R=0.038, L=0.019 & R=0.033, L=0.019 & R=0.059, L=0.02 & R=0.008, L=0.069\\
        \hline
        500 & R=0.02, L=0.03 & R=0.023, L=0.03 & R=0.024, L=0.03 & R=0.022, L=0.031 & R=0.029, L=0.027 & R=0.005, L=0.154\\
        \hline
        1000 & R=0.029, L=0.022 & R=0.029, L=0.021 & R=0.029, L=0.022 & R=0.026, L=0.02 & R=0.039, L=0.027 & R=0.003, L=0.318\\
        \hline
        \end{tabular}
    \label{tab:noncovg1}
    \end{subtable}
\end{table}

%%%%%%%%%%%%%%%%%%%%%%%%%%%%%%%%%%%% 
\newpage
\section{Web Table 2}
\label{s:table2}
\begin{table}[h!]
    \centering 
    \caption{Coverage and non-Coverage at the tails of 95\% confidence intervals based on the four bootstrapping inference approaches for $\gamma_I$ under Scenario I (normality). We set $\gamma_I$ to be 0.48 and the cluster size to be 30. There were $1000$ replicates per bootstrap. }
    \begin{subtable}\textwidth
    \centering
    \caption{Coverage of bootstrap methods with 1000 replications each bootstrap}
        \begin{tabular}[t]{p{1.5cm}p{3cm}p{3cm}p{3cm}p{3cm}p{3cm}}
        \hline
        Size & Cluster bootstrap percentile	& Two-stage bootstrap percentile & Cluster bootstrap SE	& Two-stage bootstrap SE\\
        \hline
        25 & 0.889 & 0.932 & 0.913 & 0.918\\
        \hline
        50 & 0.930 & 0.960 & 0.939 & 0.950\\
        \hline
        100 & 0.934 & 0.952 & 0.937 & 0.942\\
        \hline
        200 & 0.940 & 0.935 & 0.944 & 0.946\\
        \hline
        500 & 0.952 & 0.855 & 0.950 & 0.954\\
        \hline
        \end{tabular}
    \label{tab:bootg1extracov}
    \end{subtable}
    \begin{subtable}\textwidth
    \centering
        \caption{Non-Coverage at the tails of 95\% CIs of $\gamma_I$}
        \begin{tabular}[t]{p{1.5cm}p{3cm}p{3cm}p{3cm}p{3cm}p{3cm}}
        \hline
        Size & Cluster bootstrap percentile	& Two-stage bootstrap percentile & Cluster bootstrap SE	& Two-stage bootstrap SE\\
        \hline
        25 & R=0.072, L=0.015 & R=0.067, L=0.015 & R=0.106, L=0.005 & R=0.063, L=0.005\\
        \hline
        50 & R=0.046, L=0.015 & R=0.035, L=0.015 & R=0.064, L=0.006 & R=0.026, L=0.014\\
        \hline
        100 & R=0.037, L=0.026 & R=0.032, L=0.026 & R=0.047, L=0.019 & R=0.012, L=0.036\\
        \hline
        200 & R=0.034, L=0.022 & R=0.032, L=0.022 & R=0.044, L=0.016 & R=0.006, L=0.059\\
        \hline
        500 & R=0.019, L=0.031 & R=0.018, L=0.028 & R=0.023, L=0.025 & R=0.003, L=0.142\\
        \hline
        \end{tabular}
    \label{tab:bootg1extraNoncov}
    \end{subtable}
    \label{tab:bootg1extra}
\end{table}   
%%%%%%%%%%%%%%%%%%%%%%%%%%%%%%%%%%%% 
\newpage
\section{Web Table 3}
\label{s:table3}

\begin{table}[h!]
    \centering
    \caption{Bias, standard error (SE) and coverage of 95\% confidence intervals based on the six possible inference approaches for our estimator of $\gamma_I$ at different numbers of clusters under Scenario I (normality). We set $\gamma_I$ to be 0.09 and the cluster size to be 30. There were $200$ replicates per bootstrap.}
    \begin{subtable}\textwidth
    \centering
        \caption{Coverage of 95\% CIs}
        \begin{tabular}{p{1cm}p{2cm}p{2cm}p{2cm}p{2cm}p{2cm}p{2cm}}
        \hline
        Size & Asymptotic SE & Fisher transformation & Cluster bootstrap percentile	& Two-stage bootstrap percentile & Cluster bootstrap SE	& Two-stage bootstrap SE\\
        \hline
        25 & 0.872 & 0.873 & 0.852 & 0.973 & 0.868 & 0.937\\
        \hline
        50 & 0.912 & 0.912 & 0.900 & 0.898 & 0.907 & 0.965\\
        \hline
        100 & 0.910 & 0.911 & 0.910 & 0.724 & 0.906 & 0.970\\
        \hline
        200 & 0.942 & 0.941 & 0.930 & 0.372 & 0.942 & 0.974\\
        \hline
        500 & 0.938 & 0.938 & 0.927 & 0.018 & 0.934 & 0.977\\
        \hline
        1000 & 0.952 & 0.952 & 0.945 & 0.001 & 0.952 & 0.975\\
        \hline
        \end{tabular}
    \label{tab:tabS3.1}
    \end{subtable}
    \begin{subtable}\textwidth
    \centering
        \caption{Bias and SE of $\hat \gamma_I$}
        \begin{tabular}{p{1.5cm} p{1.5cm}p{2cm}p{2.5cm}p{3cm}p{2.5cm}}
        \hline
        Size & Percent bias (\%) & Empirical SE	& Averaged asymptotic SE & Averaged cluster bootstrap SE & Averaged two-stage bootstrap SE \\ 
        \hline
        25 & -6.199 & 0.032 & 0.030 & 0.029 & 0.036\\
        \hline
        50 & -2.340 & 0.023 & 0.022 & 0.022 & 0.027\\
        \hline
        100 & -1.160 & 0.017 & 0.016 & 0.016 & 0.019\\
        \hline
        200 & -0.311 & 0.012 & 0.011 & 0.011 & 0.014\\
        \hline
        500 & -0.151 & 0.007 & 0.007 & 0.007 & 0.009\\
        \hline
        1000 & -0.002 & 0.005 & 0.005 & 0.005 & 0.006\\
        \hline
        \end{tabular}
    \label{tab:tabS3.2}
    \end{subtable}
    \label{tab:tabS3}
\end{table}  
%%%%%%%%%%%%%%%%%%%%%%%%%%%%%%%%%%%%  
\newpage
\section{Web Table 4}
\label{s:table4}

\begin{table}[h!]
    \centering
    \caption{Bias, standard error (SE) and coverage of 95\% confidence intervals based on the six possible inference approaches for our estimator of $\gamma_I$ at different numbers of clusters under Scenario I (normality). We set $\gamma_I$ to be 0.89 and the cluster size to be 30. There were $200$ replicates per bootstrap.}
    \begin{subtable}\textwidth
    \centering
        \caption{Coverage of 95\% CIs}
        \begin{tabular}{p{1cm}p{2cm}p{2cm}p{2cm}p{2cm}p{2cm}p{2cm}}
        \hline
        Size & Asymptotic SE & Fisher transformation & Cluster bootstrap percentile	& Two-stage bootstrap percentile & Cluster bootstrap SE	& Two-stage bootstrap SE\\
        \hline
        25 & 0.960 & 0.924 & 0.816 & 0.852 & 0.975 & 0.974\\
        \hline
        50 & 0.969 & 0.949 & 0.893 & 0.921 & 0.978 & 0.977\\
        \hline
        100 & 0.956 & 0.939 & 0.902 & 0.941 & 0.963 & 0.958\\
        \hline
        200 & 0.953 & 0.952 & 0.932 & 0.949 & 0.954 & 0.953\\
        \hline
        500 & 0.951 & 0.948 & 0.931 & 0.932 & 0.944 & 0.944\\
        \hline
        1000 & 0.950 & 0.945 & 0.937 & 0.885 & 0.949 & 0.947\\
        \hline
        \end{tabular}
    \label{tab:tabS4.1}
    \end{subtable}
    \begin{subtable}\textwidth
    \centering
        \caption{Bias and SE of $\hat \gamma_I$}
        \begin{tabular}{p{1.5cm} p{1.5cm}p{2cm}p{2.5cm}p{3cm}p{2.5cm}}
        \hline
        Size & Percent bias (\%) & Empirical SE	& Averaged asymptotic SE & Averaged cluster bootstrap SE & Averaged two-stage bootstrap SE \\ 
        \hline
        25 & -2.507 & 0.039 & 0.039 & 0.044 & 0.043\\
        \hline
        50 & -1.161 & 0.024 & 0.025 & 0.027 & 0.027\\
        \hline
        100 & -0.566 & 0.017 & 0.017 & 0.018 & 0.017\\
        \hline
        200 & -0.268 & 0.012 & 0.012 & 0.012 & 0.012\\
        \hline
        500 & -0.099 & 0.007 & 0.007 & 0.007 & 0.007\\
        \hline
        1000 & -0.048 & 0.005 & 0.005 & 0.005 & 0.005\\
        \hline
        \end{tabular}
    \label{tab:tabS4.2}
    \end{subtable}
    \label{tab:tabS4}
\end{table}   
    
%%%%%%%%%%%%%%%%%%%%%%%%%%%%%%%%%%%%   
\newpage
\section{Web Table 5}
\label{s:table5}

\begin{table}[h!]
    \centering
    \caption{Bias, standard error (SE) and coverage of 95\% confidence intervals based on the six possible inference approaches and for our estimator of $\gamma_I$ at different numbers of clusters under Scenario I (normality). The cluster size followed a uniform distribution from 2 to 50. We set $\gamma_I$ to be 0.48. There were $200$ replicates per bootstrap. ``Equal clusters'' refers to assigning equal weights to clusters. ``Equal obs'' refers to assigning equal weights to observations. ``PR.'' refers to the approaches using bootstrap percentiles. ``Asym. SE'' refers to the approach using asymptotic standard error. ``Emp. SE'' refers to the empirical standard error. ``Avg. SE'' refers to the average of standard errors. ``boot.'' means bootstrap.}
    \begin{subtable}\textwidth
    \centering
        \caption{Coverage of 95\% CIs of $\gamma_I$}
       \begin{tabular}{p{0.8cm}p{0.8cm}p{0.8cm}p{0.8cm}p{0.8cm}p{0.8cm}p{0.8cm}p{0.8cm}p{0.8cm}p{0.8cm}p{0.8cm}}
        \hline
        \multicolumn{1}{c}{Size} & \multicolumn{5}{c}{Equal clusters} & \multicolumn{5}{c}{Equal obs} \\
        \multicolumn{1}{c}{ } & \multicolumn{1}{C{0.8cm}}{Asym. SE} & \multicolumn{2}{C{2.5cm}}{Cluster bootstrap} & \multicolumn{2}{C{2.5cm}}{Two-stage bootstrap} & \multicolumn{1}{C{0.8cm}}{Asym. SE} & \multicolumn{2}{C{2.5cm}}{Cluster bootstrap} & \multicolumn{2}{C{2.5cm}}{Two-stage bootstrap} \\
          &   & PR. & SE & PR. & SE &   & PR. & SE & PR. & SE\\
        \hline
        25 & 0.910 & 0.893 & 0.909 & 0.952 & 0.919 & 0.900 & 0.872 & 0.908 & 0.917 & 0.908\\
        \hline
        50 & 0.934 & 0.919 & 0.926 & 0.954 & 0.939 & 0.910 & 0.899 & 0.913 & 0.931 & 0.912\\
        \hline
        100 & 0.953 & 0.941 & 0.948 & 0.923 & 0.950 & 0.950 & 0.941 & 0.951 & 0.955 & 0.952\\
        \hline
        200 & 0.953 & 0.948 & 0.950 & 0.831 & 0.952 & 0.948 & 0.934 & 0.952 & 0.928 & 0.952\\
        \hline
        500 & 0.942 & 0.932 & 0.944 & 0.525 & 0.944 & 0.945 & 0.938 & 0.945 & 0.861 & 0.945\\
        \hline
        1000 & 0.947 & 0.935 & 0.945 & 0.218 & 0.946 & 0.938 & 0.932 & 0.938 & 0.717 & 0.942\\
        \hline
        \end{tabular}
    \label{tab:tabS5.1}
    \end{subtable}
    \begin{subtable}\textwidth
    \centering
        \caption{Bias and SE of $\hat \gamma_I$}
      \begin{tabular}{p{0.8cm}p{1.3cm}p{0.8cm}p{0.8cm}p{1cm}p{1cm}p{1.3cm}p{0.8cm}p{0.8cm}p{1cm}p{1cm}}
        \hline
        \multicolumn{1}{c}{Size} & \multicolumn{5}{c}{Equal clusters} & \multicolumn{5}{c}{Equal obs} \\
          &  Bias (\%) & Emp. SE & Avg. asym. SE & Avg. cluster boot. SE & Avg. two-stage boot. SE &  Bias (\%) & Emp. SE & Avg. asym. SE & Avg. cluster boot. SE & Avg. two-stage boot. SE\\
         \hline
        25 & -4.033 & 0.084 & 0.082 & 0.081 & 0.082 & -4.605 & 0.086 & 0.083 & 0.082 & 0.083\\
        \hline
        50 & -2.029 & 0.059 & 0.058 & 0.058 & 0.058 & -2.375 & 0.063 & 0.060 & 0.059 & 0.060\\
        \hline
        100 & -0.783 & 0.040 & 0.041 & 0.041 & 0.041 & -1.059 & 0.041 & 0.043 & 0.043 & 0.043\\
        \hline
        200 & -0.713 & 0.029 & 0.029 & 0.029 & 0.029 & -0.708 & 0.031 & 0.031 & 0.030 & 0.031\\
        \hline
        500 & -0.258 & 0.019 & 0.019 & 0.019 & 0.019 & -0.258 & 0.019 & 0.019 & 0.019 & 0.019\\
        \hline
        1000 & -0.174 & 0.013 & 0.013 & 0.013 & 0.013 & -0.181 & 0.014 & 0.014 & 0.014 & 0.014\\
        \hline
        \end{tabular}
    \label{tab:tabS5.2}
    \end{subtable}
    \label{tab:tabS5}
\end{table} 
    
%%%%%%%%%%%%%%%%%%%%%%%%%%%%%%%%%%%% 
\newpage
\section{Web Table 6}
\label{s:table6}

\begin{table}[h!]
    \centering
    \caption{Bias, standard error (SE) and coverage of 95\% confidence intervals based on the  eight possible inference approaches for our estimator of $\gamma_{I2}$ at different number of level-3 units. We set $(\gamma_{I2},\gamma_{I3})=(0.53,0.19)$, the number of level-2 units in a level-3 unit to be 15, and the number of level-1 units in a level-2 unit to be 2. There were $200$ replicates per bootstrap. ``Avg. SE'' refers to the average of standard errors.}
    \begin{subtable}\textwidth
    \centering
        \caption{Coverage of 95\% CIs of $\gamma_{I2}$}
        \begin{tabular}{p{1cm}p{2cm}p{2cm}p{1cm}p{1cm}p{1cm}p{1cm}p{1cm}p{1cm}}
        \hline
        \multicolumn{1}{p{1cm}}{Size} & \multicolumn{1}{p{1cm}}{Asymptotic SE} & \multicolumn{1}{p{2cm}}{Fisher transformation} &  \multicolumn{3}{c}{Bootstrap SE} &\multicolumn{3}{c}{Bootstrap percentiles} \\
         &  &  & One-stage & Two-stage & Three-stage & One-stage & Two-stage& Three-stage\\
        \hline
        25 & 0.935 & 0.931 & 0.925 & 0.982 & 0.875 & 0.908 & 0.975 & 0\\
        \hline
        50 & 0.950 & 0.946 & 0.946 & 0.991 & 0.887 & 0.934 & 0.985 & 0\\
        \hline
        100 & 0.940 & 0.937 & 0.939 & 0.992 & 0.869 & 0.938 & 0.983 & 0\\
        \hline
        200 & 0.942 & 0.938 & 0.940 & 0.980 & 0.867 & 0.929 & 0.977 & 0\\
        \hline
        500 & 0.958 & 0.961 & 0.958 & 0.995 & 0.899 & 0.955 & 0.989 & 0\\
        \hline
        1000 & 0.950 & 0.952 & 0.948 & 0.987 & 0.871 & 0.945 & 0.984 & 0\\
        \hline
        \end{tabular}
    \label{tab:tabS6.1}
    \end{subtable}
    \begin{subtable}\textwidth
    \centering
        \caption{Bias and SE of $\hat \gamma_{I2}$}
        \begin{tabular}{p{1cm}p{1.5cm}p{1.5cm}p{2cm}p{2cm}p{2cm}p{2cm}}
       \hline
        Size & Percent bias (\%) & Emp.SE & Avg. asymptotic SE & Avg. cluster bootstrap SE & Avg. two-stage bootstrap SE & Avg. three-stage bootstrap SE\\
        \hline
        25 & -0.981 & 0.047 & 0.045 & 0.044 & 0.058 & 0.037\\
        \hline
        50 & -0.480 & 0.031 & 0.032 & 0.032 & 0.041 & 0.026\\
        \hline
        100 & -0.381 & 0.024 & 0.023 & 0.023 & 0.029 & 0.018\\
        \hline
        200 & -0.120 & 0.017 & 0.016 & 0.016 & 0.021 & 0.013\\
        \hline
        500 & -0.025 & 0.010 & 0.010 & 0.010 & 0.013 & 0.008\\
        \hline
        1000 & -0.031 & 0.007 & 0.007 & 0.007 & 0.009 & 0.006\\
        \hline
        \end{tabular}
    \label{tab:tabS6.2}
    \end{subtable}
\label{tab:tabS6}
\end{table}  

%%%%%%%%%%%%%%%%%%%%%%%%%%%%%%%%%%%%
\newpage
\section{Web Table 7}
\label{s:table7}
    
\begin{table}[h!]
    \centering
    \caption{Bias, standard error (SE) and coverage of 95\% confidence intervals based on the  eight possible inference approaches for our estimator of $\gamma_{I3}$ at different number of level-3 units. We set $(\gamma_{I2},\gamma_{I3})=(0.53,0.19)$, the number of level-2 units in a level-3 unit to be 15, and the number of level-1 units in a level-2 unit to be 2. There were $200$ replicates per bootstrap. ``Avg. SE'' refers to the average of standard errors.}
    \begin{subtable}\textwidth
    \centering
        \caption{Coverage of 95\% CIs of $\gamma_{I3}$}
         \begin{tabular}{p{1cm}p{2cm}p{2cm}p{1cm}p{1cm}p{1cm}p{1cm}p{1cm}p{1cm}}
        \hline
        \multicolumn{1}{p{1cm}}{Size} & \multicolumn{1}{p{1cm}}{Asymptotic SE} & \multicolumn{1}{p{2cm}}{Fisher transformation} &  \multicolumn{3}{c}{Bootstrap SE} &\multicolumn{3}{c}{Bootstrap percentiles} \\
         &  &  & One-stage & Two-stage & Three-stage & One-stage & Two-stage& Three-stage\\
        \hline
        25 & 0.896 & 0.895 & 0.885 & 0.934 & 0.941 & 0.861 & 0.968 & 0.970\\
        \hline
        50 & 0.928 & 0.928 & 0.924 & 0.958 & 0.963 & 0.915 & 0.939 & 0.951\\
        \hline
        100 & 0.933 & 0.934 & 0.933 & 0.955 & 0.962 & 0.922 & 0.809 & 0.831\\
        \hline
        200 & 0.940 & 0.939 & 0.932 & 0.967 & 0.970 & 0.924 & 0.584 & 0.628\\
        \hline
        500 & 0.961 & 0.960 & 0.953 & 0.975 & 0.978 & 0.946 & 0.136 & 0.162\\
        \hline
        1000 & 0.939 & 0.939 & 0.939 & 0.958 & 0.965 & 0.935 & 0.007 & 0.008\\
        \hline
        \end{tabular}
    \label{tab:tabS6.3}
    \end{subtable}
    \begin{subtable}\textwidth
    \centering
        \caption{Bias and SE of $\hat \gamma_{I3}$}
        \begin{tabular}{p{1cm}p{1.5cm}p{1.5cm}p{2cm}p{2cm}p{2cm}p{2cm}}
       \hline
        Size & Percent bias (\%) & Emp.SE & Avg. asymptotic SE & Avg. cluster bootstrap SE & Avg. two-stage bootstrap SE & Avg. three-stage bootstrap SE\\
        \hline
        25 & -4.809 & 0.053 & 0.050 & 0.049 & 0.056 & 0.058\\
        \hline
        50 & -2.064 & 0.037 & 0.037 & 0.036 & 0.041 & 0.042\\
        \hline
        100 & -0.774 & 0.027 & 0.027 & 0.026 & 0.029 & 0.030\\
        \hline
        200 & -0.617 & 0.019 & 0.019 & 0.019 & 0.021 & 0.022\\
        \hline
        500 & -0.112 & 0.012 & 0.012 & 0.012 & 0.013 & 0.014\\
        \hline
        1000 & -0.114 & 0.009 & 0.008 & 0.008 & 0.009 & 0.010\\
        \hline
        \end{tabular}
        \label{tab:tabS6.4}
    \end{subtable}
\label{tab:tabS6-2}
\end{table}  
    
%%%%%%%%%%%%%%%%%%%%%%%%%%%%%%%%%%%%  
\newpage
\section{Web Table 8}
\label{s:table8}
\begin{table}[h!]
    \centering
    \caption{Bias, standard error (SE) and coverage of 95\% confidence intervals based on the  eight possible inference approaches for our estimator of $\gamma_{I2}$ at different number of level-3 units. We set $(\gamma_{I2},\gamma_{I3})=(0.53,0.19)$, the number of level-2 units in a level-3 unit to follow a uniform distribution between 2 and 15, and the number of level-1 units in a level-2 unit to be 2. There were $200$ replicates per bootstrap. ``Equal level-3 units'' refers to assigning equal weights to level-3 units. ``Equal level-2/1 units'' refers to assigning equal weights to level-2/1 units. ``Asym. SE'' refers to the approach using asymptotic standard error.}
    \begin{subtable}\textwidth
    \centering
        \caption{Coverage of 95\% CIs of $\gamma_{I2}$}
        \begin{tabular}{p{1cm}p{1.2cm}p{1.2cm}p{1.5cm}p{1.7cm}p{1.2cm}p{1.2cm}p{1.5cm}p{1.5cm}p{1.5cm}}
        \hline
        \multicolumn{1}{p{1cm}}{Size} & \multicolumn{4}{c}{Equal level-3 units}  & \multicolumn{4}{c}{Equal level-2/1 units}\\
         & Bias (\%) & Asym. SE & Bootstrap SE & Bootstrap percentiles  & Bias (\%) & Asym. SE & Bootstrap SE & Bootstrap percentiles \\
        \hline
        25 & -1.772 & 0.932 & 0.925 & 0.921 & -1.576 & 0.915 & 0.915 & 0.899\\
        \hline
        50 & -0.655 & 0.948 & 0.946 & 0.939 & -0.689 & 0.952 & 0.946 & 0.934\\
        \hline
        100 & -0.529 & 0.946 & 0.946 & 0.933 & -0.345 & 0.944 & 0.945 & 0.938\\
        \hline
        200 & -0.059 & 0.950 & 0.949 & 0.941 & -0.066 & 0.953 & 0.947 & 0.944\\
        \hline
        500 & -0.108 & 0.951 & 0.953 & 0.942 & -0.077 & 0.950 & 0.949 & 0.941\\
        \hline
        1000 & -0.047 & 0.946 & 0.949 & 0.940 & -0.064 & 0.954 & 0.952 & 0.942\\
        \hline
        \end{tabular}
    \label{tab:tabS7.1}
    \end{subtable}
    \begin{subtable}\textwidth
    \centering
        \caption{Coverage of 95\% CIs of $\gamma_{I3}$}
        \begin{tabular}{p{1cm}p{1.2cm}p{1.2cm}p{1.5cm}p{1.7cm}p{1.2cm}p{1.2cm}p{1.5cm}p{1.5cm}p{1.5cm}}
        \hline
        \multicolumn{1}{p{1cm}}{Size} & \multicolumn{4}{c}{Equal level-3 units}  & \multicolumn{4}{c}{Equal level-2/1 units}\\
         & Bias (\%) & Asym. SE & Bootstrap SE & Bootstrap percentiles  & Bias (\%) & Asym. SE & Bootstrap SE & Bootstrap percentiles \\
         \hline
        25 & -7.010 & 0.898 & 0.891 & 0.864 & -6.003 & 0.875 & 0.867 & 0.844\\
        \hline
        50 & -3.672 & 0.928 & 0.923 & 0.912 & -3.204 & 0.911 & 0.901 & 0.896\\
        \hline
        100 & -2.087 & 0.932 & 0.932 & 0.919 & -1.532 & 0.927 & 0.925 & 0.918\\
        \hline
        200 & -0.273 & 0.944 & 0.939 & 0.938 & 0.113 & 0.935 & 0.929 & 0.922\\
        \hline
        500 & -0.628 & 0.945 & 0.944 & 0.923 & -0.536 & 0.938 & 0.939 & 0.929\\
        \hline
        1000 & -0.439 & 0.954 & 0.953 & 0.942 & -0.239 & 0.954 & 0.950 & 0.947\\
        \hline
        \end{tabular}
    \label{tab:tabS7.2}
    \end{subtable}
    \label{tab:tabS7}
\end{table}  
%%%%%%%%%%%%%%%%%%%%%%%%%%%%%%%%%%%%%%%%%%%%%%%%%%%%%%%%%%%%%%
\newpage
\section{Web Figure 1}
\label{s:figure1}
\begin{figure}[h!]
     \centering
     \begin{subfigure}[b]{0.4\textwidth}
         \centering
         \includegraphics[width=0.7\textwidth]{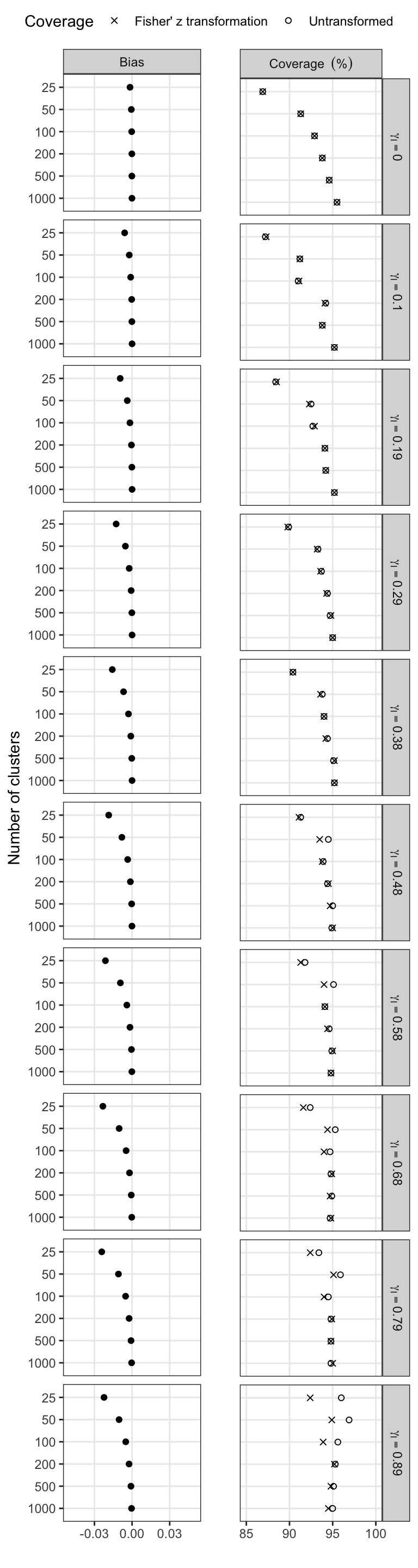}
         \caption{Scenario I and II}
         \label{fig:s1.1s}
     \end{subfigure}\hfill
     \begin{subfigure}[b]{0.4\textwidth}
         \centering
         \includegraphics[width=0.7\textwidth]{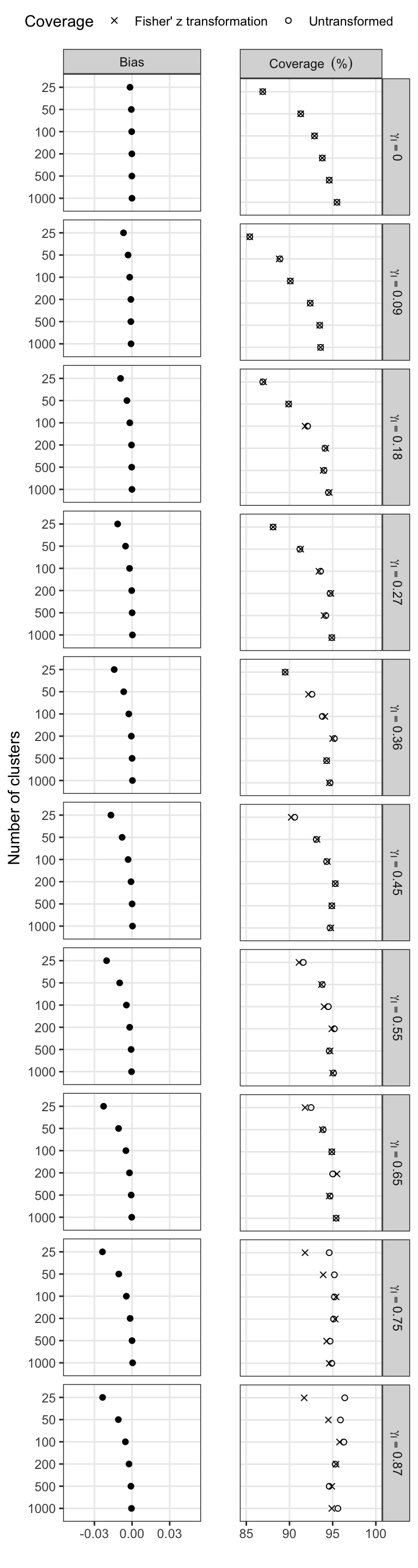}
         \caption{Scenario III}
         \label{fig:s3.1s}
     \end{subfigure}
        \caption{\small{Bias and coverage of 95\% confidence intervals (i.e., based on the asymptotic SE only, based on the asymptotic SE and Fisher' z transformation) for our estimator of $\gamma_I$ at different true values of $\gamma_I$ and different numbers of clusters under Scenarios I (normality), II (exponentiated outcomes), and III (exponentiated cluster means). The number of observations per cluster was set at 30.}}
        \label{fig:fig1s}
\end{figure}
%%%%%%%%%%%%%%%%%%%%%%%%%%%%%%%%%%%%%%%%%%%%%%%%%%%%%%%%%%%%%%
\newpage
\section{Web Figure 2}
\label{s:figure2}
\begin{figure}[h!]
     \centering
     \begin{subfigure}[b]{0.4\textwidth}
         \centering
         \includegraphics[width=0.7\textwidth]{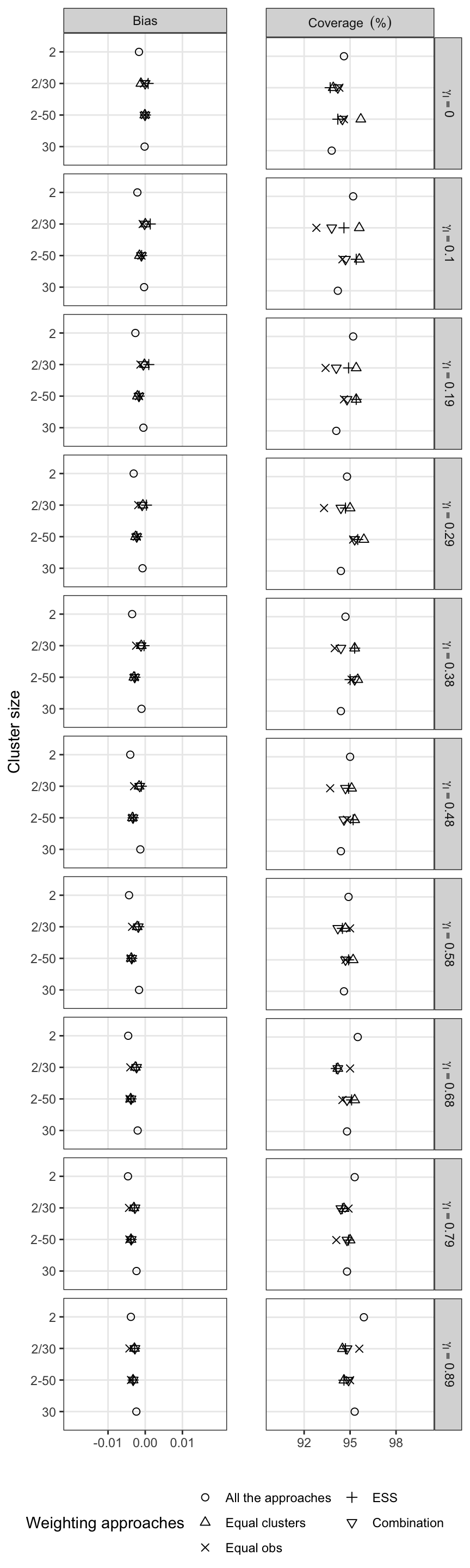}
         \caption{Scenario I and II}
         \label{fig:s1.2s}
     \end{subfigure}\hfill
     \begin{subfigure}[b]{0.4\textwidth}
         \centering
         \includegraphics[width=0.7\textwidth]{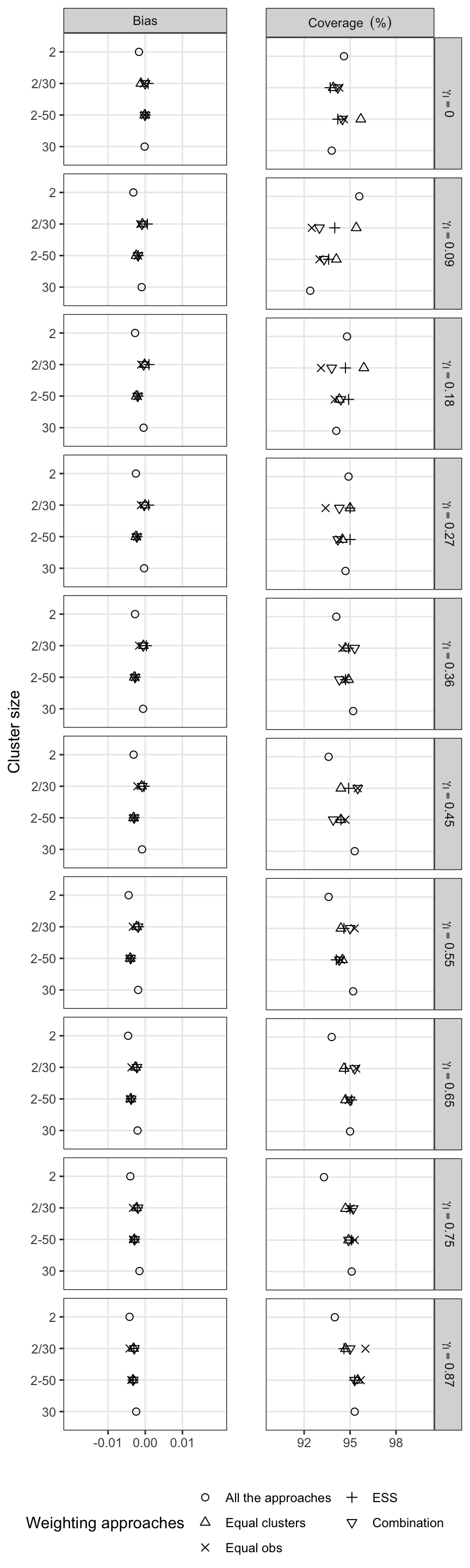}
         \caption{Scenario III}
         \label{fig:s3.2s}
     \end{subfigure}
        \caption{\small{Bias and coverage of 95\% confidence intervals for our estimator of $\gamma_I$ at different true values of $\gamma_I$ and different cluster sizes under Scenarios I (normality), II (exponentiated outcomes), and III (exponentiated cluster means). The number of clusters was set at 200. ``2-50'' means the cluster size follows a uniform distribution from 2 to 50, ``2/30'' means half of the clusters have size 2 and half have 30. ``Equal clusters'' refers to assigning equal weights to clusters, ``Equal obs'' refers to assigning equal weights to observations, ``ESS'' refers to the iterative weighting approach based on the effective sample size, and ``Combination'' refers to the iterative weighting approach based on the linear combination of equal weights for clusters and equal weights for observations. We set the tolerance of the two iterative approaches to be 0.00001. The estimates of the four approaches were identical when cluster sizes were equal.}}
        \label{fig:fig2s}
\end{figure}
%%%%%%%%%%%%%%%%%%%%%%%%%%%%%%%%%%%%%%%%%%%%%%%%%%%%%%%%%%%%%%
\newpage
\section{Web Figure 3}
\label{s:figure3}
\begin{figure}[h!]
     \centering
     \begin{subfigure}[b]{0.4\textwidth}
         \centering
         \includegraphics[width=0.72\textwidth]{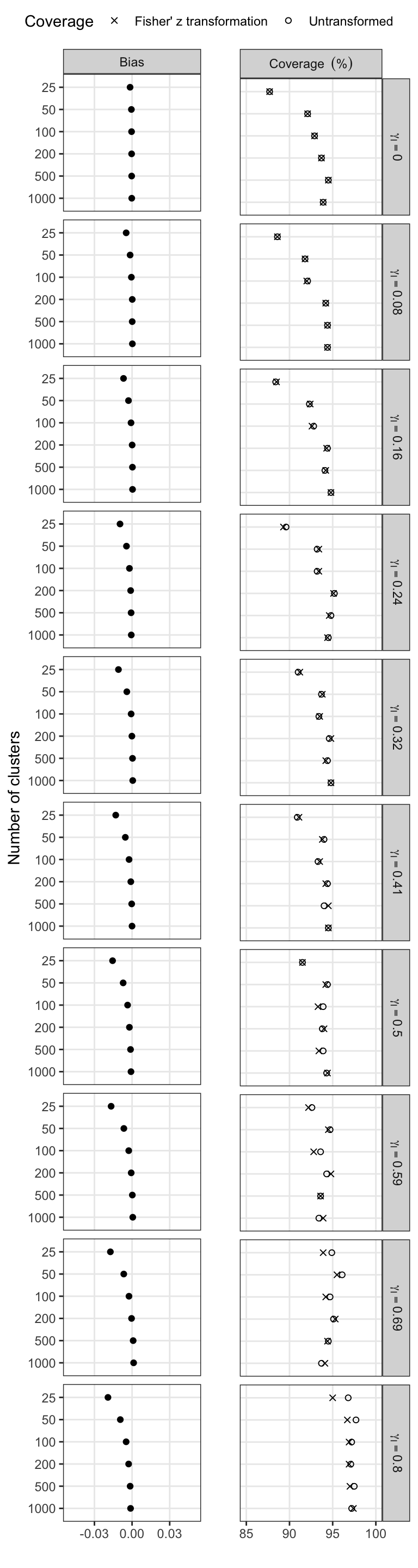}
         \caption{3-level}
         \label{fig:s7.2s}
     \end{subfigure}\hfill
     \begin{subfigure}[b]{0.4\textwidth}
         \centering
         \includegraphics[width=0.72\textwidth]{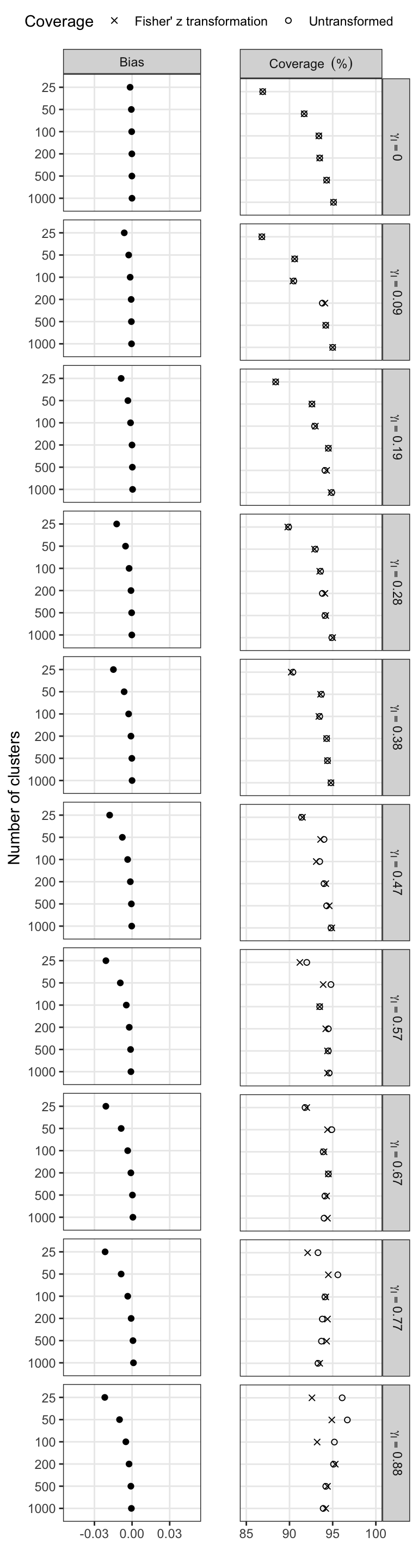}
         \caption{10-level}
         \label{fig:s7.2s}
     \end{subfigure}
        \caption{\small{Bias and coverage of 95\% confidence intervals (i.e., based on the asymptotic SE only, based on the asymptotic SE and Fisher' z transformation) for our estimator of $\gamma_I$ at different true values of $\gamma_I$ of 3-level and 10-level ordered categorical variables. The number of observations per cluster was set at 30.}}
        \label{fig:fig7s}
\end{figure}

%%%%%%%%%%%%%%%%%%%%%%%%%%%%%%%%%%%%%%%%%%%%%%%%%%%%%%%%%%%%%%
\newpage
\section{Web Figure 4}
\label{s:figure4}

\begin{figure}[h!]
    \centering
    \includegraphics[width=0.75\textwidth]{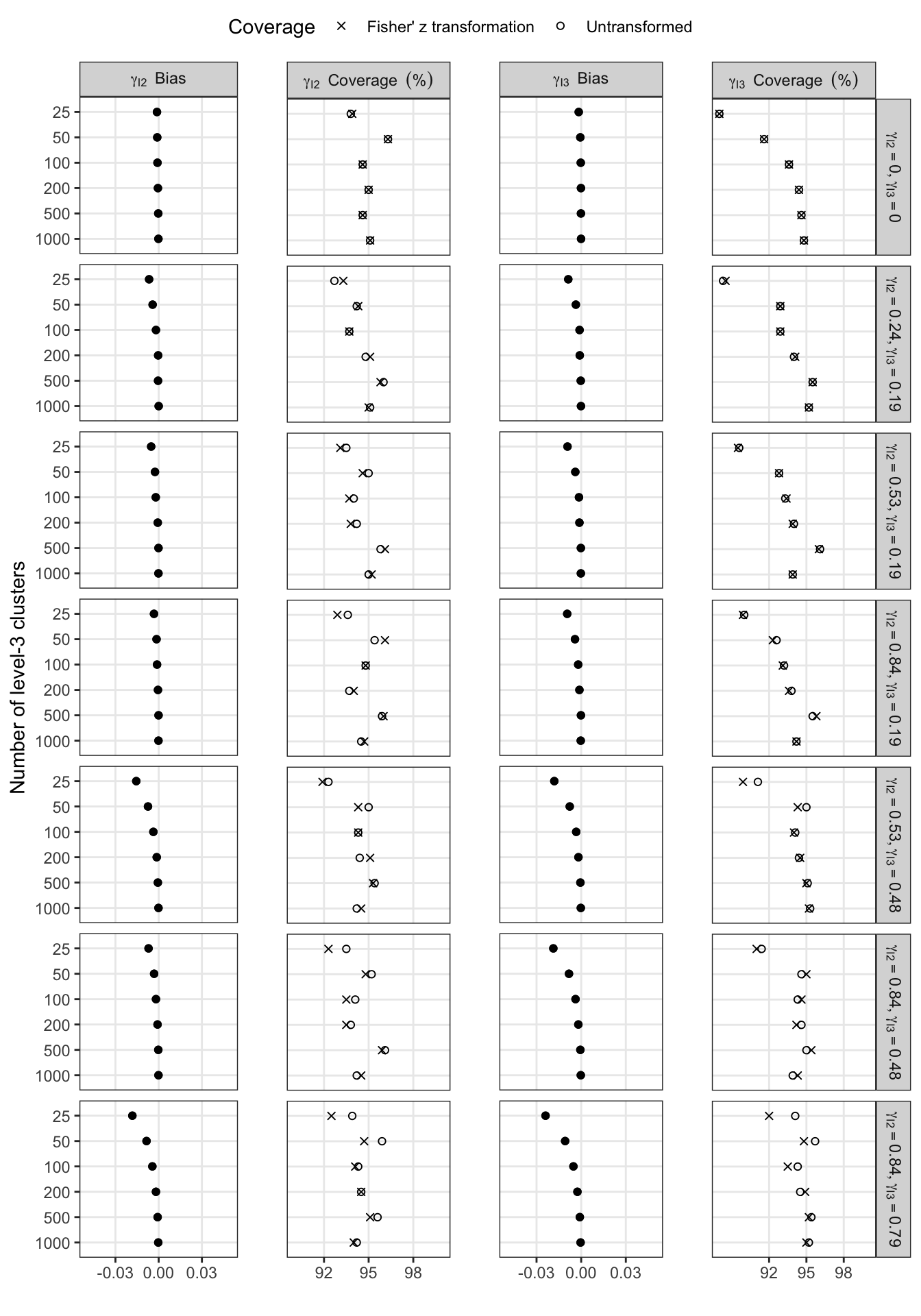}
    \caption[]%
    {\small{Bias and coverage of 95\% confidence intervals (i.e., based on the asymptotic SE only, based on the asymptotic SE and Fisher' z transformation) for our estimators of $\gamma_{I2}$ and $\gamma_{I3}$ at different true values of $\gamma_{I2}$ and $\gamma_{I3}$ and different numbers of level-3 units. The number of level-2 units in a level-3 unit was set at 15. The number of level-1 units in a level-2 unit was set at 2.}}
     \label{fig:fig3.1s}
\end{figure}
%%%%%%%%%%%%%%%%%%%%%%%%%%%%%%%%%%%%%%%%%%%%%%%%%%%%%%%%%%%%%%
\newpage
\section{Web Figure 5}
\label{s:figure5}

\begin{figure}[h!]
        \centering
        \includegraphics[width=0.8\textwidth]{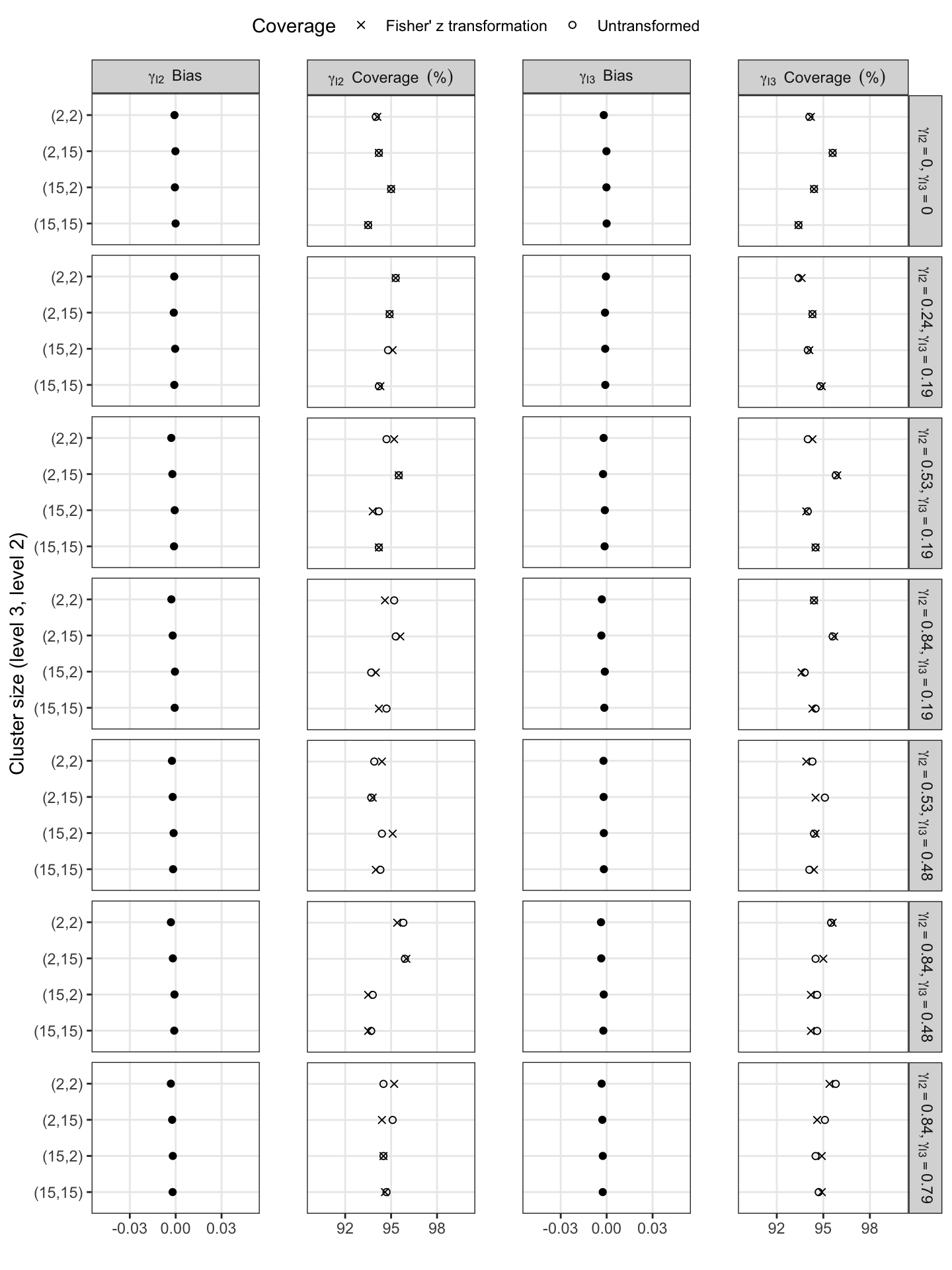}
        \caption[]%
        {\small{Bias and coverage of 95\% confidence intervals (i.e., based on the asymptotic SE only, based on the asymptotic SE and Fisher' z transformation) for our estimators of $\gamma_{I2}$ and $\gamma_{I3}$ at different true values of $\gamma_{I2}$ and $\gamma_{I3}$ and different sizes of level-3 units and level-2 units. The number of level-3 units was set at 200.}}
        \label{fig:fig3.2s}
\end{figure}

%%%%%%%%%%%%%%%%%%%%%%%%%%%%%%%%%%%%%%%%%%%%%%%%%%%%%%%%%%%%%%
\newpage
\section{Web Figure 6}
\label{s:figure6}

\begin{figure}[h!]
     \centering
     \includegraphics[width=0.7\textwidth]{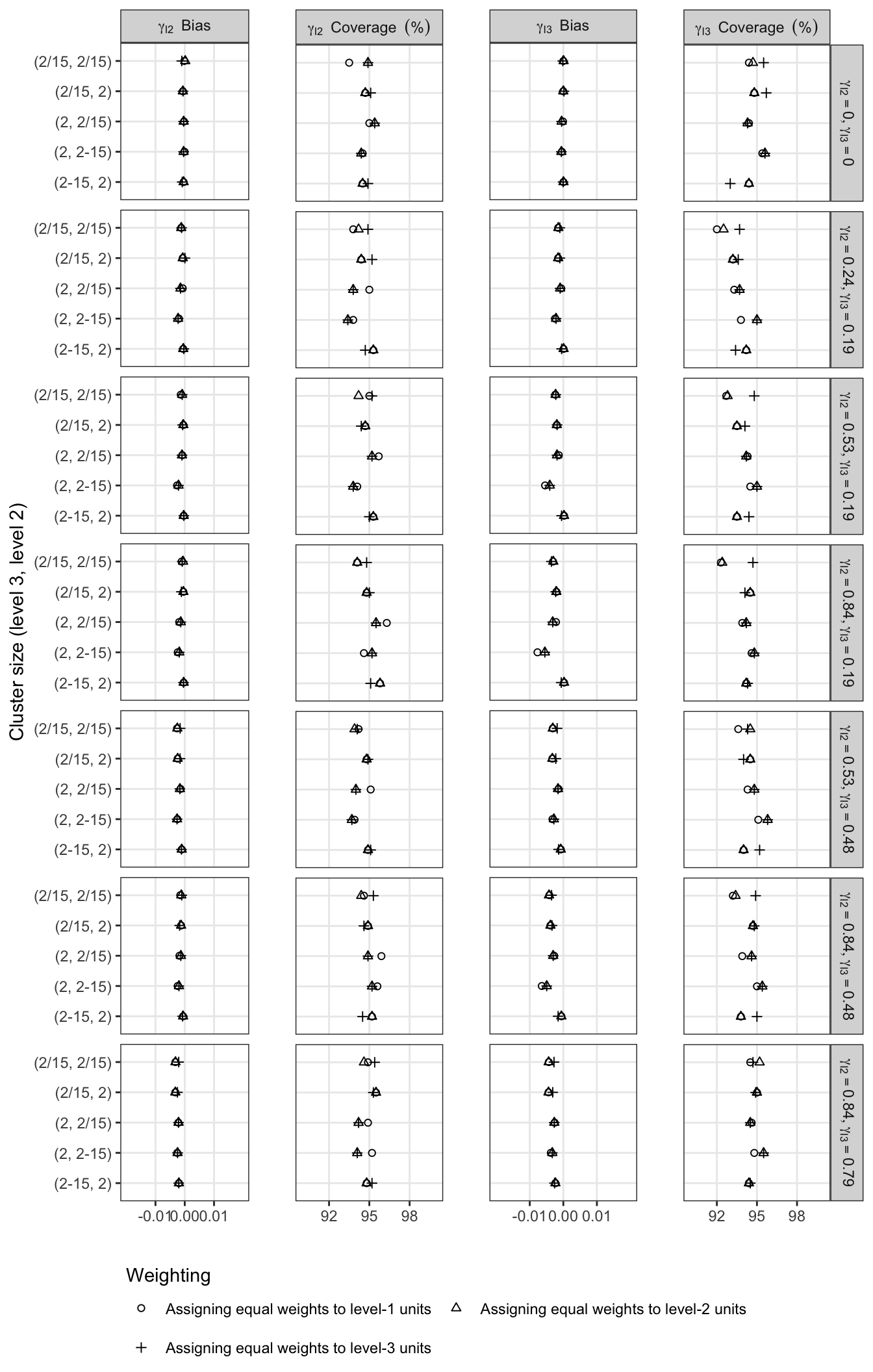}
             \caption{\small{Bias and coverage of 95\% confidence intervals for our estimators of $\gamma_{I2}$ and $\gamma_{I3}$ at different true values of $\gamma_{I2}$ and $\gamma_{I3}$ and different unequal sizes of level-3 units and level-2 units. The number of level-3 units was set at 200. ``2-15'' means the cluster size follows a uniform distribution from 2 to 15, ``2/15'' means half of the clusters have size 2 and a half have 15. }}
        \label{fig:fig3.3s}
\end{figure}

\label{lastpage}